\documentclass[12pt]{article}
\usepackage{epsfig,graphicx}

\begin{document}

\title{A NEW APPROACH TO PHYSICS OF NUCLEI}
\author{E. G. Drukarev, M. G. Ryskin, V. A. Sadovnikova\\
Petersburg Nuclear Physics Institute,\\
Gatchina, St.~Petersburg 188300, Russia
}
\date{}
\maketitle

\begin{abstract}\noindent

We employ the QCD sum rules method for description of nucleons in
nuclear matter. We show that this approach provides a consistent formalism
for solving various problems of nuclear physics. Such nucleon characteristics as the Dirac effective
mass $m^*$ and the vector self-energy $\Sigma_V$ are expressed in terms
of the in-medium values of QCD condensates. The values of these
parameters at saturation density and the dependence on the baryon
density and on the neutron-to-proton density ratio is in agreement with
the results, obtained by conventional nuclear physics method. The
contributions to $m^*$ and $\Sigma_V$ are related to observables and do
not require phenomenological parameters.  The scalar interaction is
shown to be determined by the pion--nucleon $\sigma$-term.  The
nonlinear behavior of the scalar condensate may appear to provide  a
possible mechanism  of the saturation. The approach provided reasonable
results for renormalization of the axial coupling constant, for the
contribution of the strong interactions to the neutron--proton mass
difference and for the behavior of the structure functions of the
in-medium nucleon.  The approach enables to solve the problems which
are difficult or unaccessible for conventional nuclear physics methods.
The method provides guide-lines for building the nuclear forces. The
three-body interactions emerge within the method in a natural way. There
rigorous calculation will be possible in framework of self-consistent
calculation in nuclear matter of the scalar condensate
and of the nucleon effective mass $m^*$.
 \end{abstract}

PACS numbers 21.65.-f, 21.65.Mn, 24.85.+p

\newpage

\begin{center} \bf C o n t e n t s   \end{center}

\begin{enumerate}

\item {\large Introduction}

\item{\large Nucleon QCD sum rules in vacuum}

2.1.~~ General ideas\\
2.2.~~ Explicit form of the SR equations\\
2.3. Inclusion of radiative corrections

\item{\large QCD sum rules in nuclear matter}

3.1.~~ Choice of the variables\\
3.2.~~ Operator product expansion\\
3.3.~~ Model of the spectrum

\item{\large Nucleon self-energies in the lowest orders of OPE}

4.1.~~ General equations\\
4.2.~~ Left-hand sides of the sum rules\\
4.3.~~ Approximate solution\\
4.4.~~ Gas approximation\\
4.5.~~ Asymmetric matter\\
4.6.~~ Possible mechanism of saturation\\

\item{\large Other characteristics of the in-medium nucleons}

5.1.~~ Axial coupling constant \\
5.2.~~ Charge-symmetry breaking forces\\
5.3.~~ Nucleon deep inelastic structure functions\\

\item{\large Intermediate summary}

6.1.~~ Reasons for optimism\\
6.2.~~ Reasons for scepticism\\
6.3.~~ Response to the sceptical remarks

\item{\large Four-quark condensates}

7.1.~~ General equations for contribution of the four-quark condensates
\\
7.2.~~ Approximations for the four-quark condensates\\
7.3.~~ Perturbative Chiral Quark Model\\
7.4.~~ Four-quark condensates in the PCQM\\

\item{\large Contribution of the higher order terms}

8.1.~~ Symmetric matter with the four-quark condensates\\
8.2.~~ Radiative corrections\\
8.3.~~ Asymmetric matter\\
8.4.~~ Many-body interactions

\item{\large Self-consistent scenario}

\item{\large Summary}

\item{\large Epilogue}

\end{enumerate}

\section{Introduction}
In the present paper we review our approach to description of a
nucleon, placed into nuclear matter. The in-medium characteristics of
nucleon can tell us much about the medium itself. On the other hand,
the introduction of nuclear matter enables to separate the problems of
nucleon interactions from those, connected with individual features of
specific nuclei. Thus investigation of nucleon characteristics in
nuclear matter is an important step for studies of physics of nuclei.

 Until mid 70-th the studies of nuclear matter were based on the
nonrelativistic approach.  Since the publication of the paper \cite{1}
description of the in-medium nucleon was based on Dirac phenomenology.
It was successful in describing most of characteristics of nucleons
both in nuclear matter and in finite nuclei \cite{2,3}. In the meson
exchange picture the vector and scalar fields correspond to exchange by
the vector and scalar mesons between a nucleon and the nucleon of the
matter. This picture is called Quantum Hadrodynamics (QHD). In the
simplest version (QHD-I) only vector $\omega$ and scalar $\sigma$
mesons are involved. In more complicated versions some other mesons are
included \cite{2,4}. While the studies in framework of the
nonrelativistic approach are going on, i.e. the applications of the
Nuclear Density Functional Method provided very accurate description of
the data \cite{5,6}.

On the other hand, many efforts have been made to improve the QHD.
There were many reasons for this. First of all, the QHD itself has
several weak points.  It is not clear, if the scalar $\sigma$ meson
does exist, the experimental data are controversial. It is rather an
effective way of describing the two-pion exchange. The mass of this
effective state is about 500~Mev. The mass of vector $\omega$ meson is
 about 780~MeV.  Hence, the exchange by theses mesons takes place at
the distances, where the nucleon can not be treated as a point -like
particle.  Thus the QHD inherited the problems of nonrelativistic
phenomenology, connected with description of the interaction at small
distances. The other weak points are discussed in \cite{7,8}. Also, it
is desirable to match the QHD with the Quantum Chromo\-dynam\-ics (QCD )
which is believed to be a true theory of strong interactions. Another
reason is that there are various problems in nuclear physics. It is
desirable to have an approach, which would enable to calculate:

\begin{itemize}
\item The nucleon single-particle potential energy $U(\rho)$, where
$\rho$ in the density of the baryon quantum number.  This enables one
to find the saturation density $\rho_0$ and the single-particle binding
energy $\varepsilon(\rho)$.

\item Parameters of interaction with external fields. These are
magnetic moments $\mu(\rho)$ and the axial coupling constant
$g_A(\rho)$. The latter is important for understanding of the chiral
properties of the matter.

 \item Neutron--proton mass splitting in isotope-symmetric matter. It
was observed for a number of nuclei, being known as the Nolen--Schiffer
anomaly.

\item Structure functions of the deep inelastic scattering. They
describe the internal structure of nucleons. Investigation of the
latter is important for construction of the quark models of nucleons,
for studies of the confinement, etc. The reasons for some of the
in-medium modifications of the structure functions are still obscure.
They become a subject for discussions from time to time.

\item The single particle potential energy for hyperons in nuclear
matter. Can a system of hyperons be stable? In other words, can a
strange matter exist\,?
 \end{itemize}

As we said earlier, the first problem was solved in framework of Dirac
phenomenology \cite{1,2} for the values of density close to the
saturation value. The nucleon was considered as a relativistic
particle, moving in superposition of vector and scalar fields $V_{\mu}$
and $\Phi$. In the rest frame of the matter $v_{\mu}=V_0\delta_{\mu
0}$. The dynamics of the nucleon  is described by equation
\begin{equation}
(\hat q-\hat V)\psi\ =\ (m+\Phi)\psi\,,
\label{1} \end{equation}
with $\hat A= A_{\mu}\gamma^{\mu}$, $q_{\mu}=-i\partial_{\mu}$.
In nuclear matter the fields
$V$ and $\Phi$ depend only on the density $\rho$, and do not depend on
the space coordinates. The values of the fields are adjusted to
reproduce either the data on nucleon--nucleon scattering or the nuclear
data -- see, e.g. \cite{3}. However, each of the other listed problems
requires additional improvements of the QHD. Turning to the other
problems from the above list, note that the axial coupling constant
changes due to  polarization of medium by pions \cite{9}, with a
crucial role of the delta-isobar excitations \cite{10}. Thus, in order
to solve the second problem one should introduce additional degrees of
freedom into the QHD. Also, the third problem requires introduction of
the nuclear forces, which break the isospin invariance \cite{11}.
Finally, the fourth problem is just unaccessible for traditional
methods of nuclear physics.

There were several attempts to combine the QHD and the quark structure
of nucleons. The nucleon-nucleon forces, constructed within such models
enabled to reproduce semi-quantitatively or quantitatively the nucleon
characteristics in nuclear matter. In the Quark--Meson Coupling (QMC)
model \cite{12} the nucleon was considered as a three-quark system in a
bag. The quarks were directly coupled to $\sigma$ and $\omega$ mesons.
The values of the nucleon effective mass $m^*$ and of the in-medium
coupling constant $g_A$ appeared to be somewhat smaller than in QHD
\cite{13}. This model is employed nowadays as well \cite{14}.

In another class of works the short-range interaction between the
nucleons was treated as interaction between their quarks. The latter
were described in framework of some QCD motivated models. The
long-range NN interaction was described in terms of nucleons and pions.
These works were reviewed in \cite{15}.

During the two latest decades much work was done on development of the
Effective Field Theory (EFT). The starting point is the most general
Lagrangian, which includes nucleons and pions as the degrees of freedom
and respects all the symmetries of QCD, i.e. the Lorentz invariance,
chiral symmetry, etc. \cite{16}. The applications of the EFT is usually
combined with the expansion in powers of the pion mass (Chiral
Perturbation Theory). Expansion in powers of low momenta is also
usually carried out.  The works, based on the EFT are reviewed in
\cite{17}, for some of recent developments - see \cite{18}. The EFT
approach does not touch the quark degrees of freedom of the nucleons.

All the traditional nuclear physics approaches face difficulties in
attempts to describe  the nucleon-nucleon (NN) interaction at small
distances. On the other hand, the Quantum Chromo\-dynamics (QCD) is
believed to be a true theory of strong interactions. It has many
unsolved problems at large distances. However, it becomes increasingly
simple at small distances due to the asymptotic freedom. The
interaction of quarks and gluons, which are the ingredients of QCD can
be treated perturbatively.  This is known as the asymptotic freedom
\cite{18a}. It is tempting to use this feature of QCD in building the
nucleon forces.  One should take into account, however, that due to
spontaneous breakdown of the chiral symmetry of the QCD, the vacuum
expectations of some QCD operators (condensates) have nonzero values.

In medium with a nonzero value of the density of the baryon quantum
number the condensates change their values. Also, some other
condensates, which vanish in vacuum, obtain nonzero values.

The QCD sum rules (SR) method describes the vacuum hadron parameters
basing on the quark dynamics at short distances, where the asymptotic
freedom works. In other words, the dynamics of the quark system at the
distances of the order of the confinement  radius  is described basing
on that at small distances, where it is determined by the QCD
condensates.  Thus, the SR method enables to describe the hadron
parameters in terms of the QCD condensates.

The approach was worked out in \cite{19}, where it was used for
mesons. It is based on the dispersion relation for the function,
describing the system which carries the quantum numbers of the hadron.
The SR method was successfully applied for nucleons in vacuum \cite{20},
describing  all their static and some of dynamical characteristics
-- see \cite{20a,21} for a review.  Thus it looks reasonable to try to
apply the QCD SR method for the description of nucleons in nuclear
matter.  It was suggested in \cite{22,23,24} that the parameters of
nucleon in nuclear matter can be expressed in terms of the in-medium
values of QCD condensates. In medium with a nonzero value of the
density of the baryon quantum number the condensates change their
values. Also, some other condensates, which vanish in vacuum, obtain
nonzero values.

The generalization of the SR method for the case of finite densities
was not straightforward. One of the main problems was the choice of
variables, which enabled to separate the singularities connected with
the in-medium nucleon from those connected with the medium itself. This
was done in \cite{22}--\cite{24}.

It was found also in these papers that the nucleon characteristics
(effective mass $m^*$ and the vector self-energy $\Sigma_V$) can be
presented in terms of the vector and scalar condensates
\begin{equation}
v(\rho)=\langle M|\sum_i\bar q^i\gamma_0 q^i|M\rangle; \quad
\kappa(\rho)=\langle M|\sum_i\bar q^iq^i|M\rangle\,.
\label{2} \end{equation}
 Here $\rho$ and $\langle M|$ are the density and vector of the ground state of the matter, $q^i$ is the quark field, the summation over flavors $i=u,d$ is carried out. The vector condensate is written in the rest frame of the matter.
Due to conservation of the vector current the vector condensate is a linear function of $\rho$
\begin{equation}
v(\rho)=v_N\rho; \quad  v_N=\langle N|\sum_i \bar q^i\gamma_0 q^i|N\rangle=3
\label{3} \end{equation}
is just the number of valence quarks in a nucleon. The scalar
condensate can be represented as \cite{22,23}
\begin{equation}
\kappa(\rho)=\kappa(0)+\kappa_N\rho +S(\rho), \quad \kappa_N=\langle
N|\sum_i \bar q^i q^i|N\rangle\,,
\label{4} \end{equation}
with $S(\rho)$ caused by interaction of the nucleons of the matter.
Since $\kappa_N$ can be expressed in terms of the pion--nucleon
 $\sigma$ term \cite{25}, while the latter is related to observables
\cite{25a,26}, one can obtain the values of the nucleon parameters at
 least in the gas approximation. Since $S(\rho)$ is small for the densities, close to the
 saturation point (see below), the values obtained in such approach are
 close to the physical ones. They appeared to be  $\Sigma_V\approx200\,$MeV,
and $m^*-m\approx-300\,$MeV close to the saturation
 point.  Thus, including only the condensates of the lowest dimension
 and neglecting the radiative corrections, we found that the  QCD SR
 method reproduce the main features of the QHD \cite{1}.

However, the role of the condensates of higher dimension remained
 obscure. The contribution of the gluon condensate $\langle
M|\frac{\alpha_s}{\pi}G^{a\mu \nu}G^a_{\mu \nu}|M\rangle$ is rather
small.  However, an estimation for the value of the four-quark
condensate $\langle M|\bar qq \bar qq|M\rangle$, which suggests itself,
destroys the agreement with the Walecka model. Also, the lowest order
radiative corrections are numerically large. This took place for the
vacuum as well. This caused doubts in possibility to expand the SR
method for the case of finite densities.

The four-quark condensates are the most important among those of the
higher dimension. Their calculation requires some model assumption on
the quark structure of nucleon. We employed the Perturbative Chiral
Quark Model (PCQM), suggested originally in \cite{27}. We calculated
these condensates \cite{28} and found that previous naive estimation of
its contribution was wrong, due to some cancelations which take place
in any reasonable model.We demonstrated that  the SR method provides
the dependence of the  nucleon characteristics $m^*$ and $\Sigma_V$ on
the density $\rho$ and on the neutron-to-proton density ratio
\cite{29,30}, which is consistent with the results obtained by using
traditional nuclear physics methods.

We analyzed the role of radiative corrections for the nucleon SR in
vacuum and demonstrated that their influence on the value of the
nucleon mass is small \cite{31}. Also, in nuclear matter the radiative
corrections do not change much the nucleon characteristics $m^*$ and
$\Sigma_V$ \cite{32}.

We found that the nonlinear contribution to the scalar condensate
$S(\rho)$ is determined mainly by the pion contribution to the
self-energy of the nucleon of the matter. Simple estimations show that
this term may provide a saturation mechanism in our approach \cite{33}.
However, a rigorous treatment requires renormalization of the pion
propagator by the particle-hole excitations. The renormalized pion
propagator depends on the nucleon effective mass $m^*$ and on the
in-medium value of the pion--nucleon coupling constant $g_s$.The latter
can be obtained by the SR method. This brings us to a self-consistent
scenario. At the present level of our knowledge it requires some more
phenomenological assumptions \cite{34}.

Note also, that the finite density SR method provided reasonable
results for the axial coupling constant \cite{35}, for the
neutron--proton mass splitting \cite{36} and for the difference between
the deep inelastic structure of nucleus and that of sum of those of
free nucleons\,\cite{37}. However, in these calculations only the
condensates of the lowest dimensions have been included.

Now we give the details.

\section{Nucleon QCD sum rules in vacuum}

This approach is described in details in many publications. Earlier
papers are reviewed in \cite{38}. The nowadays state of art is
presented in \cite{20a} and \cite{21}, see also the book \cite{18a}.
However, to make the text self-consistent we recall the main points of
the approach. We emphasize the points, which we shall need for the
extension of the SR method for the case of finite density of the baryon
quantum number.

\subsection{General ideas}

The nucleon QCD sum rules  succeeded in describing  of the nucleon
characteristics in vacuum in terms of the vacuum expectation values of
 the products of quark or (and) gluon operators (QCD condensates)
 \cite{7,8}. This  approach is based on the dispersion relation for the
 function
$$
 \Pi_0(q)=\hat q \Pi^q_0(q^2)+I\Pi^I_0(q^2),
 $$
describing the propagation of the system which carries the quantum
 numbers of the proton, $I$ is the unit $4\times 4$ matrix. In the simplest form the dispersion relations
 are
 \begin{equation}
\Pi^i_0(q^2)=\frac1\pi\int dk^2\frac{\mbox{Im}\Pi^i_0(k^2)}{k^2-q^2}\,;
 \quad i=q,I.
\label{5}
\end{equation}
As we shall see below, we do not need to worry about possible subtractions.

In quantum mechanics $\Pi_0(q^2)$ is just the proton propagator. In
the field theory different degrees of freedom are important in
different regions of the value of $q^2$.

One can consider the proton as a system of three strongly interacting
quarks. Due to asymptotic freedom of QCD \\cite{18a} the description becomes
increasingly simple at $q^2\to-\infty$. This means that at
$q^2\to-\infty$ the function $\Pi_0(q^2)$ can be presented as
a power series of $q^{-2}$ and of the QCD coupling constant $\alpha_s$.
The coefficients of the expansion in powers of $q^{-2}$ are the QCD
condensates. Such presentation known as the Operator Product Expansion
(OPE) \cite{39} provides the perturbative expansion of the short
distance effects, while the nonperturbative physics is contained in the
condensates.  In QCD SR approach the left-hand side (LHS) of
Eq.(\ref{5}) is considered at $q^2\to-\infty$, and several
lowest order terms  of the OPE are included. In this and in the next
Subsections we neglect the radiative corrections, i.e. we do not
include interactions and self-interactions (self-energy insertions) of
the quarks, putting $\alpha_s=0$.

Turning to the right-hand side (RHS) of Eq.~(\ref{5}) note that
Im$\Pi_0(k^2)=0$ at $k^2<m^2$ with $m$ being the position of the lowest
lying pole, i.e. $m$ is the proton mass. There are the other
singularities at larger values of $k^2$. These are the cuts
corresponding to the systems ``proton+pions", the pole N(1440), etc. The
next to leading singularity is the physical branching point
$k^2=W_{phys}^2$, and one can write
\begin{equation}
\mbox{Im}\Pi_0^i(k^2)=\lambda_N^2\xi^i
\delta(k^2-m^2)+f^i(k^2)\theta(k^2-W_{phys}^2); \quad i=q,I, \label{6}
\end{equation}
with $\xi^q=1$, $\xi^I=m$; $\lambda_N^2$ -- the residue at
the pole. Now Eq.~(\ref{5}) takes the form
\begin{equation}
\Pi^{i~OPE}_0(q^2)=\frac{\lambda_N^2\xi^i}{m^2-q^2}+\frac1\pi
\int\limits_{W^2_{phys}}^{\infty} dk^2\frac{f^i(k^2)}{k^2-q^2}; \quad
i=q,I.
\label{7} \end{equation}
The upper index OPE means that several lowest OPE terms are included.

The SR approach is focused on studies of the lowest state. Thus we
keep the first term on the RHS of Eq.~(\ref{7}) and try to write
approximate expression for the contribution of the higher states. The
detailed structure of the spectral function $f(k^2)$ in the second term
on the RHS of Eq.~(\ref{7}) cannot be obtained by the SR
method. However, at $k^2 \gg |q^2|$
\begin{equation}
f^i(k^2)=\frac{\Delta \Pi_0^{i~OPE}(k^2)}{2i}\,,
\label{8} \end{equation}
with $\Delta$ denoting the discontinuity.

The standard ansatz consists in extrapolation of Eq.~(8) to the lower
values of $k^2$, assuming that the cut starts at certain unknown point
$W^2$. In other words
$$
\frac1\pi\int_{W^2_{phys}}^{\infty} dk^2\frac{f^i(k^2)}{k^2-q^2}=
\frac{1}{2\pi i}\int_{W^2}^\infty dk^2
\frac{\Delta\Pi_0^{i~OPE}(k^2)}{k^2-q^2}\,,
$$
and thus Eq. (5) takes the form
\begin{equation}
\Pi^{i~OPE}_0(q^2)=\frac{\lambda_N^2\xi^i}{m^2-q^2}+\frac1{2\pi i}
\int\limits_{W^2}^\infty dk^2\frac{\Delta\Pi_0^{i~OPE}(k^2)}{k^2-q^2};
\quad i=q,I.  \label{9}
\end{equation}
Recall that $\xi^q=1$,\ $\xi^I=m$.

This ``pole+continuum"  presentation of the RHS makes sense only if its
first term, treated exactly is larger than the second term, which
approximates the higher states.  The position of the lowest pole $m$,
its residue $\lambda^2$ and the continuum threshold $W^2$ are the
unknowns in  Eqs.~(\ref{9}).

In the next step one usually applies the Borel transform. It is defined as
\begin{eqnarray}
BF(Q^2)&=& \lim_{Q^2,n\to\infty}=\frac{(Q^2)^{n+1}}{n!}
\Big(-\frac{d}{dQ^2}\Big)^n F(Q^2)\equiv\tilde F(M^2);
\nonumber\\
&& Q^2=-q^2, \quad M^2=Q^2/n,
\label{10}
\end{eqnarray}
converting a function of $q^2$ into the Borel transformed function of
the Borel mass $M^2$. The reasons for applying the transform are
\begin{itemize}

\item  Since $B (Q^2)^k =0$ for any integer $k$, it kills the
polynomials of $q^2$. Hence, it eliminates the divergent terms in the
function $\Pi^i_0$. This explains, why we wrote the dispersion relation
(5) without subtractions.

\item It emphasizes the contribution of the lowest state, since
$$
B \frac{1}{Q^2+m^2}= e^{(-m^2/M^2)}\,.
$$

\item It improves the convergence of the OPE series, since
$$ B[(Q^2)^{-n}]=\frac{(M^2)^{1-n}}{(n-1)!}\,.
$$
\end{itemize}

The Borel transformed SR take the form
\begin{equation}
\tilde\Pi^{i~OPE}_0(M^2)=\lambda_N^2 \xi^i e^{(-m^2/M^2)}+
\frac1{2\pi i}\int\limits_{W^2}^\infty dk^2\Delta\Pi_0^{i~OPE}(k^2)
e^{(-k^2/M^2)}.
\label{11} \end{equation}
If both LHS and RHS of Eq.~(\ref{11}) were calculated exactly, the
relation would be independent on $M^2$. However, certain approximations
are made on both sides. The analytical dependence of both sides on
$M^2$ is quite different. The OPE on the LHS becomes increasingly true
at large $M^2$. The accuracy of the ``pole+continuum" model used for
the RHS increases at small $M^2$. An important assumption is that there
is an interval of the values of $M^2$, where the two sides have a good
overlap, approximating also the true functions $\tilde \Pi_0^i(M^2)$.

Thus our task is to find the region of the values of $M^2$, where the
overlap can be achieved and to obtain the characteristics of the lowest
state $m$ and $\lambda^2$ and the value of the continuum threshold
$W^2$.

\subsection{Explicit form of the SR equations}

The general definition of the function $\Pi_0(q^2)$, which is
sometimes called ``correlator"  or the "correlation function"  is
\begin{equation}
\Pi_0(q^2)=i\int d^4xe^{i(q\cdot x)} \langle 0|T[j(x)\bar j(0)]|0 \rangle,
\label{12}
\end{equation}
where $j$ is the local operator with the proton quantum numbers. It is
often refereed to as ``current". It was shown in \cite{33} that there
are three independent currents
\begin{equation}
j_1=(u^T_aC\gamma_{\mu}u_b)\gamma_5 \gamma^{\mu}d_c \varepsilon^{abc},
\quad j_2=(u^T_aC\sigma_{\mu,\nu}u_b)\gamma_5 \sigma^{\mu,\nu}
d_c\varepsilon^{abc},
\label{13} \end{equation}
$$
j_{3\mu}=\left[(u^T_aC\gamma_{\mu}u_b)\gamma_5 d_c-(u^T_aC\gamma_\mu
d_b) \gamma_5 u_c\right]\varepsilon^{abc}.
$$
Here $a,b,c$ are the
color indices, $C$ is the charge conjugation matrix,  while $T$ denotes
the transpose in the Dirac space.  It was shown in \cite{40} that the
operators  $j_{2}$ and $j_{3 \mu}$ provide strong admixtures of the
states with negative parity and of the states with spin 3/2
correspondingly. Thus, the calculations with  the operator $j_1$ are
most convincing. We shall assume $j=j_1$ in further studies.

Expansion of the matrix element on the RHS of Eq.~(\ref{12}) in powers
of $x^2$ corresponds to expansion of its LHS in powers of $q^{-2}$. In
the lowest orders of $x^2$ expansion the $T$ product on the RHS of
Eq.~(\ref{12}) can be written in terms of those of two quark operators.
The latter can be written, following the Wick theorem
\begin{equation}
\langle0|T\left[q_i^a(x)\bar q_j^b(0)\right]|0 \rangle=g_q(x)-\frac
{1}{12}\sum _{A}\Gamma_{ij}^A \delta^{ab}\langle 0|\bar q\Gamma^A
q|0\rangle +O(x^2).
\label{14} \end{equation}
Here
\begin{equation}
g_q(x)\ =\ \frac{i}{2\pi^2}\frac{(\hat x-im_qx^2/2)_{ij}}{x^4}
\delta^{ab} \label{15}
\end{equation}
is the
free quark propagator, $m_q$ is the quark mass, $\Gamma^A$ are the
basic $4\times4$ matrices with the scalar, pseudoscalar, vector,
pseudovector  and tensor structures, i.e. $I, \gamma_5,
\gamma_{\alpha}, \gamma_5 \gamma_{\alpha}$ and $\sigma_{\alpha \beta}$.
For the fields which respect the chiral invariance all the expectation
values on the RHS of Eq.~(\ref{14}) vanish. However in the QCD the
expectation value $\langle 0|\bar q q|0\rangle$ ($\Gamma_A=I$) has a
nonzero value. All the other condensates $\langle 0|\bar q\Gamma^A
q|0\rangle$ vanish due to invariance of vacuum. Thus
\begin{equation}
\langle 0|T\left[q_i^a(x)\bar q_j^b(0)\right]|0 \rangle=g_q(x)-\frac
{1}{12}I_{ij}\delta^{ab}\langle 0|\bar q q|0\rangle +O(x^2).
\label{16} \end{equation}

Now we present
$$
\Pi^q_0(q^2)=\sum_n A_n\,; \quad  \Pi^I_0(q^2)=\sum_n B_n\,,
$$
with $n$ denoting the dimension of the condensate, contributing to the
term $A_n$ or $B_n$.

If all expectation values of the $T$ products of the quark fields are
described by the free propagators (\ref{15}), we find the leading OPE
contribution $A_0$ to the structure  $\Pi^q_0(q^2)$, corresponding to
the free three-quark loop. If one of the quark pairs is described by
the second term on the RHS of Eq.~(14), while the others are given by
the first term, we find the leading contribution $B_3$ to the structure
$\Pi^I_0(q^2$). In this case two quarks form a free loop, while the
current exchanges by a quark--antiquark pair with vacuum.

Direct calculation provides \cite{20}
\begin{equation}
A_0=-\frac{q^4\ln(-q^2/L^2_q)}{64 \pi^4}\,; \quad
B_3=\frac{\langle 0|\bar dd|0\rangle q^2\ln(-q^2/L^2_q)}{4\pi^2}\,.
\label{17}
\end{equation}
Here $L_q$ is the cutoff of the integral over $x$ in Eq.~(\ref{12}).
Its value is not important, since the terms containing $L_q$ will be
eliminated by the Borel transform.

Going beyond the single-particle presentation (\ref{14}) one finds
also the next to leading OPE terms.  For example, the contribution
$A_4$ is due to the lowest order interaction of the quark system with
the gluon condensate
$$
\langle 0|\frac{\alpha_s}{\pi}G^{a\mu \nu}G^a_{\mu \nu}|0\rangle=
-2\langle 0|\frac{\alpha_s}{\pi}(E^2-B^2)|0\rangle,
$$
with $E$ and $B$ the color-electric and color-magnetic fields. This
condensate also has a nonzero value only due to the violation of the
chiral symmetry in the ground state of the QCD. The higher terms
contain the four-quark condensate
\begin{equation}
A_6\ =\ -\frac{2\langle 0|\bar q q \bar q q|0\rangle}{3q^2}
\label{18} \end{equation}

Finally, the Borel-transformed sum rules can be written as
\begin{equation}
{\cal L}_0^q(M^2, W^2)=R^q(M^2); \quad {\cal L}_0^I(M^2, W^2)=R^I(M^2).
\label{19}
\end{equation}
Here
\begin{equation}
R^q(M^2)=\lambda^2e^{-m^2/M^2}; \quad R^I(M^2)=m\lambda^2e^{-m^2/M^2},
\label{20}
\end{equation}
with $\lambda^2=32\pi^4\lambda_N^2$. The factor $32\pi^4$ is
introduced in order to deal with the values of the order of unity (in
GeV units). The contribution of continuum is moved to the LHS of
Eqs.~(\ref{19}) (see Eq.~(\ref{22}) below). Following \cite{20, 41} we
can write
\begin{equation}
{\cal L}^q=\tilde A_0(M^2, W^2)+ \tilde A_4(m^2,W^2)+\tilde A_6(M^2) ;
\quad {\cal L}^I=\tilde B_3(M^2, W^2)+ \tilde B_7(M^2).
\label{21}
\end{equation}
The terms $\tilde A_n$ and $\tilde B_n$ are the Borel transforms of
the contributions $A_n$ and $B_n$ to the functions $\Pi_0^{q,I}(q^2)$,
with subtraction of the corresponding contributions of continuum.
Recall that the lower index denotes the dimension of the condensate.
The terms, proportional to condensates of higher dimension $A_6$ and
$B_7$ do not contain the logarithmic loops and thus do not contribute
to continuum.  Actually the  calculations are carried out in the chiral
limit $m_q=0$. The explicit form of the contributions is  \cite{20,41}
\begin{eqnarray}
&& A_0=M^6E_2\left(\frac{W^2}{M^2}\right);\quad A_4=\frac{bM^2E_0
(\frac{W^2}{M^2})}4;\quad  A_6=\frac43 a^2;
\nonumber\\
&& B_3=2aM^4E_1\left(\frac{W^2}{M^2}\right); \quad
B_7=-\frac{ab}{12}\,.  \label{22}
\end{eqnarray}
Here
\begin{equation}
E_0(x)=1-e^{-x}, \quad E_1(x)=1-(1+x)e^{-x}, \quad E_2(x)=1-(1+x+x^2/2)e^{-x},
\label{23} \end{equation}
while
\begin{equation}
a=-(2\pi)^2\langle 0|\bar q q|0\rangle ; \quad b=(2\pi)^2\langle 0|\frac{\alpha_s}{\pi}G^{a\mu \nu}G^a_{\mu \nu}|0\rangle.
\label{24} \end{equation}
The contributions are illustrated by Fig.~1.
Note that we provided these equations mostly as illustration of the
main ideas and did not include several numerically not very important
terms.

The term $\tilde A_6$ presents the contribution of the four-quark
condensates $\langle 0|\bar q\Gamma^A q\bar q\Gamma^A q|0\rangle$,
which, generally speaking, obtain nonzero values for all structures
$\Gamma_A$.  It is evaluated under the factorization approximation
\cite{6}
$$
\langle 0|\bar q\Gamma^A q\bar q\Gamma^A q|0\rangle=\frac1{16}
(\langle 0|\bar q q|0\rangle)
\left[(\mbox{Tr}\Gamma_A)^2-\frac13\mbox{Tr}(\Gamma_A^2)\right].
$$

One can find numerical values of the main QCD condensates presented by
Eq.~(\ref{24}). There is the well known Gell-Mann--Oakes--Renner
relation (GMOR) for the scalar condensate \cite{42}
\begin{equation}
\langle 0|\bar uu+\bar dd|0\rangle\ =\ -\frac{2f_\pi^2m_\pi^2}{m_u+m_d}.
\label{25} \end{equation}
Here $f_{\pi}$ and $m_{\pi}$ are the decay constant and the mass of
the $\pi$ meson, $m_u$ and $m_d$ are the current masses of $u$ and $d$
quarks. Its numerical value is $\langle 0|\bar uu+\bar
dd|0\rangle=2(-240\rm\,MeV)^3$. The value of the gluon condensate
$\langle 0|\frac{\alpha_s}\pi G^{a\mu \nu}G^a_{\mu\nu}|0\rangle
\approx(0.33\rm\,GeV)^4$ was extracted from the analysis of leptonic
decay of $\rho$ and $\phi$ mesons \cite{43}. This data were supported
by the QCD sum rules analysis of charmonium spectrum \cite{19}.

Now one must find the set of parameters $m, \lambda^2, W^2$, which
insure the most accurate approximation of ${\cal L}^i(M^2)$ by the
functions $R^i(M^2)$ and also the interval of the values of $M^2$,
where this can take place.The set of parameters $m, \lambda^2, W^2$,
which minimize the function
\begin{equation}
\chi^2(m,\lambda^2,W^2)=\sum_j\sum_{i=q,I}
\Big(\frac{{\cal L}^i(M_j^2)-R^i(M^2_j)}{{\cal L}^i(M_j^2)}\Big)^2,
\label{26} \end{equation}
will be referred to as a solution of Eqs.~(\ref{11}).

The appropriate interval
\begin{equation}
0.8~GeV^2<M^2<1.4 GeV^2
\label{27} \end{equation}
and the values of nucleon parameters
\begin{equation}
m=0.93~GeV ; \quad \lambda^2=1.8~GeV^6; \quad
W^2=2.1~GeV^2 \label{28}
\end{equation}
were found in \cite{20, 41}. If we fix $m=0.94\,$GeV, the SR provide
\begin{equation}
\lambda^2=2.0~GeV^6,\quad W^2=2.2~GeV^2.  \label{29}
\end{equation}
These values will be employed in the present paper.

It was shown also in \cite{20} that the nucleon mass vanishes if there is no spontaneous breakdown of the chiral symmetry, e.g. if $\langle 0|\bar q q|0 \rangle=0$. Numerically \cite{20, 41}
\begin{equation}
m^3\ \approx\ -2(2\pi)^2\langle 0|\bar q q|0 \rangle\,.
\label{30} \end{equation}
In the QCD SR approach the mass of the nucleon is formed due to
exchange by quarks between the nucleon and vacuum.

The role of instantons in QCD SR was analyzed in \cite{d1}, \cite{d2}.
It was shown that for the current $j_1$ the contribution of instantons
vanishes for the $\hat q$ structure of the SR. In the scalar structure
instantons form a factor, which change the LHS of SR by about $15\%$.
Following \cite{20a} we include the contribution into uncertainties of
the numerical value of the vacuum expectation value  $\langle 0|\bar q
q|0 \rangle$. However, a rigorous analysis requires investigation of
the $M^2$ dependence of the contribution.

Note that the solutions (28), (29) are not stable with respect to
modification of the values of the condensates \cite{44}.  Even for the
small changes the absolute minimum of the RHS of Eqs.~(19) is provided
by another solution, i.e. $m=0.6\,$GeV, $\lambda^2=0.79\rm\,GeV^6$,
$W^2=1.0\rm\,GeV^2$ \cite{45}. We treat this solution as an
unphysical one, since the contribution of the continuum exceeds more
than twice that of the lowest pole. This contradicts the key assumption
of the ``pole+continuum"  model for the spectrum -- see Eq.~(9).

The nucleon SR with another form of nuclear current were obtained in
\cite{46}. In \cite{47} it was used also for the description of delta
isobars. Further we shall mention some other applications.

\subsection{Inclusion of radiative corrections}

A typical radiative correction is shown in Fig.~2.
In the analysis, carried out in \cite{20,41} the most important radiative corrections of the order $\alpha_s\ln{Q^2}$ have been included. These contributions were summed to all orders of $(\alpha_s\ln{Q^2})^n$. This is called the Leading Logarithmic Approximation. The LLA corrections are expressed in terms of the factor \cite{48}
\begin{equation}
L(Q^2)\ =\ \frac{\ln Q^2/\Lambda^2}{\ln \mu^2/\Lambda^2}\,,
\label{31} \end{equation}
where $\Lambda\approx150\,$MeV is the QCD scale, while $\mu$ is the
normalization point, the standard choice is $\mu=500\,$MeV.
\begin{equation}
A_0^r=A_0/L^{4/9}; \quad B_3^r=B_3\,.
\label{32} \end{equation}
Here the upper index $r$ indicates inclusion of the radiative corrections.

The radiative corrections to the OPE terms of the function $\Pi_0$ have been calculated beyond the LLA in
the lowest order of the $\alpha_s$ expansion
\cite{49} (see also \cite{50}). The results are
\begin{equation}
A_0^r=A_0\Big(1+\frac{71}{12}\frac{\alpha_s}\pi-\frac12\frac{\alpha_s}\pi
\ln\frac{Q^2}{\mu^2}\Big); \quad
A_6^r=A_6 \Big(1-\frac56\frac{\alpha_s}\pi-\frac13\frac{\alpha_s}\pi
\ln\frac{Q^2}{\mu^2}\Big);
\label{33} \end{equation}
$$
B_3^r=B_3\Big(1+\frac32\frac{\alpha_s}{\pi}\Big).
$$

The running coupling constant in the one-loop approximation, with
inclusion of three lightest quarks is
\begin{equation}
\alpha_s(k^2)\ =\ \frac{4\pi}{9\ln k^2/\mu^2}\,.
\label{34} \end{equation}
The calculations in \cite{49,50} have been carried out for
$\alpha_s=const$. Since the linear momenta in the loops,
corresponding to radiative corrections are of the order $q$, it is
reasonable to assume that in Eq.~(\ref{34}) $\alpha_s=\alpha_s(Q^2)$,
converting to $\alpha_s(M^2\approx1\,$GeV$^2)$. Assuming
$\alpha_s(1\,\rm GeV^2)\approx0.37$ (somewhat larger values are
often used nowadays \cite{51}) we find that the radiative correction to
the contribution $A_0$ changes its value by about $50\%$. This
``uncomfortably large" correction was often claimed as the most weak
point of the SR approach \cite{52}.

In \cite{31} we investigated the influence of radiative corrections on
the values of characteristics, obtained in framework of the Borel
transformed nucleon SR in vacuum. We demonstrated that inclusion of the
radiative corrections in various ways (the radiative corrections are
totaly neglected, are included in LLA, are taken into account beyond
the LLA in the lowest order) alter the value of nucleon mass $m$ by
about $5\%$. The radiative corrections modify mainly the value of the
nucleon residue.

Note also that inclusion of the radiative corrections diminishes the
role of the unphysical solution, mentioned above. Once they are
included, minimization of the function $\chi^2$ defined by
Eq.~(\ref{26}) is provided by the physical solution in a broader
interval of the values of the condensates \cite{44,45}.

\section{QCD sum rules in nuclear matter}
\subsection {Choice of the variables}

Now we shall try to use the SR approach for calculation of the nucleon parameters in nuclear matter. The propagation of the system which has a four-momentum $q$ and carries the quantum numbers of the proton is determined by the equation
\begin{equation}
\Pi_m=i\int d^4xe^{i(q\cdot x)}\Xi(x); \quad \Xi(x)=\langle M|T[j(x)\bar j(0)]|M \rangle,
\label{35} \end{equation}
with $|M\rangle$ the ground state of the nuclear matter. It is just an
analog of Eq.~(12). We consider nuclear matter as a system of $A$
nucleons with momenta $p_i$, introducing
\begin{equation}
P\ =\ \frac{\sum p_i}{A}\,.
\label{36} \end{equation}
In the rest frame of the matter $P \approx (m,0)$.

The spectrum of the function $\Pi_m(q,P)$ is much more complicated
than that of the vacuum function $\Pi_0(q^2)$.  Our main task is to
separate the singularities connected with the nucleon in the matter
from those connected with the matter itself. Include in the first step
only the two-nucleon interactions. The singularities connected with the
matter manifest themselves as singularities in variable $s=(P+q)^2$.
Thus the separation can be done by considering $\Pi_m(q^2,s)$ and
keeping $s=const$. Thus we consider the dispersion relations
\begin{equation}
\Pi^i_m(q^2, s)=\frac1\pi\int dk^2\frac{\mbox{Im}
\Pi^i_m(k^2,s)}{k^2-q^2} \label{37}
\end{equation}
for the three  structures $(i=q,P,I)$ of the function
\begin{equation}
\Pi_m(q^2,s)=\hat q \Pi^q_m(q^2,s)+\hat P\Pi^P_m(q^2,s)+I \Pi^I_m(q^2,s).
\label{38}
\end{equation}
Each contribution $\Pi^i$ can be viewed as the sum of the vacuum term
$\Pi^i_0$ and that provided by the nucleons of the matter
$\Pi^i_{\rho}$
\begin{equation}
\Pi_m^i=\Pi_0^i+\Pi_{\rho}^i; \quad \Pi_0^P=0.
\label{39} \end{equation}
This notation will be used also for the other functions.

We clarify the value of $s$ putting
\begin{equation}
s\ =\ 4E_{0F}^2\,,
\label{40} \end{equation}
with $E_{0F}$ being the relativistic value of the nucleon energy on
the Fermi surface. This insures that the nucleon pole on the RHS of
Eq.~(\ref{37}) describes the nucleon, added to the Fermi surface. For the
analysis, carried out in this section we can neglect the bound, thus
putting
\begin{equation}
s\ =\ 4m^2.
\label{41} \end{equation}

This was the choice of variables in our papers \cite{9}--\cite{11},
\cite{29,30} and \cite{32}--\cite{37}. It was used also in \cite{53},
where the approach was used for calculation of the nucleon--nucleus
scattering amplitude.

Note that the vacuum dispersion relation (5) can be viewed as a
relativistic generalization of the nonrelativistic dispersion relation
in time component (energy) $q_0$, known as the Lehmann representation
\cite{54}.  The latter is based on casuality. It converts into a
dispersion relation in $q^2$ after being combined with the symmetric
relation in negative values of $q_0$. The reasoning does not work in
medium, since the Lorentz invariance is lost. To prove the dispersion
relation (37) we must be sure of the possibility of the contour
integration in the complex $q^2$ plane. A strong argument in support of
this possibility is the analytical continuation from the region of real
$q^2\to-\infty$. At these values the asymptotic freedom of
QCD enables one to find an explicit expression  for the integrand. The
integral over the large circle may have a nonvanishing contribution.
However, the latter contains only polynomials in $q^2$ which are killed
by the Borel transform. Thus we consider dispersion relations in $q^2$
to be a reasonable choice.

On the contrary, dispersion relations in $q_0$ contain all possible
excited states of the matter on its RHS.  To illustrate the latter
point, consider photon propagation in medium (see, i.e. \cite{54a}).

In vacuum the propagator of the photon which carries the energy
$\omega$ and linear momentum ${\bf k}$ is $D_0=(\omega^2-k^2)^{-1}$. It
has a pole at $q^2=\omega^2-k^2=0$. In medium it takes the form
$D_m=(\omega^2\varepsilon(\omega, k)-k^2)^{-1}$. Here $\varepsilon
(\omega, k)$ is the dielectric function, related to the amplitude of
the photon scattering on the ingredients of the medium. If the photon
energies are small enough, dependence of  $\varepsilon$ on $k$ can be
neglected (this is known as the dipole approximation). However,
$\varepsilon(\omega, 0)$ is a complicated function of $\omega$. It
depends on the eigen energies  of the medium.  The same refers to the
function $D_m(\omega)$. However, the function $D_m(q^2)$ still has a
simple pole at the point $q^2=q^2_m=\omega^2(1-\varepsilon(\omega))$,
reflecting properties of the in-medium photon. Straightforward
calculation of the value $q_m^2$ is a complicated problem. The same is
true for the in-medium proton. The SR are expected to provide the
in-medium position of the pole in some indirect way.

The approach, in which the dispersion relation in $q_0$ at fixed value
of the three-dimensional momentum $|{\bf q}|$ is the departure point
was developed in \cite{55}.  It was used for description of nucleons
\cite{55,56}, delta-isobars \cite{56a} and hyperons \cite{57} in nuclear matter. The results are
reviewed in \cite{58}. However, the possibility to separate  the
singularities connected with the nucleon in the matter from those
connected with the matter itself in this approach looks to us to be
obscure.

Thus we expect the Borel transformed dispersion relations in $q^2$
with fixed value of $s$ to be more reliable.

\subsection {Operator product expansion}

Note that the condition $s=const$ enables to use the OPE of the LHS of
Eq.~(37). Indeed, we find
\begin{equation}
2(Pq)\ =\ s-m^2-q^2\,,
\label{42} \end{equation}
and thus in the rest frame of the matter
$$
q_0\ =\ \frac{s-m^2-q^2}{2m}\,.
$$

Hence,
\begin{equation}
q^2/q_0\to const=c\sim2m \quad  \mbox{at}  \quad |q^2|\to\infty\,.
\label{43} \end{equation}
The exponential factor on the RHS of Eq.~(\ref{35}) can be written as
$$
e^{i(qx)}=e^{i({q^2x^2}/{2cx_z}+{x_zc}/2+{q^2x_t^2}/{2cx_z})},
$$
with $z$ -- the direction of momentum $\bf{q}$, $x^2=x_0^2-x_z^2-x_t^2$.
Hence, the integral in Eq.~(37) is determined by $x^2 \sim q^{-2}$,
$x_z \sim c^{-1}$, and the function $\Xi(x)$  can be expanded in powers
of $x^2$, corresponding to expansion of $\Pi_m(q^2,s)$ in powers of
$q^{-2}$. Note that condition (43) is the same as that for the validity
of the OPE for the structure functions of the deep inelastic scattering
\cite{18a}.

In the case of vacuum the expansion of quark fields $q(x)$ in powers
of $x$ was indeed an expansion in powers of $x^2$, corresponding to a
power series  of $q^{-2}$ for the vacuum function $\Pi_0(q^2)$. In
medium the fields $q(x)$ can be expanded also in powers of $(Px)$. This
leads to expansion in powers of $(Pq)/q^2$ of the function
$\Pi_m(q^2,s)$, and thus, generally speaking, to infinite number of
condensates in each OPE term. Fortunately, due to the logarithmic $q^2$
dependence of the quark loops, the leading OPE terms contain only
finite number of condensates. We shall give details in the next
Section.

\subsection{Model of the spectrum}

Now we turn to the RHS of Eq.~(\ref{37}). The function $\Xi(x)$ defined by
Eq.~(\ref{35}) can be written as
\begin{eqnarray}
\Xi(x) &=& \langle M|T{j(x)\bar j(0)}|M\rangle=\langle M_A|j(x)|M_{A+1}\rangle
\langle M_{A+1}|\bar j(0)|M_{A}\rangle \theta(x_0)
\nonumber\\
&&-\ \langle M_A|\bar j(0)|M_{A-1}\rangle \langle
M_{A-1}|j(x)|M_A\rangle \theta(-x_0), \label{44}
\end{eqnarray}
with
$|M_{K}\rangle$ standing for the system with the baryon number $K$.
Here $|M\rangle=|M_{A}\rangle$ is the ground state, summation over all
states with $K=A\pm 1$ is assumed. This equation is illustrated by
Fig.~3

The matrix element $\langle M_{A+1}|\bar j|M_{A}\rangle$ contains the
term $\langle N|\bar j|0\rangle$, which adds the nucleon (the ``probe
nucleon") to the Fermi surface of the state $| M_{A}\rangle$, while the
rest $A$ nucleons are spectators. If interactions of this nucleon with
the matter are neglected, this contribution to $\Xi(x)$ has a pole at
$q^2=m^2$. Interactions of the probe nucleon with the matter shift the
position of the pole. In the mean field approximation shown in Fig.~4a
the shift does not depend on $s$. Going beyond the mean field
approximation we find the Hartree self-energy diagrams (Fig.4b), which
have singularities in $s$. Due to condition (41) they do not add new
singularities in variable $q^2$ to the function $\Pi_m(q^2,s)$. The
exchange (Fock) self-energy diagram is shown in Fig.~4c.

The matrix element $\langle M_{A+1}|\bar j|M_{A}\rangle$ contains also
the terms  $\langle B_1|\bar j|0\rangle$ with $B_1$ standing for the
system, containing the nucleon and mesons. The current $j(x)$ creates
the nucleon and a meson with the  mass $m_x$, which is absorbed by the
nucleons of the matter -- see Fig.~5. This contribution has a cut in
$q^2$ complex plane, at $q^2>m^2+2m_mm_x$. It has also a cut in $s$,
which is not important for us, since $s$ is fixed.

The matrix element $\langle M_{A-1}|j(x)|M_{A}\rangle$  contains the
terms $\langle B_0|j(0)|N\rangle$, with $B_0$ describing a system with
the baryon quantum number equal to zero. The other $A-1$ nucleons are
spectators. The system $B_0$  can be a set of $\pi$ mesons, $\omega$
meson, etc. These contributions depend on the variable
$u(q^2)=(P-q)^2$, providing singularities at $u>m_x^2$, with $m_x$ the
mass of state $B_0$ -- see Fig.~6.  Thus the lowest singularity in $u$
corresponds to the branching point $q^2=m_m^2+2m_{\pi}^2$,
corresponding to the real two-pion state in the $u$ channel. A
single-particle meson state with the mass $m_x$ generates a pole at
$q^2=m^2+m_x^2/2$.
The diagram, shown in Fig.~4c. is also one of the
contributions, which has singularities in the $u$ channel.

Note that the antinucleon state corresponding to $q_0=-m$
generates the pole $q^2=5m^2$, shifted far to the right from the lowest
lying state.

Thus the spectrum of the function $\Pi_m(q^2,s)$ consists of the pole
at $q^2=m^2$, a set of higher laying poles, generated by the $u$
channel and a set of branching points. The lowest lying branching point
is separated from the position of the pole $q^2=m_m^2$ by a much
smaller distance than in the case of vacuum ($q^2=m^2+2mm_\pi$ in the
latter case). Note, however, that at the very threshold the
contribution is quenched since the vertices contain linear moments of
intermediate pions. Thus the higher singularities can be considered as
separated from the pole $q^2=m_m^2$.

The situation becomes more complicated if we include the interaction
of the probe nucleon with $n>1$ nucleons of the matter. The
corresponding amplitudes depend on the variables $s_n=(nP+q)^2$. This
causes the cuts, running to the left from the point $q^2=m^2$. Its
contribution thus is not quenched by the Borel transform. However, as
we shall see in Sec.~8, such multinucleon interactions require
inclusion of the condensates of high dimensions on the LHS of (37).
Such contributions do not have logarithmic loops, thus contributing
only to the pole terms on the RHS of Eq.~(37). Hence, this
singularities should be disregarded in our approach, and the three-body
forces are included in the mean-field approximation in our approach.

To summarize the results of this Section, we use Borel transformed
dispersion relations of the function $\Pi_m(q^2,s)$ at fixed $s$. The
OPE is used on the LHS. The "pole+ continuum" model is used for the
RHS. Now we have three  SR equations
\begin{eqnarray}
&& \hspace*{-0.5cm}
\tilde\Pi_m^{i~OPE}(M^2,s)\ =\ \lambda_{Nm}^2 \xi^i e^{(-m_m^2/M^2)}
\nonumber\\
&&+\ \frac{1}{2\pi i}\int_{W_m^2}^{\infty} dk^2\Delta_{k^2}
\Pi_m^{i~OPE}(k^2,s) e^{(-k^2/M^2)};\quad i=q,P,I.
\label{45}
\end{eqnarray}
Here $m_m$ and $\lambda_{Nm}^2$ are the position of the nucleon pole
in medium and the value of its residue, $W_m^2$ is the in-medium value
of the effective threshold. The meaning of the parameters $\xi^i$ will
be clarified in next Section.

\section{Nucleon self-energies in the lowest orders of OPE}

In the Subsections $4.1-4.4$ we consider the symmetric nuclear matter,
with equal number of protons and neutrons. Also, due to isotope
invariance
\begin{equation}
\langle M|\bar dd|M\rangle=\langle M|\bar uu|M\rangle
=\langle M|\bar qq|M\rangle\,.
\label{46} \end{equation}

\subsection {General equations}

Start with description of the nucleon pole. We can write for the
inverse nucleon propagator in medium
\begin{equation}
G_N^{-1}\ =\ (G_N^0)^{-1}-\Sigma
\label{47} \end{equation}
with $G_N^0=(\hat q-m)^{-1}$ being the propagator of the free nucleon,
while
\begin{equation}
\Sigma\ =\ \hat q\Sigma_q+\frac{\hat P}{m}\Sigma_P+\Sigma_I
\label{48} \end{equation}
is the general expression for the self-energy in the nuclear matter.
In the kinematics, determined by Eq.~(41) we find
\begin{equation}
G_N\ =\ Z\frac{\hat q-\hat P(\Sigma_V/m)+m^*}{q^2-m_m^{2}}
\label{49} \end{equation}
with
\begin{equation}
\Sigma_V=\frac{\Sigma_P}{1-\Sigma_q}\,; \quad
m^{*}=\frac{m+\Sigma_I}{1-\Sigma_q}\,.
\label{50} \end{equation}
For the new position of the nucleon pole we find
\begin{equation}
m_m^2\ =\ \frac{(s-m^2)\Sigma_V/m-\Sigma_V^2+m^{*2}}{1+\Sigma_V/m}\,,
\label{51}
\end{equation}
while
\begin{equation}
Z\ =\ \frac{1}{(1-\Sigma_q)(1+\Sigma_V/m)}\,.
\label{52} \end{equation}

Thus in Eq. (\ref{45})
\begin{equation}
\xi^q=1; \quad \xi^P=-\Sigma_V; \quad \xi^I=m^*.
\label{53} \end{equation}

Note that Eqs.~(\ref{50}) correspond to those for the vector self-energy and
for the effective mass in nuclear physics -- see, e.g. \cite{60}. Also
in accordance with definition accepted in nuclear physics the scalar
self-energy is
\begin{equation}
\Sigma_s\ =\ m^{*}-m\,.  \label{54}
\end{equation}

In the nonrelativistic limit the proton dynamics is determined by the potential energy
\begin{equation}
U\ =\ \Sigma_V+\Sigma_s\,.
\label{55} \end{equation}
Keeping the only terms, which are linear in the density $\rho$ we find
a simple expression for the shift of the position of the nucleon pole
\begin{equation}
m_m-m\ =\ U\,.
\label{56} \end{equation}

Instead of Eqs. (19) we have now
\begin{equation}
{\cal L}_m^i(M^2, W_m^2 )=R^i(M^2); \quad i=q,P,I\,.
\label{57} \end{equation}
Here ${\cal L}_m^i$ are the Borel transforms of the LHS of Eq. (37),
multiplied by $32\pi^4$ -- see the previous Section, while
\begin{eqnarray}
&& R^q(M^2) =\lambda_m^{2}e^{-m_m^2/M^2}; \quad
R^P(M^2)=-\Sigma_V\lambda_m^{2}e^{-m_m^2/M^2};
\nonumber\\
&& R^I(M^2)\ =\ m^*\lambda_m^{2}e^{-m_m^2/M^2},
\label{58}
\end{eqnarray}
where $\lambda_m^2$ is the effective value of the in-medium value of
the residue, i.e. $\lambda_m^2=Z\cdot 32\pi^4\lambda_{Nm}^2$, following
Eqs. (49) and (52).

Employing Eqs. (57) and (58) one finds
\begin{equation}
-\frac{{\cal L}_m^P(M^2, W_m^2 )}{{\cal L}_m^q(M^2, W_m^2 )}=\Sigma_V;
\quad \frac{{\cal L}_m^I(M^2, W_m^2 )}{{\cal L}_m^q(M^2, W_m^2 )}=m^*.
\label{59}
\end{equation}

   \subsection{Left-hand sides of the sum rules}
\subsubsection{Contribution of the condensates of lowest dimension}

The calculation is based on the presentation of the single-quark propagator in nuclear matter
\begin{eqnarray}
&& \hspace*{-0.5cm}
\langle M|T[q_i^a(x)\bar q_j^b(0)]|M \rangle = \frac{i}{2\pi^2}
\frac{(\hat x-im_qx^2/2)_{ij}}{x^4}\delta^{ab}
\nonumber\\
&&-\ \frac1{12}\sum_A\Gamma_{ij}^A \delta^{ab}
\langle M|\bar q(0)\Gamma^A q(x)|M\rangle\,,
\label{60}
\end{eqnarray}
just analogous to Eq.~(14). The bilocal operators in the last term on
the RHS of Eq.~(60) are not gauge invariant.The gauge invariant
expression
\begin{equation}
q(x)=\left(1+x_{\alpha}D_{\alpha}+\frac{1}{2}x_{\alpha}
x_{\beta}D_{\alpha}D_{\beta}+...\right)q(0),
\label{61}
\end{equation}
with $D_{\alpha}$ standing for the covariant derivatives, provides the
infinite series of the local condensates.  Anyway, here we are looking
for the condensates of the lowest dimension. Hence, we put $q(x)=q(0)$
in the last term of Eq.~(60). Thus
\begin{eqnarray}
&& \hspace*{-0.5cm}
\langle M|T[q_i^a(x)\bar q_j^b(0)]|M \rangle=\frac{i}{2\pi^2}
\frac{(\hat x-im_qx^2/2)_{ij}}{x^4}\delta^{ab}-
\nonumber\\
&&-\
\frac{1}{12}\delta_{ij}\delta^{ab}\langle M|\bar q(0)q(0)|M\rangle
-\frac1{12}\gamma^\mu_{ij}\delta^{ab}\langle M|\bar q(0)\gamma_\mu
q(0)|M\rangle.
\label{62}
\end{eqnarray}
The first term presents the singular part of the propagator. It is
just the same as in vacuum-Eq.(14).  We include it in the chiral limit
$m_q=0$. Recall that the calculations in vacuum have been carried out
in the chiral limit as well. The second term also has the same form as
in vacuum. However, it includes the operator $\bar q q$ averaged over
the ground state of the matter. The last term contains a new vector
condensate, which vanishes in vacuum. In the rest frame of the matter
it is
\begin{equation}
v^q_\mu=\frac12(v_p^q +v_n^q)\rho\delta_{\mu0}; \quad
v_N^q= \langle N|\bar q(0)\gamma_{0} q(0)|N\rangle\,.
\label{63} \end{equation}

In the leading order of the OPE two of the quarks are described by the
free propagators $g_q$, while the rest one is presented by one of the
two last terms on the RHS of Eq.~(62). Present, similar to the vacuum
case
\begin{equation}
\Pi^q_m(q^2,s)=\sum_n A_n; \quad  \Pi^I_m(q^2,s)=\sum_n B_n; \quad \Pi^P_m(q^2,s)=\sum_nP_n
\label{64} \end{equation}
with $n$ denoting the dimension of the condensates. We find immediately that the contribution to $\Pi_m^I$ can be expressed by Eq.(17) for $B_3$ with the matrix element $\langle 0|\bar dd|0\rangle$ replaced by
$\langle M|\bar dd|M\rangle$. Thus
\begin{equation}
B_3=\frac{\langle M|\bar dd|M\rangle q^2\ln(-q^2/L^2_q)}{4\pi^2}\,.
\label{65} \end{equation}
The leading contributions to $\Pi^q$ and $\Pi^P$ are determined by the
vector condensate
$$
\Pi_\rho=\frac{4i}{\pi^4}\int\frac{d^4x}{x^8}\left(
\frac{x^2}2(\hat v^u+\hat v^d)+\hat x(x,v^u+v^d)\exp{(i(q \cdot
x))}\right).
$$
Direct calculation provides
\begin{equation}
A_{3\rho}=\frac{1}{6\pi^2}\frac{(P\cdot
q)}{mL^{4/9}}\ln{(-q^2/L_q^2)}v(\rho); \quad  P_{3 \rho}
=\frac{q^2}{3\pi^2}\frac{\ln{(-q^2/L_q^2)}}{L^{4/9}}v(\rho), \label{66}
\end{equation}
where $v(\rho)=3\rho$ is the vector condensate -- see
Eq.~(3). The contributions of these terms to the LHS of Eqs.~(57) are
\begin{equation}
\tilde A_{3\rho}=-\frac{8\pi^2}{3}\frac{(s-m^2)M^2E_0(M^2, W_m^2)-M^4E_1(M^2, W_m^2)}{mL^{4/9}}v(\rho);
\label{68}\end{equation}
$$
\tilde B_{3 \rho}=-4\pi^2M^4E_1(M^2, W_m^2)\kappa_{\rho}(\rho); \quad
\tilde P_{3 \rho}=-\frac{8\pi^2}{3}\frac{4M^4E_1(M^2, W_m^2)}{L^{4/9}}v(\rho).
$$
The functions $E_n$ are defined by Eq.~(23). Following our notations
\begin{equation}
\kappa_{\rho}(\rho)=\kappa(\rho)-\kappa(0)
\label{69}
\end{equation}

We illustrate Eq.~(\ref{68})  by Fig.~7. It describes the exchange by
noninteracting  quarks with the quarks of the matter. Of course, there
are strong interactions between the quarks of the condensate.  The
corresponding contributions to the nucleon pole (the RHS of
Eq.~(\ref{57})) account for the exchanges by the systems of strongly
correlated quarks (mesons) with the same quantum numbers. They have
standard Lorentz structures, i.e. the terms proportional to $v(\rho)$
and $\kappa(\rho)$ contribute to vector and scalar structures of
Eq.~(\ref{1}) correspondingly. This is illustrated by Fig.~8.

\subsubsection{Next to leading OPE terms}

Now we include the contribution of the condensates of dimension $d=4$.
These are the contributions caused by nonlocality of the vector
condensate $\langle M|\bar q( 0)\gamma_0 q(x)|M\rangle$ -- see Eq.~(60)
with $q(x)$ defined by Eq.~(61) to $q$ and $P$ structures on the LHS of
Eqs.~(57) and also the contribution of the gluon condensate to the $q$
structure.

Recall that in the case of vacuum there was a contribution $A_4$
caused by the vacuum gluon condensate -- see Eqs. (22) and (24). In the
nuclear matter the vacuum gluon condensate should be replaced by its
in-medium value. The corresponding contribution is
\begin{equation}
\tilde A_{4g\rho}\ =\ \pi^2\frac{M^2E_0(M^2,
W_m^2)}{L^{4/9}}g_\rho(\rho)\,. \label{70}
\end{equation}
Here we denoted
\begin{equation}
g(\rho)=\langle M|\frac{\alpha_s}{\pi}G^{a\mu
\nu}G^a_{\mu \nu}|M\rangle=g(0)+g_{\rho}(\rho).  \label{71}
\end{equation}

Now we consider the contribution of the nonlocal vector condensate.
For the sake of simplicity we shall present the results in terms of the
nucleon matrix elements, assuming thus that the matrix element
$\langle M|\bar q(0)\gamma_0 q(x)|M\rangle$ is a linear function of
$\rho$.  We can write
\begin{equation}
\theta_{\mu}^f(x)=\langle M|\bar q(0)\gamma_0
q(x)|M\rangle=\frac{P_\mu}m\Phi_a^f\Big((Px),x^2\Big)
+ix_\mu m\Phi_b^f\Big((Px),x^2\Big).
\label{72} \end{equation}
Presenting \cite{61}
\begin{equation}
\Phi_{a,b}^f\Big((Px),x^2\Big)=\int^1_0d\alpha e^{-i\alpha (P\cdot
x)}f_{a,b}^f(\alpha, x^2), \label{73}
\end{equation}
we can expand $f_{a,b}^q(\alpha,x^2)=\eta^q_{a,b}(\alpha)
+x^2m^2\xi^q_{a,b}(\alpha)/8+O(x^4)$.  Here
$\eta^f_{a}(\alpha)=f_{a}^f(\alpha,0)$ is the contribution of the
valence quark with flavor $f$ to the asymptotics of the nucleon
structure function $\eta_a=\eta^u_a+\eta^d_{a}$, normalized by
the condition
\begin{equation}
\int^1_0d\alpha\eta_a(\alpha)\ =\ 3\,.
\label{74}
\end{equation}
The moments of the function $\eta_b$ can be presented in terms of
those of $\eta_a$ and $\xi_a$\cite{33} , e.g.
\begin{equation}
\int^1_0d\alpha\eta_b(\alpha)=\frac14\int^1_0d\alpha\alpha\eta_a(\alpha).
\label{75} \end{equation}

In the equations, displayed in this Subsection we omit some terms, which are not important numerically.

Direct calculation provides the contribution of the nonlocal condensate to ${\cal L}^P$
$$
P_{4 \rho}=  P_{4 \rho}^{(1)}+P_{4 \rho}^{(2)};
\quad P_{4\rho}^{(1)}= -\frac{1}{6\pi^2m}\ln{\frac{Q^2}{L_p^2}}
\int\limits^1_0d\alpha[(5\alpha(P\cdot q')+2\alpha^2m^2)
\eta_a(\alpha)+9m^2\eta_b(\alpha)]\rho;
\label{76}
$$
\begin{equation}
P_{4 \rho}^{(2)} =\frac{1}{6\pi^2m}\int^1_0d\alpha
\ln{\frac{Q^{'2}}{Q^2}}[(-\alpha(Pq')+2q'^2)\eta_a(\alpha)
-9m^2\alpha\eta_b(\alpha)]\rho,
\end{equation}
with $q'=q-P\alpha$, $Q'^2=-q'^2$.

To evaluate the contribution  $P_{4\rho}^{(2)}$ we rearrange the logarithmic term
$$
\ln{\frac{Q^{'2}}{Q^2}}\ =\ \ln{\frac{(1+\alpha)(Q^2+X^2(\alpha))}{Q^2}},
$$
with
$$
X^2(\alpha)\ =\ \frac{\alpha(s-m^2(1+\alpha))}{1+\alpha}\,.
$$

Performing the Borel transform and using the numerical values of the
moments of the structure function $\eta_a$ (actually we used those,
presented in  \cite{62}) we find all the coefficients of the $M^{-2}$
expansion to be of the same order of magnitude. Hence, following
\cite{33} we do not use this expansion here. We employ the explicit
expression \cite{33}, providing for the contribution to the LHS of
Eq.~(57)
\begin{eqnarray}
\tilde P_{4\rho}^{(2)}&=&\frac{8\pi^2}{3}\int^1_0d\alpha
\Big[\Big(-5(s-m^2)\alpha+6m^2\alpha^2\Big)G_0(M^2, \alpha)
\nonumber\\
&&+\ (4+5\alpha)G_1(M^2,
\alpha))\eta_a(\alpha)-18m^2\alpha\eta_b(\alpha)\Big]\rho.  \label{77}
\end{eqnarray}
Here $G_n(M^2, \alpha)=M^{2(n+1)}E_n(X^2(\alpha))$. We
included only the lowest moments of the function $\eta_b(\alpha)$,
which are related to the moments of the structure function $\eta_a$ by
expression, analogous to Eq.~(\ref{75}) -- see \cite{33}. The values of
these moments enable to expect the convergence of the latter expansion.
It will be useful (see next Subsection) to present the contribution in terms of those of lowest nonvanishing (second) moments of the structure functions. For the $P$ structure it is
\begin{equation}
 (\tilde P_{4\rho})_2=\frac{8\pi^2}{3}5\frac{(s-m^2)M^2E_0(M^2, W_m^2)
-M^4E_1(M^2, W_m^2)}{m}{\cal M}_2\rho,
\label{78} \end{equation}
where
\begin{equation}
{\cal M}_2\ =\ \int^1_0d\alpha\alpha\eta_a(\alpha).
\label{79} \end{equation}
The conventional notation is ${\cal M}_n=\int^1_0d\alpha\alpha^{n-1}\eta_a(\alpha)$. Numerically ${\cal M}_2=0.32$ \cite{34}.

To avoid the cumbersome formula, we do not present the expressions for the contribution of the nonlocal vector condensate to the $q$ structure of Eq.(\ref{57}), addressing the readers to our papers \cite{29} and \cite{38}. The contribution of the second moment is
\begin{equation}
(\tilde A_{4\rho})_2 =  \frac{16\pi^2}{3}mM^2E_0(M^2, W_m^2){\cal M}_2\rho.
\label{80} \end{equation}

Note that the nonlocality of the scalar condensate vanishes in the
chiral limit \cite{33}, due to equations of motion. Thus~ $\tilde
B_{4\rho}=0$.

The contribution of the condensate
$\langle M|\bar q\lambda^a_{\mu,nu}G^{a\mu\nu}\sigma_{\mu\nu}q|M\rangle$
of dimension $d=5$ vanishes in medium \cite{24}, as well as in vacuum
\cite{20} if we are using $j=j_1$ -- see Eq.~(\ref{13}). There is no such
cancellation for the currents $j_2$ and $j_3$. This is one more
argument in favor of using the current $j_1$, since the in-medium value
of  this condensate is not known.

\subsection{Approximate solution}

In this subsection we find an approximate solution of Eq.~(\ref{57}).
The nucleon self-energies are expressed explicitly as functions of
in-medium QCD condensates.

This solution can be obtained by replacing the in-medium threshold
value $W_m^2$ by its vacuum value $W^2$ in Eqs.~(57).  This can be
done, since the variation of the threshold $\delta W^2$ contains a
small factor $e^{-W^2/M^2}$.  The accuracy of the approximate solution
will be discussed in the next Subsection.

We put $W_m^2=W^2$ on the LHS of  Eqs.~(\ref{59}) and multiply both
numerators and  denominators  by $e^{m^2/M^2}$.  Taking into account
only the terms of dimension $d=3$ in the medium contributions
${\cal L}_{\rho}^i$, we can write these equations as
\begin{eqnarray}
&& \frac{\frac{32\pi^2}{3}f_P(M^2)v(\rho)}{{\cal L}_0^q(M^2)e^{m^2/M^2}
- \frac{8\pi^2}{3m}f_q(M^2)v(\rho)}\ =\ \Sigma_V\,;
\nonumber\\
\label{81}
&& \frac{{\cal L}_0^I(M^2)e^{m^2/M^2}-4\pi^2f_I(M^2)\kappa_\rho
(\rho)}{{\cal L}_0^q(M^2)e^{m^2/M^2}-\frac{8\pi^2m}3
f_q(M^2)v(\rho)}\ =\ m^*.
\end{eqnarray}
Here
the $M^2$ dependence of the contributions, provided by medium is
contained in the functions
\begin{equation}
f_q(M^2)=\frac{[(s-m^2)M^2E_0(M^2)-M^{4}E_1(M^2)]e^{m^2/M^2}}{L^{4/9}}; \quad
f_P(M^2)=\frac{M^{4}E_1(M^2) e^{m^2/M^2}}{L^{4/9}}; \quad
\label{82}     \end{equation}
$$f_I(M^2)=M^{4}E_1(M^2) e^{m^2/M^2}$$

One can see that they exhibit only weak dependence on $M^2$ in the
interval (\ref{27}) -- see Fig.~9.  They can be approximated as
$$ \overline{f_q(M^2)}\approx c_0=3.29 ; \quad \overline{f_P(M^2)}\approx c_P= 1.30 ;
\quad \overline{f_I(M^2)} \approx c_I= 1.60.$$
Here the overline means that the function of $M^2$ is replaced by a
constant value, corresponding to the lowest value of the function
$\chi^2$ defined by Eq.~(26) in the interval (27) at the vacuum value
of the continuum threshold $W^2$. The RHS of these equations are
written in powers of GeV (their dimensions are GeV$^4$).

Now one can replace the functions $f_i(M^2)$ by the parameters $c_i$
in Eqs.~(\ref{81}). Also, due to Eqs. (\ref{20}) and (\ref{21}) one can replace the
functions  ${\cal L}_0^q(M^2, W^2)e^{m^2/M^2}$  and ${\cal L}_0^I(M^2,
W^2)e^{m^2/M^2}$ by the constant values $\lambda^2$ and $m\lambda^2$
correspondingly. We define also for the contributions induced by the
matter
$$
{\cal F}^i\ =\ \frac{\overline{{\cal L}_\rho^i(M^2,W^2))e^{m^2/M^2}}}{\lambda^2}.
$$
Here the overline has the same meaning as in Eq.~(\ref{82}).

Now Eqs. (\ref{59}) can be written as
\begin{equation}
\Sigma_V=\frac{{\cal F}^P}{1+{\cal F}^q}\,; \quad
m^{*}=\frac{m+{\cal F}^I}{1+{\cal F}^q}\,.
\label{83} \end{equation}
This is a direct analog of Eq.~(50). We can identify
\begin{equation}
\Sigma_q\ =\ -{\cal F}^q.
\label{84} \end{equation}

We denote also
\begin{equation}
{\cal F}^i\ =\ \sum_n{\cal F}_{n}^i\,,
\label{85} \end{equation}
where ${\cal F}_{n}^i$ is the contribution of the condensates with the
dimension $d=n$.

Employing Eq.~(68) we find the contribution of the condensates with
$d=3$. In Eqs. (\ref{83}) and (\ref{84})
\begin{equation}
{\cal
F}_3^q=-47v(\rho); \quad {\cal F}_3^P=66v(\rho); \quad {\cal
F}_3^I=-32\kappa_{\rho}(\rho),
\label{86} \end{equation}
with all values written in powers of GeV.  Using Eq.~(3) we immediately
find the values of $\Sigma_q$ and $\Sigma_V$ at any value of the
 density $\rho$. For example, at phenomenological value of saturation
 density
\begin{equation}
\rho=\rho_0=0.16\rm\, Fm^{-3}=1.3 \cdot10^{-3}\, GeV^3
\label{87} \end{equation}
we obtain
\begin{equation}
\Sigma_V=314\,\mbox{MeV}, \quad \Sigma_q=0.18\,.
\label{88} \end{equation}

Now we include the condensates of dimension $d=4$. The contribution of
the gluon condensate can be obtained by employing Eq.~(\ref{70}) at
$W^2_m=W^2$. We present the contributions of the nonlocal vector
condensate at $W^2_m=W^2$ through the contributions ${\cal F}_4^{iL}$,
corresponding to inclusion of only the
second moments of the structure functions ${\cal M}_2$,
given by Eqs. (\ref{78}) and (\ref{80}). Direct calculation
provides ${\cal F}_{4\rho}^{PL}=290{\cal M}_2\rho$, and ${\cal
F}_{4\rho}^{qL}=55{\cal M}_2\rho$ in GeV units, while
\begin{equation}
{\cal F}_{4}^P=0.37{\cal F}_{4}^{PL}\,; \quad
{\cal F}_{4}^q=0.62{\cal F}_{4}^{qL}\,.
\label{89} \end{equation}

Thus, with inclusion of the condensates of dimension $d=3,4$
\begin{equation}
{\cal F}_q=-47v(\rho)+8.6g_{\rho}+34{\cal M}_2\rho\,.
\label{90} \end{equation}
Also, from Eq.~(78) we find
\begin{equation}
{\cal F}_P\ =\ 66v(\rho)-106{\cal M}_2\rho\,,
\label{91} \end{equation}
while
\begin{equation}
{\cal F}_I\ =\ -32\kappa_{\rho}(\rho)\,,
\label{91a} \end{equation}
with the RHS written in GeV units. The value of  ${\cal F}^I$ remained unchanged.
Recall that ${\cal M}_2=\int_0^1 d\alpha \alpha \eta_a(\alpha)=0.32$ is the second moment of the nucleon deep inelastic structure function.

\subsection{Gas approximation}

In the lowest order of OPE expansion the nucleon scalar self-energy
depends on the scalar condensate $\kappa(\rho)$ defined by Eq.~(4). The
next to leading order terms contain the gluon condensate $g(\rho)$,
defined by Eq.~(\ref{71}). Below we  shall consider the problem of
finding the density dependence of these condensates. In this Subsection
we include them in the framework of the gas approximation \cite{22,23}
\begin{equation}
\kappa(\rho)=\kappa(0)+\kappa_N\rho\,; \quad
\kappa_N=\langle N|\sum_i \bar q^i q^i|N\rangle\,,
\label{94} \end{equation}
and
\begin{equation}
g(\rho)=  g(0)+g_N\rho; \quad
g_N=\langle N|\frac{\alpha_s}{\pi}G^{a\mu \nu}G^a_{\mu \nu}|N\rangle.
\label{95} \end{equation}
Thus in this approximation $\kappa_{\rho}=\kappa_N\rho$, $g_{\rho}=g_N\rho$.

The scalar quark condensate is numerically more important. Numerous
model calculations -- see, e.g. \cite{14, 63}, and \cite{64} for
earlier papers, provide relatively small nonlinear term $S(\rho)$ on
the RHS of Eq.~(4) at $\rho$ close to the saturation value $\rho_0$.
Thus, employing of the gas approximation is reasonable.

Now we must find the values of $\kappa_N$ and $g_N$.

\subsubsection{The value of \boldmath$\langle N|\bar uu+\bar
dd|N\rangle$}

The expectation value $\kappa_N$ is connected to the pion-nucleon sigma-term $\sigma$ by the relation \cite{25a}
\begin{equation}
\kappa_N=\langle N|\bar uu+\bar dd|N\rangle=\frac{2\sigma}{m_u+m_d}\,.
\label{96} \end{equation}
On the other hand, the $\sigma$ term can be expressed in terms of the
pion--nucleon elastic scattering amplitude $T=T(s,t,k^2,k'^2)$
\cite{25a}. Here $k$ and $k'$ are the pion momenta before and after
scattering, $s$ and $t$ are the Mandelstam variables $s=(p+k)^2$ and
$t=(k'-k)^2$, with $p$ denoting the momentum of the nucleon. The
$\sigma$ term $\sigma$ is connected to the amplitude $F$ in the
unphysical point
\begin{equation}
T(0,0,0,0)\ =\ -\frac{\sigma}{f_{\pi}^2}\,,
\label{97} \end{equation}
where $f_{\pi}$ is the pion decay constant.
The experiments provide the data on the physical amplitude
$T_{phys}=T((m+m_{\pi})^2, m_{\pi}^2, m_{\pi}^2, m_{\pi}^2)$. The
method of extrapolation of observable  physical amplitude to the
unphysical point was worked out in \cite{26}. The $\Sigma$ term
$\Sigma^{\pi N}=-T_{phys}f_{\pi}^2$ differs from the $\sigma$ term by
\begin{equation}
\Sigma^{\pi N}-\sigma\ =\ 15\rm\, MeV.
\label{98} \end{equation}
Note that $(\Sigma^{\pi N}-\sigma)/\sigma \sim m_{\pi}$, in the chiral limit we must put $\Sigma=\sigma$.

For many years the value
\begin{equation}
\sigma\ =\ (45\pm8)\rm\,MeV,
\label{99} \end{equation}
based on the results of \cite{65} was assumed to be true
\cite{66}. This corresponds to $\Sigma^{\pi N}\approx60\,$MeV. However,
some of the latest results support rather the value of $\Sigma^{\pi N}$
close to 80~MeV \cite{67,68,69}, corresponding rather to
\begin{equation}
\sigma\ =\ (60-70)\rm\,MeV.
\label{100} \end{equation}
This is somewhat closer to some of results, claimed earlier \cite{70}.

There is large discrepancy between the theoretical results -- see
\cite{64} for references.  Note that some of the latest ones \cite{71,
71a} support the value $\sigma=45\,$MeV.

\subsubsection{The value of \boldmath
$\langle N|\frac{\alpha_s}{\pi}G^{a\mu \nu}G^a_{\mu \nu}|N\rangle$}

This expectation value was calculated in \cite{72} by averaging the
trace of the QCD energy-momentum tensor $\Theta$ over the nucleon
state. The trace is
\begin{equation}
\Theta^\mu_\mu\ =\ \sum_i m_i\bar q q_i-\frac{b\alpha_s}{8\pi}G^2,
\label{101} \end{equation}
where $m_i$ the mass of the quark with the flavor $i$, $G^2=G^{a\mu
\nu}G^a_{\mu \nu}$, $b=11-2n/3$, $n$ is the total number of flavors.
There is a remarkable cancellation \cite{72}
\begin{equation}
\langle N|\sum_j m_j\bar q_j q_j|N\rangle-\frac{2n_h}{3}\frac{1}{8}
\langle N|\frac{\alpha_s}{\pi}G^2|N\rangle=0.
\label{102} \end{equation}
Here $j$ stand for the ``heavy" quarks with the masses $m_j$ much
larger that the inverse confinement radius $\mu$ (the accuracy of
Eq.~(\ref{102}) is $(\mu/m_j)^2)$; $n_h$ is the number of heavy quarks.
Hence, the charmed quark is the lightest one among those, which
contribute to the LHS of Eq.~(\ref{102}), and
\begin{equation}
g_N=-\frac{8}{9}(m-\sum_im_i\langle N|\sum \bar q^iq^i|N\rangle),
\label{103} \end{equation}
with $i=u,d.s$. The simplest expression for
$g_N$ can be obtained in the chiral $SU(3)$ limit
\begin{equation}
g_N=g_N^{(1)}=-\frac{8}{9}m\,.  \label{104}
\end{equation}
One can see that
\begin{equation}
0>g_N>g_N^{(1)}\,.  \label{105}
\end{equation}
Various estimations for the strange condensate in nucleon  $\langle
N|\bar ss|N\rangle$ provide controversial results (see, e.g.
\cite{65}). However, since contribution of the gluon condensate is
small even for $g_N=g_N^{(1)}$, the estimation (\ref{105}) is sufficient for
our analysis.

\subsubsection{Numerical results in the gas approximation}

The results of previous subsection enable us to obtain the complete
set of numerical results. We compare the approximate solution given by
Eqs. (\ref{86}), (\ref{90}), (\ref{91}), (\ref{91a}) and ``exact"
solution of the system of Eqs.~(57) at the saturation value of
density $\rho=\rho_0$.

Start with the value $\kappa_N=8$, which is consistent with Eq.(\ref{99}).
Including only the condensates of dimension $d=3$ we find for
approximate solution $\Sigma_V=314\,$MeV, $m^*-m=-205\,$MeV. Note that
due to relatively large value $\Sigma_q=0.18$ there is a contribution
of about $+200\,$MeV to $m^*-m$, which is  proportional to the {\it
vector} condensate, and  the one of about $-400\,$MeV, proportional to
the {\it scalar} condensate. Solving Eqs.~(57) we find
\begin{eqnarray}
&&\Sigma_V=323\,\mbox{MeV}, \quad m^*-m=-201{\rm\, MeV},
\nonumber\\
&& \lambda_m^2-\lambda^2=-0.077\,\mbox{GeV}^6,\quad  W_m^2-W^2=0.12\,
\mbox{GeV}^2.  \label{106}
\end{eqnarray}
Thus, indeed the approximate solution is a very accurate one.

Turning to the condensates of dimension $d=4$ and using Eq.(\ref{105}) we see the gluon condensate provides a small contribution of 0.009 to $\Sigma_q$ in the approximate solution.  Thus it adds about $3$ MeV to both $\Sigma_V$ and $m^*$. The nonlocal vector condensate subtracts $0.014$ from the value of $\Sigma_q$. The nonlocality of the vector condensate subtracts also $54$ MeV from $\Sigma_V$. The approximate solution provides now $\Sigma_V=257$ MeV, $m^*-m=-202$ MeV.  Inclusion of the condensates with dimension $d=4$ changes the   solution of Eq.(\ref{57}) to
\begin{eqnarray}
&&\Sigma_V=253\,\mbox{MeV}, \quad m^*-m=-213\,\mbox{MeV},
\nonumber\\
&&\lambda_m^2-\lambda^2=-0.23\,\mbox{GeV}^6,\quad
 W_m^2-W^2=0.04\rm\, GeV^2.
\label{107}
\end{eqnarray}

For $\kappa=11$, corresponding to $\sigma=60$ MeV (\ref{100}) we find
$$
\Sigma_V=238\,\mbox{MeV}, \quad m^*-m=-368\,\rm MeV.
$$
Fixing the
continuum threshold $W^2_m=W^2$ we would obtain
$$
\Sigma_V=276\,\mbox{MeV}, \quad m^*-m=-336\rm\,MeV.
$$
Thus the approximate solution, discussed above
becomes less precise at larger values of $\kappa$.

We show the density dependence of the nucleon parameters and of the
effective threshold value $W_m^2$ for $\kappa_N=8$ in Fig.~10. The
approximate solution provided by Eqs. (\ref{90})--(\ref{91a}) is shown
by dashed line in Fig.10 a. In Fig.~11 we show the dependence of these parameters
on $\kappa_N$ at saturation value of nuclear density $\rho_0$.

\subsection{Asymmetric matter}

\subsubsection{Condensates of the lowest dimension}

Now we employ the SR approach for calculation of the nucleon self-energies in the nuclear matter, composed of neutrons and protons, distributed with the different densities $\rho_n$ and $\rho_p$, i.e. in the $SU(2)$ asymmetric matter. We calculate the dependence on the total density
$\rho=\rho_n+\rho_p$ and on asymmetry parameter
\begin{equation}
\beta\ =\ \frac{\rho_n-\rho_p}{\rho_n+\rho_p}\,.
\label{162}
\end{equation}

We shall calculate the parameters for the proton. Those for the
neutron can be obtained by changing $\beta$ to $-\beta$.  As well as in
the case of symmetric matter the lowest order OPE terms depend on the
vector and scalar condensates. However, in addition to the isospin
symmetric condensates $v(\rho)$ and $\kappa(\rho)$, defined by
Eqs. (3), (4), we have  two $SU(2)$ asymmetric condensates of dimension
$d=3$. These are the vector condensate
\begin{equation}
v^{(-)}_\mu(\rho)\ =\ \langle M|\bar u\gamma_\mu u-\bar d\gamma_\mu
d|M\rangle,
\label{163} \end{equation}
and a scalar one
\begin{equation}
\zeta_m(\rho,\beta)\ =\ \langle M|\bar uu-\bar dd|M\rangle\,.
\label{164} \end{equation}
The lower index $m$ is introduced in order to show that for the
$SU(2)$ invariant vacuum $\zeta_m(0)\!=\!0$.

The vector isotope asymmetric condensate can be immediately
calculated. In the rest frame of the matter
\begin{equation}
v^{(-)}_\mu(\rho)=v^{(-)}_{0}(\rho,\beta)\delta_{0\mu}\,; \quad
v^{(-)}(\rho, \beta)=-\beta\rho\,.
\label{165} \end{equation}

The situation with the scalar condensate is more complicated.
Even in the gas approximation, where
\begin{equation}
\zeta_m(\rho, \beta)=-\beta \rho\zeta_p; \quad
\zeta_p= \langle p|\bar uu-\bar dd|p\rangle
\label{166} \end{equation}
cannot be directly related to observables.
Thus we need some model assumptions.

Considering the nucleon as a system of three valence quarks the sea of
the quark--antiquark pairs we can write
$$
\zeta_p\ =\ \zeta_p^v+\zeta_p^s\,,
$$
where the upper indices denote the contributions of the valence and
sea quarks correspondingly.  In nonrelativistic quark models
$\zeta_p^v=1$, while $\zeta_p^v<1$ for the relativistic models.  The
experimental data on the deep inelastic scattering  provide $\zeta_p^s
\approx-0.15$ \cite{K1,K2}.

\subsubsection{Nucleon self-energies}

In this Subsection we include only the condensates of the lowest
dimension $d=3$. We can write the contributions to the LHS in a form
similar to Eqs.~(\ref{68}). The term $\tilde A^q_{3\rho}$ depends only
on the vector condensate $v(\rho)$, and does not depend on $\beta$.
Thus it remains the same as for the symmetric matter, while
\begin{eqnarray}
&& \tilde B_{3 \rho}=-4\pi^2M^4E_1(M^2, W_m^2)\theta(\rho,\beta);
\nonumber\\
&& \tilde P_{3 \rho}=-\frac{8\pi^2}{3}\frac{4M^4E_1
(M^2, W_m^2)}{L^{4/9}}w(\rho, \beta).
\label{168}\end{eqnarray}
The functions $E_n$ are given by Eq.~(23). Here we defined
$\theta(\rho,\beta)=(\kappa_N+\beta \zeta_N)\rho$ and
$w(\rho,\beta)=3\rho(1-\beta/4)$. Comparing these expressions with
Eq.(\ref {68}) we find a simple solution in which the proton self
energies are expressed in terms of the solution for symmetric matter \cite{30}.
\begin{equation}
\Sigma_V(\rho, \beta)=\Sigma_V(\rho, 0)(1-\beta/4); \quad m^*(\rho, \kappa_N, \beta,\zeta_p)=m^*(\rho, \kappa_N+ \beta\zeta_p,0,0);
\label{169} \end{equation}
$$
W_m^2(\rho, \beta)=W_m^2(\rho,0).
$$

For example, employing Eq.(\ref{106}) we find that in the neutron
matter ($\beta=1$) the proton vector self-energy is 242~MeV, while
for neutron it is 403~MeV. We shall see that inclusion of the higher
condensates will modify the values.

\subsection{Possible mechanism of saturation}
\subsubsection{A special role of the pion cloud}

Note that the simplest account of the nonlinear effects signals on a
possible saturation mechanism in our approach \cite{22}--\cite{24}. The
nucleons of the matter interact by meson exchange, and the nonlinear
contribution to the scalar condensate $S(\rho)$ -- see Eq.~(4) comes
from the meson cloud, created by nucleons. The contributions of each
meson is thus proportional to the value of the operator $\bar q q$
averaged over the meson states.

We expect exchanges by pions to dominate the value $S(\rho)$.
The matrix element $\langle h|\sum \bar q^iq^i|h\rangle$ counts the
total number $n_{\bar q q}$  of quarks and antiquarks in a hadron $h$
\cite{76,77,78} under certain reasonable assumptions on the quark wave
function of the hadron \cite{77,78}. Thus one can expect $\langle
\mu_j|\bar q^iq^i|\mu_j\rangle\sim2$ for a meson $\mu$. However for
the pion, which is a collective Goldstone excitation
\begin{equation}
\langle\pi|\sum\bar q^iq^i|\pi\rangle=\frac{2m_{\pi}^2}{m_u+m_d}
=2m_{\pi}n_{\bar q q}\,,
\label{112} \end{equation}
(the factor $2m_{\pi}$ comes from normalization of the pion vector of state). Thus $n_{\bar q q} \approx 12$ for pions, and we expect their exchanges to provide the leading contribution to $S(\rho)$.

\subsubsection{Lowest order in Fermi momentum series}

Here we consider the case of symmetric matter.
We shall calculate the contribution the nonlinear term $S(\rho)$ in the lowest order of expansion in powers of Fermi momentum. The latter is related to the density as
\begin{equation}
\rho\ =\ \frac{2}{3\pi^2}p_F^3\,.
\label{109} \end{equation}
The leading contribution is provided by single-pion exchange term (known also as the Fock term or the Pauli blocking term). It can be written as
\begin{equation}
S(\rho)=-\int\frac{d^3p_1}{(2\pi)^3}\frac{ d^3p_2}{(2\pi)^3}\Gamma(k)D^{(0)2}(k)\Gamma(k)
\langle \pi|\bar uu+\bar dd|\pi\rangle,
\label{114} \end{equation}
where  $p_{1,2}\leq p_F$ are the nucleon momenta, ${\bf k}={\bf
p_1}-{\bf p_2}$, $\Gamma$ stand for the $\pi NN$ vertices, $
D^{(0)}=1/(\omega^2-k^2-m_{\pi}^2+i\varepsilon)$ is the propagator of
free pion carrying the energy $\omega$. Summation over spin and isospin
variables is assumed.  The corresponding contribution to the correlator
is illustrated by Fig.~12.  The function $S(\rho)$ is obtained
analytically \cite{33}. The expression becomes especially simple in the
chiral limit $m_{\pi}^2=0$
\begin{equation}
S(\rho)\ =\ -\frac{9}{32\pi^2}\frac{p_F\rho}{f_\pi^2}
\frac{2m_\pi^2}{m_u+m_d}\,,
\label{116} \end{equation}
where the last factor is the pion expectation value Eq.~(\ref{112}).
Thus $S(\rho) \sim \rho^{4/3}$.  Note that the heavier mesons do not
contribute to the leading term of the Fermi momentum expansion. Their
contribution is of the order $\rho^2$.

The difference between the RHS of Eq.~(\ref{116}) and the result,
obtained with the use of the physical value of $m_{\pi}^2$  is rather
large. However, in the chiral limit one should use the value of
$\Sigma$ term, rather than that of $\sigma$-term for the calculation of
$\kappa_N$ -- see Eq.~(96), since the difference $\Sigma-\sigma$
contains additional powers of $m_{\pi}$. This diminishes the
difference. As a result, the value of $\kappa$ with $\kappa_N$ replaced
by
\begin{equation}
\kappa_N^{ch}\ =\ \frac{2\Sigma}{m_u+m_d} \label{117}
\end{equation}
appears to be close to that, obtained for the physical
pion mass \cite{27}. Some additional arguments supporting the use of
the chiral limit at $\rho$ close to $\rho_0$ were given in \cite{79}.

Now we expand the nucleon potential energy $U(\rho)$, defined by
Eq.~(55) in powers of Fermi momentum. We take into account the
contributions of the order $\rho$ and $\rho p_F \sim \rho^{4/3}$,
neglecting, however, the terms of the order $\rho^2$. Including only the condensates of the lowest dimension
$d=3$ and using
Eq.~(\ref{86}) we find (in GeV units)
\begin{equation}
\Sigma_V=66v(\rho); \quad \Sigma_s=46v(\rho)-32\kappa_{\rho}(\rho).
\label{118}
\end{equation}
Here $\kappa_{\rho}(\rho)=\kappa_N^{ch}\rho+S(\rho)$ with $S(\rho)$
given by Eq.~(\ref{116}). Thus the nucleon energy on the Fermi surface
can be written as
\begin{equation}
\mu(z)=\Big[38 z^{2/3}+(430-42\kappa_N^{ch})z+126z^{4/3}\Big]\rm\,MeV.
\label{119} \end{equation}
Here $z=\rho/\rho_0$, with $\rho_0$ being the phenomenological value of
saturation density -- Eq.~(\ref{87}).  The first term on the RHS is the
kinetic energy.

\subsubsection {Equation of state}

The saturation value $\rho$ can be found by minimization of the
binding energy per particle
\begin{equation}
{\epsilon}(\rho)\ =\ \frac1\rho\int_0^{\rho}dx\mu(x)\,.
\label{120} \end{equation}
In our case
\begin{equation}
\epsilon(z)=\Big[23z^{2/3}+(215-21\kappa_N^{ch})z+54z^{4/3}\Big]\rm\,MeV.
 \label{121} \end{equation}
One can see that ${\epsilon}(z)'=0$ at $z=1$ if
$\kappa_N^{ch}=14.4$. This corresponds to $\Sigma^{\pi N}=79\,$MeV, in
agreement with the experimental data. The nucleon energy at the Fermi
surface is $\mu=-11$ MeV. The vector self-energy is $\Sigma_v \approx 260$ MeV,
while $m-m^* \approx -310$ MeV. The value of incompressibility $K=9\rho_0^2
d^2{\epsilon}/d\rho^2=170\,$MeV. A standard value obtained in the
 nuclear physics approaches is $K\approx230\,$MeV \cite{79a}.
 Inclusion of the condensates with $d=4$ modifies these
 values slightly.

Note that somewhat similar saturation mechanism take place in the
 model, suggested later in \cite{80a}. The approach of \cite{80a} was
 based on the chiral effective Lagrangian. The authors calculated the
 nonlinear contribution to the scalar condensate, caused by the pion
 exchange. The value of $S(\rho)$ appeared to be close to the ours one.

Of course, one should not take these results too seriously. We tried
 to obtain the values of the self-energies.  The calculations of the
 potential energy require somewhat higher accuracy, since it is a
 result of subtraction of two large values. Thus, reasonable values for
 the potential energy is a surprise. Of course, we can not expect
 accurate values for the binding energy. Also, the results are very
 sensitive to the exact value of $\kappa_N^{ch}$. For example, in this
 approach the saturation is achieved at $\rho=2\rho_0$ if
$\kappa_N^{ch}=14.7$, corresponding to $\Sigma=83\,$MeV. Also, the
 contribution of two-pion exchange to $S(\rho)$ is too large to be
 neglected \cite{80}.

However, the results of this subsection can be treated as a sigh of a
possible mechanism for saturation of nuclear matter. It is due to the
 nonlinear behavior of the scalar condensate. Thus in our approach the
 saturation is possible only if the three-body interactions
 (strictly speaking, many-body interactions) are included.

The potentials of the form $U(z)=C_1z+C_{4/3}z^{4/3}$ were studied
earlier in nuclear physics. Such potential with $C_1\approx-210\,$MeV,
$C_{3/4}\approx160\,$MeV
was analyzed in \cite{80b}. Note that our values in Eq.(\ref{119}) are
$C_1=-175$ MeV (for $\kappa_N^{ch}=14.4$ )and $C_{4/3}=126$ MeV.

Note that this mechanism differs from that of the Walecka model
 \cite{1,2}. Recall that in the latter case the saturation is due to
 the different behavior of the vector and scalar fields on density. The
 vector and scalar fields can be expressed as \cite{1}
\begin{equation}
V=\int\frac{d^3p}{(2\pi)^3}N_{v(s)}(p)g_{v}\theta(p_F-p); \quad \Phi=\int\frac{d^3p}{(2\pi)^3}N_{v(s)}(p)g_{s}\theta(p_F-p)
\label{122} \end{equation}
with the sum over spin and isospin variables being assumed. Here
 $g_v$ and $g_s$ are the coupling constants. In the vector case  $N_v=\bar
 u_N(p)\gamma_0u_N(p)/2E(p)$  with $u_N$ standing for the nucleon
 bispinor, while $E(p)=V(\rho)+(m^*(\rho)+p^2)^{1/2}$. One finds
immediately that $N_v=1$, and $V(\rho)=g_v\rho$ is exactly proportional
to $\rho$. In the scalar case $N_s(p)=\bar u_N(p)u_N(p)/2E(p)$, leading
to a more complicated dependence $\Phi(\rho_s)=g_s\rho_s(\rho)$, where
the scalar density
\begin{equation}
\rho_s =\int\frac{d^3p}{(2\pi^3}N_{(s)}(p)g_\theta(p_F-p)
\label{123} \end{equation}
is a nonlinear function of $\rho$.
Thus, the saturation in Walecka model is a relativistic effect. It
takes place in framework of the two-body interactions. On the contrary,
in our approach it does not require relativistic treatment of the
nucleons of the matter, being caused by the many-body interactions.

We shall return to this problem in Sec.~9.

\section{Other characteristics of the in-medium nucleons}

\subsection{Axial coupling constant}

The SR method have been used for calculation of the nucleon coupling
constants for isovector and isoscalar axial currents in vacuum
\cite{81}-\cite{83}. The approach is based on  considering the function
$\Pi_0$ defined by Eq.~(\ref{12}) and the dispersion relation (\ref{9})
in external axial field $A_{\mu}$, coupled to the quarks by
interaction $V=g_qA_{\mu}\gamma^{\mu}\gamma_5$, and included in
the lowest order of the perturbation theory.  In the isovector case,
directly related to the neutron $\beta$ decay $g_u=1, g_d=-1$. The
corresponding nucleon coupling constant in vacuum is $g_A=1.27$
\cite{51}. Recall that the vector coupling constant is $g_V=1$, and the
nonzero difference $g_A-g_V$ is due to the nonconservation of the
chiral current.

The function $\Pi_0$ contains several structures, which are linear in
$A$. The structure $\hat q(A\!\cdot\!q)\gamma_5$.  appeared to be the
most convenient one for the SR analysis. The LHS contains now several
expectation values, which vanish in the absence of the axial field. The
condensate $\iota_{\alpha}=\langle 0|\bar q\tilde G_{\alpha
\beta}\gamma_{\beta}q|0\rangle_A=A_{\alpha}\nu\langle 0|\bar q q
|0\rangle$, with $\nu$ usually referred to as the field induced
susceptibility appeared to be numerically the most important one.  The
numerical value of this expectation value was obtained in \cite{84}.

On the RHS of the SR  the lowest lying singularity is now the double
pole, corresponding to the contribution $\langle 0|\bar
j|N\rangle\langle N|\hat A\gamma_5 |N\rangle\langle N|\bar j|0\rangle$.
The SR reproduced the experimental value of the isovector coupling
constant $g_A\approx1.25$.

The first attempt to calculate the value of $g_A^m$ for the nucleon in
nucleon matter was made in \cite{85}. The authors  employed Eq.~(30),
combining it with several phenomenological assumptions.

In \cite{35} the renormalization of the nucleon coupling constant
$\delta g_A=g_A^m-g_A$ in nuclear matter was calculated by extending of
the method developed in \cite{22}--\cite{23} for the case of the
external axial field. The LHS of the SR was dominated by the
configuration, in which one of the $\bar qq$ pairs was exchanged with
the condensate $\iota_\alpha$, and the other one-with the quarks of
the nucleons of the matter. The  RHS of the SR included the
contribution of the isobar-hole ($\Delta N$) excitations in terms of the
single-pole terms.

The result of calculation depend on the value of the scalar condensate
in nuclear matter. Thus it depends on the actual value of the $\sigma$
term. It was obtained that $\delta g_A=-0.24$ at the saturation density
if $\sigma=60\,$MeV. For $\sigma=45\,$MeV the shift is $\delta
g_A=-0.22$.  This is in agrement with the experimental data
\cite{86,87}, providing $g_A^m(\rho_0)\approx1.0$. If the
single-pole terms are not included, the result is $\delta g_A=-0.05$.
Hence, the $\Delta-N$ excitations is the main mechanism of the process,
in agreement with \cite{9,10}.

However, the approximations for the LHS of the SR, made in \cite{35}
are too crude. A more rigorous  analysis, which includes the four-quark
condensates is still needed.

\subsection{Charge-symmetry breaking forces}

\subsubsection{Neutron--proton mass difference}

It is known \cite{51} that the difference between the neutron and
proton masses in vacuum is $m_{np}=1.3\,$MeV, while the
electromagnetic interactions contribute  to this value is $m_{np}^e
\approx -0.7\,$MeV \cite{11}. Thus the contributions of the strong
interactions is $m_{np}^s\approx2.0\,$MeV. If the isospin symmetry
(known also as the charge symmetry) in the hadron interactions  is
assumed, the value of $m_{np}^s$ would not change in nuclear matter.
Thus, it is interesting to  study the density dependence
$m_{np}^s(\rho)$.

One can expect the mass difference of mirror nuclei to be
\begin{equation}
\Delta M\ =\ E^e+m_{np}\,,
 \label{124} \end{equation}
where $E^e$ is the electromagnetic energy difference. The values of
$\Delta M$ and $E^e$ can be, correspondingly, measured and calculated
with high accuracy \cite{88}. However, this equation is not satisfied
if $m_{np}$ is equaled to its vacuum value \cite{89}. The discrepancy
between the two sides of Eq.(124) increases with the atomic number $A$,
reaching 0.9~KeV for $A=208$. This is called the Nolen--Schiffer
anomaly (NSA).

The phenomenological charge-symmetry breaking (CSB) potentials of the
nucleon interactions were discussed in \cite{11}. Some of them
described the NSA, but contradicted the data on the CSB effects in NN
scattering.  A possible resolution of the NSA is the assumption that
the value of $m_{np}$ in a nuclei is smaller than that in vacuum. In
the case of nuclear matter this assumption would manifest itself in the
decrease of $m_{np}(\rho)$ while $\rho$ increases.  The
renormalization of $m_{np}$ was calculated in numerous papers (see,
e.g. \cite{90}--\cite{94}) for nuclear matter and for the finite
nuclei.  However, the NSA is not fully understood yet.

\subsubsection{QCD sum rules view}

The QCD SR provide a consistent formalism for the CSB effects.  In QCD
the charge-symmetry is broken due to a nonzero value of the difference
of current quark masses
\begin{equation}
m_d \approx7\,\mbox{MeV}; \quad  m_u\approx4\,\mbox{MeV}; \quad
\mu=m_d-m_u\approx\,3\,\mbox{MeV}.  \label{125}
\end{equation}
Besides the quark mass difference the CSB  manifests
itself also through a nonzero expectation value of the operator $\bar
uu-\bar dd$. In vacuum the ratio
$$
\gamma_0\ =\ \frac{\langle 0|\bar dd-\bar uu|0\rangle}{\langle 0|
\bar uu |0\rangle}
$$
was obtained by the
QCD SR approach \cite{95,96,96a} basing on the
experimental values of isospin breaking mass splitting of nucleons and
hyperons.

In nuclear matter one can expect the  charge-symmetry breaking effects
to be expressed in terms of the quark mass difference $\mu$ and of the
isospin symmetry breaking condensate
$$
\gamma_m\ =\ \frac{\langle M|\bar dd-\bar uu|M\rangle}{\langle M|
\bar uu |M\rangle}
$$

The first attempt to calculate $m_{np}(\rho)$ was carried out in
\cite{97}. The analysis was based on the vacuum SR, obtained in
\cite{95}. Only the terms, containing scalar condensate were included.
The vacuum scalar condensate in equations of \cite{95} was replaced by
its in-medium value, obtained under certain phenomenological
assumptions. Calculations of \cite{98,99} included also the density
dependence of the CSB parameter $\gamma_m$.

The SR calculations of \cite{36} present the direct extension of the
approach, developed in \cite{33}. Now we must calculate the
contribution of the strong interactions to the difference of the
binding energies $\varepsilon_{np}=\varepsilon_n-\varepsilon_p$ of the
neutron and proton in symmetric nuclear matter. One can write
$\varepsilon_i= U_i+T_i$, with $U_i(T_i)$- the potential (kinetic)
energy of the nucleon $i=p,n$. The potential energies are expressed in
terms of the self-energies by Eq.~(55), while $T_i=p_F^2/2m_i^*$.

Now the LHS of Eqs.~(57) should be calculated for the neutron and
proton separately. The explicit dependence of the free quark
propagators $g_q$, defined by Eq.~(15) on the quark masses should be
taken into account. Also, the term ${\cal L}_m^I$ on the LHS of Eq.~(57)
contains the CSB condensate  $\langle M|\bar uu-\bar dd|M\rangle$.
Thus the difference of the binding energies can be written as
\begin{equation}
\varepsilon_{np}(\rho)\,=\,\mu b_1(\rho)+\gamma_m(\rho)b_2(\rho)\,.
\label{126} \end{equation}
The functions $b_{1,2}(\rho)$ can be obtained by the SR method. The
density dependence of the condensate $\gamma_m$ can be obtained in
framework of certain models.

Start with the calculation of $b_{1,2}$
The terms, proportional to the quark mass difference contribute to the
difference of the scalar self-energies
\begin{equation}
\delta_1=0.17\mu\frac{v(\rho)}{\rho_0}\,\rm ,
\label{127}
\end{equation}
and to the difference of the vector ones
\begin{equation}
\delta_2=-0.029\mu\frac{\kappa_\rho(\rho)}{\rho_0}
\,\rm.
\label{128} \end{equation}
Recall that $v(\rho)=3\rho$ is the vector condensate,
$\kappa_{\rho}(\rho)=\kappa(\rho)-\kappa(0)$, with the scalar
condensate $\kappa$ defined by Eq.~(4), $\rho_0$ is the value of
saturation density, given by Eq.(\ref{87}). There is also a contribution to the difference of the scalar
self-energies, containing the CSB condensate $\gamma_m$
\begin{equation}
\delta_3 =C\Big(\gamma_m(\kappa(\rho)-\kappa(0))
+(\gamma_m-\gamma_0)\kappa(0)\Big); \quad C=32~\mbox{GeV}^{-2}.
\label{129} \end{equation}

One should include also the strong-interaction part to the
neutron-proton difference of the vacuum parameters
$m_{np}$, $\lambda^2_{np}$ and $W_{np}^2$. The two last ones were
obtained in \cite{96} by the SR method, basing on the empirical value
of $m^s_{np}$. These parameters, as well as the CSB condensate
$\gamma_0$ were expressed through the quark mass difference $\mu$.

We present the values of $b_{1,2}$ at the saturation value of the density
\begin{equation}
b_1(\rho_0)=-0.73; \quad b_2(\rho_0)=-1.0\rm\, GeV.
\label{130} \end{equation}

The density behavior
$\gamma_m(\rho)=\gamma_0(\kappa(\rho)/\kappa(0))^{1/3}$, based on the
Nambu--Jona--Lasinio model was suggested in \cite{98}. Under this
assumption and employing $\gamma_0=-2\cdot10^{-3}$ \cite{96} we obtain
$\varepsilon_{np}(\rho_0)=-0.9\,$MeV. Our final result is
\begin{equation}
\varepsilon_{np}(\rho_0)\ =\ (-0.9 \pm 0.6)\rm\,MeV,
\label{131} \end{equation}
with the errors caused mostly by uncertainties of the value of $\gamma_0$.

Thus at least qualitative explanation of the NSA is achieved.

\subsubsection{Consequence for conventional nuclear physics}

At least two points of the QCD SR analysis can be useful for nuclear
physics.

In conventional nuclear physics the $\omega-\rho$ mixing in the vector
channel is usually believed to be responsible for the largest part of
the Nolen--Schiffer anomaly . However, the SR analysis  predicts the
scalar channel to be important as well. Neglecting the terms,
containing  the scalar condensate, we would obtain
$\varepsilon_{np}(\rho_0)>0$, This contradicts the experimental data
and general theoretical expectations.

The other point  is the Lorentz  structure of the nucleon
interactions. In the QHD Eq.~(\ref{1}) the terms $\hat V$ and $\Phi$
are caused by the vector and scalar interactions correspondingly. In
the SR analysis described in Sec.~4. these terms are determined by the
vector and scalar condensates correspondingly.  Such separation is
violated by inclusion of the mass terms in the quark propagators $g_q$
determined by Eq.~(\ref{15}).  Relation between the Lorentz structures
of Eq.~(\ref{1}) and the condensates becomes more complicated. Now
there are contributions to $\Phi$, which are proportional to the vector
condensate and contributions to $\hat V$, proportional to the scalar
condensate. These admixtures are small, being proportional to the
current quark masses. This anomalous Lorenz structure are shown in
Fig.~13.

In the case of the CSB forces they manifest themselves in the leading
terms. One can see that  $\delta_1$ given  by Eq.~(\ref{127})
contributes to the structure $\Phi$ of Eq.~(1), being determined by the
vector interaction. On the other hand, the  contribution $\delta_2$
presented by Eq.~(\ref{128}) is determined by interaction in the scalar
channel, but contributes to the vector structure $V$. This correspond
to anomalous structure of the nucleon vertex functions.

These points can be helpful in constructing the CSB nucleon forces.

\subsection{Nucleon deep inelastic structure functions}

The SR approach enables to investigate internal structure of the
nucleon. Such problems are unaccessible for nuclear physics.

\subsubsection{Nucleon structure functions in QCD sum rules}

In deep inelastic scattering (DIS) the electrons transfer large
energies and momenta to the target. This process is a well known tool
for investigation of the internal structure of the latter. The cross
section of DIS can be expressed through the imaginary part of the
amplitude of the elastic scattering of the virtual photon. The latter
has a large and negative four-momentum squared $k^2<0$, $-k^2\gg m^2$
($m$ is the rest mass of the target) and the energy $k_0\gg m$, while
$-k^2/k_0 \sim m$. The investigation of DIS enables to study the
momentum distribution of the quarks.  The structure functions $F_2(x)$
measured in the DIS are determined by the quark distributions $xq(x)$,
where $x$ is the nucleon momentum fraction carried by a quark.

The QCD SR  method was applied for investigation of the DIS on the
proton. The second moments of the structure functions were obtained in
\cite{100} and \cite{101}. The next to leading terms of the asymptotic
expansion in powers of $k^{-2}$ were found in \cite{102}. The structure
functions $F_2(x)$ at moderate values of $x$ were calculated in
\cite{103}. Another  presentation of the structure functions was
obtained in \cite{104}. We shall employ the approach, developed in
\cite{104} since it can be extended for the case of finite density in a
natural way. On the other hand, such generalization is the extension of
approach described above.

In order to obtain the distribution of the valence quarks, the authors
of \cite{104} considered  the correlation function $G$ , which
describes the system with the quantum numbers of the proton,
interacting twice with strongly virtual hard photons
\begin{equation}
G(q,k)=i^2\int d^4zd^4y e^{(i(qz)+i(ky))}\langle
0|Tj(z)H(y,\Delta)\bar j(0)|0\rangle.  \label{132}
\end{equation}
Here
$q$ and $q+k$ are the momenta of the nucleon in the initial and final
states, $k=k_1-k_2$ is the momentum, transferred to the system by the
photon scattering. The incoming (outgoing) photon carries momentum
$k_1(k_2)$, interacting with the quarks in the point $y-\Delta/2$
($y+\Delta/2$). The quark--photon interaction is described by the
function $H(y, \Delta)$.

In \cite{104} the correlation function $G(q,k)$ was calculated in
terms of the QCD condensates. The  double dispersion relation in
variables $q_1^2=q^2$ and $q_2^2=(q+k)^2$ was employed. The crucial
point was the OPE in terms of nonlocal operators depending on the
light-like vector $\Delta$ ($\Delta^2=0$). The Borel transform in
$q_1^2$ and $q^2_2$ was carried out. The equal Borel masses
$M_1^2=M^2_2$ are considered. The Fourier transform in $\Delta$
provided the momentum distribution of the valence quarks.

Thus, in \cite{104} the momentum distributions of the valence quarks
were expressed in terms of the QCD condensates.

Before discussing the extension for nuclear matter, we describe the
most intriguing problem of the DIS on nucleus.

\subsubsection{EMC effect}

The experimental data obtained first by the EMC collaboration
\cite{105} showed that the DIS function$F^A_2(x)$ of nucleus with the
atomic number $A$ differs from the sum of those of free nucleons. The
structure function was compared to that of deutron ($A=2$), which
imitates the system of free nucleons. The deviation of the ratio
\begin{equation}
R^A(x)\ =\ \frac{F^A_2(x)}{A}\Big/\frac{F^2_2(x)}{2}
\label{133} \end{equation}
from unity characterizes the deviation of a nucleus from the system of
free nucleons.

The ratio $R^A(x)$ appeared to be the function of $x$ indeed.
Most of the early data were obtained for iron (Fe). Exceeding unity at
$x<0.2,$ the ratio drops at large $x$, reaching the minimum value
$R^{56}\approx 0.85$ at $x \approx 0.7$. This behavior was  called EMC
effect.  The same tendency was traced for other nuclei.  Both
experimental \cite{106} and theoretical \cite{107}  investigations of
the effect are going on nowadays.

There are several mechanisms which may cause the deviation of $R^A(x)$
from unity. We shall try to find how the difference of the quark
distributions inside the in-medium and free nucleon changes the ratio
$R^A(x)$.

\subsubsection{EMC effect in QCD sum rules}

In \cite{37} the approach of \cite{104} was combined with that of
\cite{33} for calculation of the quark distributions in the proton,
placed into the nuclear matter. The expression  for the correlation
function took the form of Eq.~(\ref{132}), with the averaging over
vacuum replaced by that over the ground state of the matter. The two
types of contributions to the correlator were considered -- Fig.~14a,b.
In the diagram of Fig.~14a. the photon interacted with the quark of the
free loop. In the diagram of Fig.~14b. it interacts with the quark,
exchanging with the matter. The modification of the distribution of the
quarks was expressed in terms of the vector condensate, with its
nonlocal structure being included, and through the shift of the scalar
condensate. The results are true only for moderate values of $x$. They
can not be extended to the region $x\ll1$, since the OPE diverges in
that region \cite{103}.

Omitting the details of calculation, provided in \cite{37}, we present
the results in Fig.~15. We carry out calculations for nuclear matter.
The ratio $R(x)$ can be viewed as the limiting value of $R^A(x)$ at $A\rightarrow \infty$.
One can see that the distributions of $u$ and
$d$ quarks in fraction of the target momentum $x$ are modified in a
different way. The fraction of the momentum carried by $u$ quarks
decreases by about $4\%$. The ratio $R$, determined by Eq.~(\ref{133})
has a typical EMC shape.

Note that while for a free nucleon $x\leq 1$, for the nucleus it can
be as large as $x=A$. The developed approach enables to calculate the
quark distributions at $x>1$, describing thus the cumulative aspects of
the problem.

\subsubsection{Swelling of the nucleon}

One of the possible explanations of the EMC effect is the ``swelling"
of the nucleon inside the nucleus.  Under this assumption the structure
function of the in-medium nucleon $F_2^m(x)$ is written in terms of
that for a free nucleon $F_2^0(x)$ as $F_2^m(x)=F_2^0(xr_m/r_0)$, with
$r_m<r_0$ \cite{108}. On the other hand, the value of the proton
residue $\lambda^2$ is proportional to the square of the three-quark
wave function at the origin. Assuming that there is only one length
scale $r$ for the nucleon, we find $\lambda^2\sim r^{-6}$. Employing
Eq.~(\ref{107}) we find $(r_m-r_0)/r_0 =2\%$, in agreement with estimations
of \cite{109}.

\section{Intermediate summary}
\subsection{Reasons for optimism}

One can see that we have some reasons for optimism. The QCD sum rules
analysis confirmed that the nucleon in nuclear matter can be treated as
moving in superposition of strong vector and scalar fields of the order
of about 300~MeV.  The vector field is positive, while the scalar
field is negative, and a partial cancelation takes place.  We obtained
this result without employing a controversial conception of interaction
of the point-like nucleons.

We obtained a picture for  formation of the nucleon self energies. The
effective mass $m^*$ is formed by the exchange of noninteracting quarks
between the system of the three noninteracting quarks described by the current $j_1$ determined by Eq.(\ref{13})
which  carries the
proton quantum numbers  and the scalar condensate, which differs from
that in vacuum.  Of course, there is strong interactions between the
quarks, which form the condensate.  The vector self energy $\Sigma_V$
is formed as the exchange by noninteracting quarks between the system
of the three noninteracting quarks, carrying the proton quantum numbers
and the valence quarks of the matter, forming the vector condensate.
The  quarks, which form the condensate interact strongly between
themselves and with the other quarks of the matter.

The vector and scalar self-energies are calculated in terms of the
in-medium vector and scalar condensates, the nonlocality of the vector
condensate is included. The vector condensate can be calculated easily.
The contributions, corresponding to the nonlocal structure of the
vector condensate can be  presented in terms of the moments of the
structure functions. Thus, it is also related to observables. The
in-medium scalar condensate in the gas approximation, which has a good
accuracy near the saturation point  can be related to the observable
pion- nucleon $\Sigma$ term.  Hence, the nucleon self-energies are
expressed in terms of the condensates, which can be either calculated
in a model-independent way or related to observables.

The approach was used for calculation of other nucleon parameters,
providing reasonable results.

\subsection{Reasons for scepticism}

However, the results, described above leave some reasons for
scepticism. The most important one is the obscure role of the
four-quark ($4q$) condensates. A simple estimation for the value of the
scalar $4q$ condensate is $\langle M|\bar qq\bar qq |M \rangle=2\langle
0|\bar qq|0 \rangle \langle N|\bar qq|N \rangle\rho$. Employing this
estimation one would find both vector and scalar self-energies to be
large (hundreds MeV) and positive. This would contradict to the QHD
phenomenology.

Another obscure point is the role of the radiative corrections. As we
have seen in Sec.2, in the lowest order radiative correction to the
vacuum SR the coupling constant $\alpha_s$ is multiplied by a
numerically large coefficient. The role of radiative corrections in the
vacuum case was clarified in \cite{31}. However, the case of finite
densities require a separate analysis.

Some of the results obtained in \cite{110} are sometimes viewed as the
grounds for additional sceptical statements. One of them is a possible
strong shift of the hadron mass $m_h$ due to the resonances in the $Nh$
system, which is not included to the SR analysis. Another one is about
the importance of the large distances for the formation of the nucleon
mass, while the SR actually deal with small distances of the order
1\,GeV$^{-1}$.

\subsection{Response to the sceptical remarks}

Start with the latter statement. The nucleon wave function is indeed
formed at large distances from the center of nuclei. The case of
deuteron is a bright example. However, the NN interactions take place
at much smaller distances.  One can obtain a rather good description of
the deuteron by employing the potential $V(r)=C\delta(r)$. In Walecka
model \cite{1} the $NN$ interaction radius is of the order of $\omega$
meson inverse mass $1/m_\omega$. The duality interval (27), where the
SR are used, just corresponds to the distances responsible for the $NN$
interactions. Hence, the SR approach should provide an adequate
description of the nucleon in nuclear matter.

Resonances in $Nh$ system are described by singularities in variable
$s$ of the function $\Pi_m(q^2,s)$, defined by Eq.~(\ref{38}). We avoided these
singularities by considering $\Pi_m(q^2,s)$ at fixed value of $s$.

Contribution of the $4q$ condensates and those of the radiative
corrections should be included into our analysis. This will be done in
next Sections. Note that the calculations of the scalar $4q$ condensate
in nucleon in framework of the Nambu--Jona--Lasinio model \cite{111}
provided encouraging results. The estimation, which we gave in the
beginning of this subsection overshoot the value of condensate since
there are certain cancelations.

\section{Four-quark condensates}

\subsection{General equations for contribution of \newline the
four-quark condensates}

We shall need the expectation values
\begin{equation}
T^{XY,f_1,f_2}=\langle M|:\bar q^{f_1a}\Gamma^Xq^{f_1a'}\bar q^{f_2b}
\Gamma^Yq^{f_2b'}:|M\rangle(\delta_{aa'}\delta_{bb'}-\delta_{ab'}
\delta_{ba'}),
\label{134} \end{equation}
where the colon signs denote the normal ordering of quark operators,
$f_{1,2}$ stand for the quark flavors, $a,a',b,b'$ are the color
indices. A nonzero contribution to the function $\Pi_m$ is provided by
the antisymmetric  combination of colors - see Eq.~(\ref{13}). The basic
$4\times4$ matrices $\Gamma^{X,Y}$, acting on the Lorentz indices of
the quark operators  are
\begin{equation}
\Gamma^S=I; \quad
\Gamma^{Ps}=\gamma_5; \quad \Gamma^V=\gamma_{\mu};  \quad
\Gamma^A=\gamma_{\mu}\gamma_5; \quad
\Gamma^T=\frac{i}{2}(\gamma_\mu\gamma_\nu-\gamma_\mu\gamma_{\nu}).
\label{135} \end{equation}
Thus, they describe the scalar, pseudoscalar, vector, pseudovector
(axial) and tensor cases correspondingly.

In order to simplify the formulas, we shall not display the color
indices, keeping in mind that the quark operators are
color-antisymmetric.

The contribution of the four-quark condensates to the LHS of the sum
rules is \cite{29}
 \begin{equation}
(\Pi_m-\Pi_0)_{4q}=(\Pi_{\rho})_{4q}\ =\ \frac{1}{q^2}\Big(\sum_{X,Y}\mu_{XY}H^{XY}+\sum_{X,Y}\tau_{XY}
R^{XY}\Big)\,.
\label{136} \end{equation}
Here
\begin{equation}
H^{XY}=\langle M|\bar u\Gamma^X u \bar u \Gamma^Y u|M\rangle; \quad R^{XY}=
\langle M|\bar d\Gamma^X d \bar u \Gamma^Y u|M\rangle,
\label{137}
\end{equation}
while
\begin{eqnarray}
\mu_{XY} &=& \frac{\theta_Y}{16}\mbox{Tr}(\gamma_\alpha\Gamma^X
\gamma_\beta\Gamma^Y)\gamma_5\gamma^\alpha\hat q\gamma^\beta\gamma_5\,;
\nonumber\\
\label{138}
\tau_{XY}&=&\frac{\theta_Y}4\mbox{Tr}(\gamma_\alpha\hat q\gamma_\beta
\Gamma^Y)\gamma_5\gamma^{\alpha}\Gamma^X \gamma^{\beta}\gamma_5\,.
\end{eqnarray}
The factor $\theta_Y=1$ if $\Gamma^X$ has a vector or tensor
structure, in other cases $\theta_Y=-1$. The sign is determined by that
of the commutator between the charge conjugation matrix and $\Gamma_Y$.

Considering the $4u$ condensates, one can see that the contributions $\mu_{XY}H^{XY}$
to $\Pi_{\rho}^{4q}$ obtain nonzero values only if
the matrices $\Gamma_X$ and $\Gamma_Y$ have the same Lorentz structure.
All the structures, presented by Eq.~(\ref{135}) contribute to the $4u$
condensate. In the case of $2d2u$ condensate the factor
Tr$(\gamma_\alpha\hat q\gamma_\beta\Gamma^Y)$ does not turn to zero
only if $\Gamma_Y$ has a vector or axial structure. In the latter case
$\Gamma_X$ should have axial structure as well. In the former case it
can be either Lorentz scalar or Lorentz vector.

\subsection{Approximations for the four-quark condensates}
We shall make certain approximations for the quark condensates both in vacuum and in nuclear medium.

\subsubsection{Factorization of the vacuum condensate}

The vacuum $4q$ condensate will be considered in framework of the
factorization hypothesis \cite{19}
\begin{equation}
\langle 0|\bar u\Gamma^X u \bar u \Gamma^Y u |0\rangle=\langle 0|\bar u\Gamma^X u |0\rangle\langle 0|\bar u\Gamma^X u |0\rangle.
\label{139}\end{equation}
Thus, on this approximation the condensate $\langle 0|\bar u\Gamma^X u
\bar u \Gamma^Y u|0\rangle$ does not vanish only in the scalar case
$\Gamma^X=\Gamma^Y=I$.
\begin{equation}
\langle 0|\bar u u\bar uu |0\rangle=\langle 0|\bar uu|0\rangle
\langle 0|\bar u u|0\rangle.
\label{140}
\end{equation}
This approximation has been justified in the limit of large number of
colors \cite{84}. This is a standard approximation in the SR
calculations. As it stands now, there are no indications of noticeable
violation of the factorization relations Eqs. (\ref{139}), (\ref{140}).

\subsubsection{Gas approximation}

The modification of the $4q$ condensates in nuclear matter will be
treated in the gas approximation.  We shall go beyond in Subsection~7.5.
In the gas approximation expectation value of any operator $X$ is
\begin{equation}
\langle M|X|M\rangle=\langle 0|X|0\rangle+\rho\langle N|X|N\rangle\,,
\label{141} \end{equation}
where
\begin{equation}
\langle N|X|N\rangle=\int d^3x\Big(\langle N|X(x)|N\rangle
-\langle 0|X(x)|0\rangle\Big),
\label{142} \end{equation}
is the excess of the density $X(x)$ over the vacuum value inside the nucleon.
Of course, $\langle 0|X(x)|0\rangle$ does not depend on $x$.

Assuming that at any space point $x$
$$
\langle N|\bar u\Gamma^X u\bar u\Gamma^X u|N\rangle=(\langle N|\bar
u(x)\Gamma^X u(x)|N\rangle)^2,
$$
one can write Eq.~(\ref{142}) for $4u$ condensate in another way
\begin{eqnarray}
&& \hspace*{-0.5cm}
\langle N|\bar u\Gamma^X u\bar u\Gamma^X u|N\rangle=\int d^3x
\Big(\langle N|\bar u(x)\Gamma^X u(x)|N\rangle\ -
\nonumber\\
&&-\ \langle 0|\bar u(x)\Gamma^X u(x)|0\rangle\Big)^2+2\langle 0|\bar
u\Gamma^X u |0\rangle\langle N|\bar u\Gamma^X u |N\rangle\ +
\nonumber\\
&&+\ V_N\Big((\langle 0|\bar u u|0\rangle)^2-\langle
0|\bar u u\bar uu |0\rangle \Big). \label{143}
\end{eqnarray}
Here $V_N$ is the volume of the
nucleon. The last term vanishes under the vacuum factorization
assumption -- see Eq.~(\ref{140}). For all structures but the scalar
one the second term vanishes, since $\langle 0|\bar u\Gamma^X u
|0\rangle=0$, and Eq.~(\ref{143}) takes the form
\begin{equation}
\langle N|\bar u\Gamma^X u\bar u\Gamma^X u|N\rangle=\int d^3x
\Big(\langle N|\bar u(x)\Gamma^X u(x)|N\rangle-\langle 0|\bar u(x)
\Gamma^X u(x) |0\rangle\Big)^2.
\label{144} \end{equation}
However, for the scalar case $\Gamma_X=I$ we can write
\begin{equation}
\langle N|\bar u u\bar u u|N\rangle=\int d^3x\Big(\langle N|
\bar u(x) u(x)|N\rangle-\langle 0|\bar u u |0\rangle\Big)^2+2
\langle 0|\bar u u |0\rangle\langle N|\bar u u |N\rangle.
\label{145} \end{equation}

In similar way we can write for the $2d2u$ scalar-vector condensate,
which contributes to the RHS of Eq.~(\ref{138}) for$\tau_{SV}$
\begin{eqnarray}
&& \hspace*{-0.5cm}
\langle N|\bar d d \bar u\gamma_0 u|N\rangle=\int d^3x\Big(\langle
N|\bar d (x)d(x)|N\rangle-\langle 0|\bar d d |0\rangle\Big)\ \times
\nonumber\\
&&\times\ \langle N|\bar u (x) \gamma_0 u (x)|N\rangle
+2\langle 0|\bar d d|0\rangle\langle N|\bar u\gamma_0 u|N\rangle.
\label{146}
\end{eqnarray}
In the gas approximation we can write for the condensates, defined by
Eq.~(\ref{137})
\begin{equation}
H^{XY}=h_p^{XY}\rho_p+ h_n^{XY}\rho_n ; \quad  h_N^{XY}=\langle N|\bar
u\Gamma^X u \bar u \Gamma^Y u|N\rangle
\label{137a} \end{equation}
$$
\quad R^{XY}=r^{XY}\rho; \quad r^{XY}=\langle N|\bar d\Gamma^Xd\bar u
\Gamma^Y u|N\rangle.
$$

Further calculations require certain quark model of nucleon.

\subsection{Perturbative Chiral Quark Model}

\subsubsection{Description of the model}

The Perturbative Chiral Quark Model (PCQM) originally suggested in
\cite{27}, was developed later in \cite{112}.  Further applications are
reviewed in \cite{113}.  The nucleon is considered as a system of three
relativistic valence quarks, moving in an effective static field. The
valence quarks are supplemented by a cloud of pseudoscalar mesons,
introduced in agreement with the requirements of the chiral symmetry.
In the $SU(2)$ version of the model, which will be used here, only the
pions are included. The meson cloud is included in the lowest order of
perturbation theory.

The PCQM Lagrangian is  the sum of the terms, describing the
constituent quarks, pions and their interaction
\begin{eqnarray}
&& L=L_Q+L_{\pi}+L_{int}; \quad L_Q=\bar \psi(r)[i\hat \partial-S(r)
-\gamma_0V(r)]\psi(r);
\nonumber\\
&& L_{\pi}=\frac{1}{2}(\partial_{\mu}\phi_i(r))^2; \quad
 L_{int}=-i\bar \psi(r)S(r)\gamma_5\frac{\tau^i\phi_i(r)}{f_\pi}\psi(r).
\label{147}
\end{eqnarray}
Here $\phi_i$ is the isotriplet of the pion fields. The contribution
$L_{int}$ is the lowest order expansion of the chiral interaction term
$L_{int}=-\bar \psi(r)S(r)e^{i\gamma_5\tau^i\phi_i(r)/f_{\pi}}\psi(r)$,
$f_{\pi}$ is the pion decay constant. We did not write down the quark
mass term.

Important steps in development of the model were made in \cite{114}.
In earlier applications the pions were considered as independent
point-like degrees of freedom. In \cite{114} they were treated as
containing the ``sea" quarks ($\bar qq$ pairs) of the nucleon.

Another move touched the treatment of the constituent quarks. The
authors of \cite{114} did not solve the Dirac equation for given form
of the potentials $S(r)$ and $V(r)$, but postulated the Gaussian shape
of the constituent quark density.  It was assumed that the coordinate
part of wave function of the constituent quark can be represented  as
the product of three single-particle functions
\begin{equation}
\psi(r)=\phi(r)\chi(r); \quad \phi(r)=Ne^{-r^2/2R^2};\quad
\chi(r)=\left( \begin{array}{l}
\chi_0\\
i\beta\frac{(\sigma r)}R\,\chi_0
\end{array} \right)
\label{148} \end{equation}
with the normalization condition
$\int d^3r \bar \psi(x)\gamma_0\psi(x)=1 $ expressing the conservation
of the baryon charge.  Parameters of the model $\beta$ and $R$ are
fitted to reproduce the values of the axial coupling constant and of
the proton charge radius correspondingly. Employing the Dirac equation
one finds that the wave function  represented by Eq.~(\ref{148})
corresponds to the scalar field
$$
S(r)\ =\ M+cr^2\,,
$$
where
$M=(1-3\beta^2)/2\beta R\approx230\,$MeV can be  treated as
constituent mass of the quark.

\subsection{Four-quark condensates in the PCQM}

Now each of the condensates $h_N^{XY}$ and $r^{XY}$ defined by
Eq.~(\ref{137a}) can be presented as consisting of three contributions.
All the four operators can act on the valence quarks. This will be
denoted by the lower index $val$.  All the four operators can act on
the pions. This will be denoted by the lower index $P$.  There is also
a possibility that two operators act on the constituent quarks while
the other two act on pions.  Such interference terms will be denoted by
the lower index $J$. Hence, we can write
\begin{equation}
h_N^{X}=(h_N^{X})_{val}+(h^{X})_P+(h^{X})_J; \quad
r^{XY}=(r^{XY})_{val}+(r^{XY})_P+(r^{XY})_J.
\label{149} \end{equation}
Since in the case of $4u$ condensate only the structures with $X=Y$
contribute to the SR, we used notation $h^{XX}=h^{X}$.  In the analysis
presented in the next subsubsection we omit several terms, which are
numerically not important.  Start with the contribution of the valence
quarks.

\subsubsection{Contribution of the valence quarks}

Here the products of the quark operators are averaged over the state
$|\tilde N\rangle$ of three constituent quarks, described by the wave
functions presented by Eq.~(\ref{148}). Due to the normal ordering of
the quark operators we find for neutrons $(h_n^{X})_{val}=0$, while
for protons
$$
(h_p^{X})_{val}=\langle U|\bar u\Gamma^X u \bar u \Gamma^X u|U\rangle\,.
$$
Here $|U\rangle$ denotes the vector of state of two constituent $U$
quarks.

For the scalar case the product of the PCQM operators describes the
excess of the quark density with respect to the vacuum value
\begin{eqnarray}
&& \bar u^{PCQM}(r)u^{PCQM}(r)=\bar u(r)u(r)-\langle 0 |\bar u
u|0\rangle;
\nonumber\\
\label{150}
&& \langle U|\bar u^{PCQM}u^{PCQM}|U\rangle=2\int d^3r\bar\psi(r)\psi(r).
\end{eqnarray}
Thus,
employing Eqs. (\ref{142}), (\ref{143}) we find for the scalar case
$X=Y=S$
\begin{equation}
(h_p^{S})_{val}=2\langle 0|\bar u u|0\rangle
\langle \tilde N|\bar u u|\tilde N\rangle+\int d^3r \Big(\bar
\psi(r)\psi(r)\Big)^2.  \label{151}
\end{equation}
The last factor
$\langle\tilde N|\bar u u|\tilde N\rangle$ in the first term on the RHS
has the meaning of the contribution of the valence quarks to parameter
$\kappa_N$ defined by Eq.~(\ref{4}). In the PCQM model $\langle\tilde
N|\bar u u|\tilde N\rangle=2\int d^3r \bar \psi(r)\psi(r)=1.08$.

For the other structures of $4u$ condensate we have just
\begin{equation}
(h_p^{X})_{val}\ =\ \int d^3r \Big(\bar \psi(r)\Gamma^X\psi(r)\Big)^2.
\label{152} \end{equation}

For the $2d2u$ condensate we can write
$$
(r_N^{XY})_{val}\ =\ \langle D|d \Gamma^X d|D \rangle\langle U|\bar
u\Gamma^Y u|U\rangle\, .
$$
Here $|D\rangle$ is the vector of state of
constituent $D$ quark.  Analysis, similar to that, made for the $4u$
condensates provides for the scalar-vector case $X=S, Y=V$
\begin{equation}
(r^{SV})_{val}=2\langle 0|\bar d d|0\rangle \langle \tilde N|\bar u\gamma_0 u|\tilde N\rangle+2\int d^3r
\bar \psi(r)\psi(r)\bar \psi(r)\gamma_0\psi(r).
\label{153} \end{equation}
Note that $\langle\tilde N|\bar u\gamma_0 u|\tilde N\rangle=2$. For other
structures
\begin{equation}
(r^{XX})_{val}\ =\ \int d^3r \Big(\bar\psi(r)\Gamma^X\psi(r)\Big)^2.
\label{154} \end{equation}

\subsubsection{Contribution of the sea quarks}

Now all the quark operators act on the pions. The contribution is expressed in terms of pion expectation values
of the four-quark operators
 \begin{equation}
(\Pi_{4q})_{pions}=\frac{1}{16}
\Big(\sum_{X,\alpha}\langle \pi^{\alpha}|
\mu_X\bar u\Gamma^X u\bar u\Gamma^X u
+\tau_X\bar d\Gamma^X d\bar u\Gamma^X u|\pi^{\alpha}\rangle
\Big)\frac{\partial\Sigma^{\alpha}}{\partial m_{\pi}^2}.
\label{155} \end{equation}
Here $\alpha$ denotes the pion isotopic states, $\Sigma$ stands for
the sum of the self-energy pion loop and the pion exchange
contribution. The pion expectation values can be expressed in terms of
the vacuum expectation values of the four-quark operators. This was
done in \cite{115} by employing the current algebra technique.  Using
the results of \cite{115} we
find
\begin{equation}
\sum_X\langle \pi^{\alpha}| \mu_X\bar u\Gamma^X u\bar u\Gamma^X
u +\tau_X\bar d\Gamma^X d\bar u\Gamma^X u|\pi^{\alpha}\rangle=0,
\label{156}
\end{equation}
and thus
\begin{equation}
(\Pi_{4q})_{pions}=0.
\label{157} \end{equation}

Hence, there is no such thing as ``the contribution caused by the sea
quarks only".  This is true for any model where the sea quarks are
contained in the pions.  Note that in the calculations of \cite{115}
the factorization assumption Eq.~(\ref{140}) have been used.

\subsubsection{Contribution of the interference terms}

In the case of the scalar condensate the interference terms can be
written as
\begin{equation}
(h_p^{I})_{J}=2\sum_i\langle\tilde N|H_{int}|\tilde N_i,\pi\rangle\langle
\tilde N_i|\bar u u|\tilde N\rangle\langle \pi|\bar u u|\pi\rangle\langle
\tilde N,\pi|H_{int}|\tilde N\rangle,
\label{158} \end{equation}
 where $H_{int}$ is the interaction between the constituent quark $Q$
and the pion. One can write $ H_{int}=-L_{int}$, with $L_{int}$ given
by Eq.~(\ref{147}). Here two quark operators act on the pion, while the
other two operators act on a constituent quark. The latter can be the
same as one in the matrix element of the interaction $H_{int}$ or
another one. Actually in the sum over the states of the constituent
quarks $\tilde N_i$ only those, corresponding to the nucleon are
included, i.e. $|\tilde N_i\rangle=|\tilde N\rangle$.

In the case of pseudoscalar and axial operators the interference can
take place in the first order of the $\pi Q$ interaction. This happens
because the matrix elements  $\langle \pi|\bar q\Gamma^Xq|0\rangle$
have nonzero values in  these cases. The contribution of such ``vertex
interference" is
\begin{equation}
(h_p^X)_{J}=\langle \tilde N|H_{int}|\tilde N,\pi\rangle
\langle \tilde N,\pi|\bar q\Gamma^Xq \bar q\Gamma^Xq  |\tilde N\rangle +
\langle \tilde N,\pi|\bar q\Gamma^Xq \bar q\Gamma^Xq  |\tilde N\rangle
\langle \tilde N|H_{int}|\tilde N,\pi\rangle,
\label{159} \end{equation}
with $X$ labelling an axial or pseudoscalar. Thus in one of the
vertices interaction $H_{int}$ is replaced by the amplitude of
injection of four quarks and antiquarks. For neutral pions the latter
can be evaluated  as
$$
\langle \tilde N,\pi^0|\bar q\Gamma^Xq \bar
q\Gamma^Xq  |\tilde N\rangle=\langle \pi^0|\bar q\Gamma^Xq
|0\rangle\langle \tilde N|\bar q\Gamma^Xq|\tilde N\rangle\,,
$$
$$
\langle \tilde N|\bar q\Gamma^Xq \bar q\Gamma^Xq  |\tilde N,
\pi^0\rangle=\langle 0|\bar q\Gamma^Xq|\pi^0\rangle\langle \tilde
N|\bar q \Gamma^X q|\tilde N \rangle,
$$
with analogous relations for
the charged pions.

Now we shall employ the values of the four-quark condensates, obtained in \cite{28} in the
SR equations.

\section{Contribution of the higher order terms}

\subsection{Symmetric matter with the four-quark condensates}

We can write the general equation for the contribution of the $4q$ terms
\begin{equation}
(\Pi)_{4q}=\Big (X^q_{4q}\frac{\hat q}{q^2}+X^P_{4q}\frac{(Pq)}{m^2}
\frac{\hat P}{q^2}
 +X^{I}_{4q}\frac{(Pq)}{m^2}\frac{I}{q^2}\Big)\frac{a}{(2\pi)^2}\rho.
\label{160} \end{equation}
Note that $a=-(2\pi)^2\langle 0|\bar uu|0\rangle\approx0.55\rm\,GeV^3$
(Eq.~(\ref{24})) is just a convenient scale for presentation of the results. It does not
reflect the chiral properties of $(\Pi)_{4q}$.

The coefficients $X^i_{4q}$ are obtained by using the complete set of
the nucleon four-quark condensates \cite{28} and the results of the
previous Section. In calculations of the factor  $X^{q}_{4q}$ all
contributions are numerically important due to their partial
cancellations. The coefficient $X^P_{4q}$ is determined mainly by the
$2d2u$ vector--vector condensate. The factor $X^I_{4q}$ is dominated by
the vector--scalar $2d2u$ condensate, with the first term on the RHS of
Eq.~(\ref{153}) providing the largest contribution.

The numerical values are
 \begin{equation}
X^q_{4q}=-0.11; \quad   X^P_{4q}=0.57; \quad X^I_{4q}=1.27.
\label{161} \end{equation}
We took into account the nonlocal structure of the vector condensate
in the factor $X^I_{4q}$.

Now we estimate the ratio of contributions of the $4q$ condensates to
those of the condensates with $d=3$, provided by Eqs. (\ref{65}),
(\ref{66}). The effective values of momenta are $|q^2|\sim1\,$GeV$^2$.
Putting logarithmic factors in Eqs. (\ref{65}), (\ref{66}) equal to
unity, we find that this ratio is less than 0.1 in the $\hat q$
structure of the QCD sum rules. It is about 0.13 in the $\hat P$
structure. The ratio is about $1/4$ in the scalar structure $I$ for
$\kappa_N=8$. It becomes smaller for larger values of $\kappa_N$. As
we shall see below, the gas approximation result for $(X^I)_{4q}$ provided by
Eq.~(\ref{161}) overestimates the value. Thus, the values of
contributions of $4q$ condensates are consistent with the assumption on
the convergence of the OPE series.

The contributions of dimension $d=6$ to the sum rules -- see
Eqs. (\ref{57}), (\ref{64}) are
\begin{equation}
\tilde A_6=-8\pi^2aX^q_{4q}\rho\,; \quad
\tilde P_6=-8\pi^2a\frac{s-m^2}{2m}X^P_{4q}\rho\,; \quad
\tilde B_6=-8\pi^2maX^I_{4q}\rho\,.
\label{162a} \end{equation}
Actually, $\tilde B_6$ contains also the terms of the higher
dimension, since it includes the nonlocality of the vector condensate.

The medium induced $4q$ condensates on the LHS of the sum rules
correspond the exchange by strongly correlated four-quark systems on
the RHS. This may be a local two-meson exchange, with two mesons
interaction with the nucleon at the same point -- see Fig.~16.  Another
possible interpretation is the exchanges by the four-quark mesons, if
there are any \cite{116}.

However, the leading contribution to the scalar structure, determined
by the first term on the RHS of Eq.~(\ref{153})
has another interpretation.
Here two quarks are exchanged with vacuum, while two other ones are
exchanged with the valence quarks of the matter. The contribution of this scalar-vector
condensate to $X^I_{4q}$ is
\begin{equation}
X^{SV}_{4q}\ =\ -\frac{2(Pq)}{3q^2}\frac{\langle 0|\bar dd|0\rangle
v(\rho)}{m} \label{161a}
\end{equation}
This term contains the
expectation value $\langle 0|\bar d d|0\rangle$ as a factor. On the
other hand this term is proportional to the vector condensate,
contributing, however, to the scalar Lorentz structure of the nucleon
equation of motion Eq.~(\ref{1}). On the RHS of the SR this can be
interpreted as the vector meson exchange with the anomalous structure
of the vertex. We discussed such contributions in the analysis of the
charge-symmetry breaking forces in Subsec.~5.2 -- see Fig.~13.

Note that if we go beyond the gas approximation, the discussed term
contains rather the expectation value $\langle M|\bar d
d|M\rangle=\kappa(\rho)/2$ instead of the vacuum value $\langle 0|\bar
d d|0\rangle$. As we see from Eq.~(\ref{94}),
$|\kappa(\rho)/\kappa(0)|<1$, with the reduction of about $30-50\%$ at
saturation density. Hence, the gas approximation overestimates strongly
the value of $X^I_{4q}$. In the next to leading order beyond the gas
approximation we must consider the contribution, in which two pairs of
quarks of the $4q$ condensate go to two different nucleons. Such terms
should be considered together with the other three-body contributions.

As we have seen in Sec. 4, the nucleon parameters depend on the
nucleon expectation value $\kappa_N$, which is known with large
uncertainties.  We show density dependence of the nucleon parameters
for $\kappa_N=8$ obtained in \cite{29} in Fig.~17. In Fig.~18 we present the dependence of
nucleon self-energies on $\kappa_N$ at the saturation value of density.

For $\kappa_N=8$ we find at $\rho=\rho_0$
\begin{equation}
\Sigma_V=160\,\mbox{MeV}; \quad m^*-m=-380\,\mbox{MeV}.
\label{161b} \end{equation}
Thus, the $4q$ condensates subtract 90~MeV from the vector self-energy
and 170~MeV from the effective mass $m^*$, provided by Eq.~(\ref{107}).  The potential energy
$U=-180$ MeV looks discouraging. However, inclusion of the radiative
corrections improves the situation.

\subsection{Radiative corrections}

The radiative corrections to the vector condensate, calculated
recently in \cite{123} appeared to be rather large.  Thus we have to
analyze the role of the radiative corrections more carefully.

In Subsec. 2.3 we clarified  the role of these corrections in vacuum
SR. Here we focus on the terms  $\Pi_{\rho}$, provided by medium. The
contributions of the condensates of the lowest dimension $d=3$,
provided by Eq.~(\ref{66}) are actually written in the form
\begin{equation}
\tilde A_{3 \rho}=\frac{\tilde A_{3 \rho}^{(0)}}{L^{\gamma_q}}; \quad
\tilde B_{3 \rho}=\frac{\tilde B_{3 \rho}^{(0)}}{L^{\gamma_I}};\quad
\tilde P_{3 \rho}=\frac{\tilde P_{3 \rho}^{(0)}}{L^{\gamma_P}}\,,
\label{171}\end{equation}
where the anomalous dimensions $\gamma_q=\gamma_P=4/9, \gamma_I=0$.
Here the upper index $0$ denotes that all radiative corrections are
neglected. The factors $L^{-\gamma_i}$ include the sum of the terms
$(\alpha_s \ln{q^2})^n$. This is called the Leading Logarithmic
Approximation (LLA). Actually, in our previous analysis we included the
contributions of condensates with $d=3$ with the radiative corrections
treated in LLA. Note that in computation of the next to leading order
terms we used the structure functions \cite{62}, which reproduce their
moments with proper anomalous dimensions. We did not include the
radiative corrections to the $4q$ condensates, since these terms were
obtained in framework of a model, which does not contain gluons. Also,
in comparison with the steep $q^2$ dependence of the terms, containing
the $4q$ condensates, caused by the ``normal" high dimension, the role
of anomalous dimension is relatively small.

Now we shall include the corrections of the order $\alpha_s$ beyond
the LLA. We present
\begin{equation}
\tilde A_{3\rho}=\tilde A_{3\rho}^{(0)}t_a\,; \quad
\tilde B_{3 \rho}=\tilde B_{3\rho}^{(0)}t_b\,;\quad
\tilde P_{3 \rho}=\tilde P_{3 \rho}^{(0)}t_P\,.
\label{172}\end{equation}
Here
\begin{equation}
t_i\ =\ \frac{r_i}{L^{\gamma_i}}\,,
\label{173}\end{equation}
with
$$
r_i\ =\ 1+c_i\alpha_s/\pi,
$$
where the second term on the RHS is the lowest order radiative correction beyond the LLA. While
$c_I=3/2$ \cite{49}, it was found in \cite{123} that $c_q=7/2$ and
$c_P=15/4$. Thus, employing a straightforward estimation, we can expect
a $40-50\%$ change of the nucleon parameters due to radiative
corrections.

We shall compare the three cases. All the radiative corrections are
ignored, i.e. all $t_i=1$. Corrections are included in framework of
LLA, i.e. all $r_i=1$. The third possibility is that in addition to LLA
the lowest order $\alpha_s$ correction is included beyond the LLA.

We carry out the computations for $\kappa_N=8$, assuming
$\alpha_s(1\,\rm GeV)=0.37$. The results of calculations are presented
in Table~1. We see the LLA corrections  subtracts 70 and 50~MeV
from the vector self-energy and the effective mass correspondingly.
Inclusion of the corrections beyond the LLA makes the values of
$\Sigma_V$ and $m^*$ closer to those obtained with total neglect of the
radiative corrections. The results are illustrated by Fig.~19.

Another problem is the dependence of numerical results on the actual
value of $\alpha_s$. The latter is determined by the value of
$\Lambda_{QCD}$. The authors of \cite{123} used
$\alpha_s(1\rm\,GeV)=0.47$, while $\alpha_s(1\rm\,GeV)=0.55$ was
employed in \cite{20a}. These values corresponds to
$\Lambda_{QCD}=0.23\,$GeV and $\Lambda_{QCD}=0.28\,$GeV.  These
variations of $\alpha_s$ change the nucleon self-energies by several
MeV, affecting mostly the value of the nucleon residue \cite{32}.

\subsection {Asymmetric matter}

\subsubsection{Inclusion of condensates of higher dimension}

The nonlocal structure of the vector condensate is included in the same way as in the case of the symmetric matter.

 Employing the values of the four-quark condensates obtained in the
previous Section  we find the $\beta$ dependence of the contributions
of the four-quark condensates, defined by Eq.~(\ref{160})
 \begin{equation}
X^q_{4q}=-0.11-0.21\beta\,; \quad  X^P_{4q}=0.57+0.09\beta\,; \quad
 X^I_{4q}=1.27-0.61\beta\,.
 \label{170} \end{equation}

Including also the nonlocal structure of the vector condensates, we
 solve the SR equations. In the actual numerical calculations we employ
 the PCQM value  $\zeta_p= 0.54$ for the isotope asymmetric condensate
 of dimension $d=3$ -- see Eq.~(\ref{166}). In Fig.~20, we show the
values of the proton and neutron self-energies in neutron matter ($\beta=1$),
compared to thouse in symmetric matter. In other words, these are the
proton self-energy values for the nuclear matter with $\beta=-1,0,1$ \cite{30}.

One can see that the $\beta$ dependence of self-energies has  the same
qualitative features as that determined by Eq.~(\ref{169}). In the
matter with neutron excess $\beta>0$ the neutron vector self-energy is
larger than the proton one. The proton effective mass is larger than
the neutron one. In both cases the $\beta$ dependence is relatively
weak.

\subsubsection{Comparison with results of nuclear physics}

Our results for the difference of the effective masses of neutron and
proton $m^*_{np}=m_n^*-m_p^*$ appears to be twice smaller than the
result of \cite{117}, but only $30\%$ smaller than that of \cite{118}.
However, a discrepancy from the relativistic Brueckner--Hartree--Fock
(RHFB) calculations  \cite{118} is smaller. Their results for
$\beta=0.2$ are $\Sigma_V^n-\Sigma_V^p=30\,$MeV and
$m^*_{np}=-15\,$MeV.  We obtained 20~MeV and $-15$~MeV
correspondingly for these parameters. Another RHFB calculation
\cite{119} provided results, which are very close to ours one. Their
values for $\beta =0.75$ are $\Sigma_V^n-\Sigma_V^p=80\,$MeV and
$m^*_{np}=-50\,$MeV, while we found 80~MeV and $-55~$MeV
correspondingly.

Note that the nuclear physics calculations provide $m^*_{np}<0$ for
$\beta>0$. Our values have the same sign after the four-quark
condensates are included. The lowest dimension solution expressed by
Eq.~(\ref{169}) provides  $m^*_{np}>0$.

We found unexpectedly satisfactory agreement for the potential energy
splitting $U_{np}=U^{(n)}-U^{(p)}$ and even for the $\beta$ dependence
of the average binding energy per nucleon $\varepsilon(\rho, \beta)$.
Various approaches \cite{120,121} provided $U_{np}\approx60\,$MeV,
while our result is  $U_{np}=40\,$MeV. The ``symmetry energy" $\varepsilon^{sym}=1/2\partial{\varepsilon(\rho_0,
\beta)}/\partial{\beta}$ is 29~MeV in our approach. All earlier
\cite{117}--\cite{121} and nowadays \cite{122} calculations  provide
$\varepsilon^{sym}\approx30\,$MeV. We did not expect a good  agreement
here, since these parameters contain the large terms, which cancel each
other to large extent.

\subsection{Many-body interactions}

The structure and the role of the three-body (many-body) nuclear
interactions is much studied nowadays -- see, e.g. \cite{18,124,125}.
In the SR approach such interactions emerge in a natural way.

There are two types of many-body interactions in our approach. One of
them is connected with the four-quark condensate. Another one manifests
itself in the baryon-hole excitations, taken into account by the
in-medium pion propagator.

In Sec.~6 the contribution of four-quark condensates was included in
the gas approximation. Our probe nucleon exchanged both pairs of quarks
with the same nucleon of the matter. Beyond the gas approximation two
quark pairs can be  exchanged with different nucleons of the matter.
This corresponds to the three-body interactions.

Such terms are shown in Fig.~21. Their contribution, e.g. to the LHS
of  Eq.~(\ref{57}) for the $q$ structure ${\cal L}_m^q(M^2, W_m^2 )$
can be estimated as $4\pi^4\kappa_N^2\rho^2$. At the saturation value
of density this can change  the value of ${\cal L}_m^q(M^2, W_m^2 )$
by about $10\%$. This will lead to relative changes of the same order
 of the nucleon self-energies. At larger values of density the role of
 these terms increases.

As we discussed in Sec.~3, the contributions, involving $n\geq2$
nucleons of the matter have the branching points in  variable
$S_n(q^2)=(nP+q)^2$. Generally speaking, this means that we need a
 more complicated model of the spectrum. However, we see that in the
 OPE series the three body terms do not contain branching points in
 $q^2$. In framework of our model for the higher lying states they
 contribute only to the pole term on the RHS of the sum rules.

Adding gluon interactions with the matter, i.e. including the
 four-quark-gluon condensates, we obtain the terms, involving larger
 number of nucleons of the matter, corresponding to many-nucleon
 forces. They contribute to the higher order OPE terms. Due to the
 small value of the in-medium gluon condensate we expect such
 contributions to be small.

Another type of many-body interactions manifests itself in the
 nonlinear contribution to the scalar condensate $S(\rho)$. As we saw
in Subsection~4.5, it is determined by the pion cloud. The general form
of the contribution is
\begin{eqnarray}
S &=& -\int\frac{d^3p}{(2\pi)^3}\int \frac{
d^4k}{i(2\pi)^4}\sum_B\Big(\Gamma_B(k)D^{(m)2}(k)G_B(p-k)\Gamma_B(k)\ -
\nonumber\\
\label{174}
&&-\ \Gamma_B^{(0)}(k)D^{(0)2}(k)G_B^{(0)}(p-k)\Gamma_B^{(0)}(k)\Big)\langle
\pi|\bar q q  |\pi\rangle,
\end{eqnarray}
where summation over spin and isospin
variables is assumed.  Here $G_B$ is the propagator of intermediate
baryon in nuclear matter, $D^{(m)}(k)$ is the pion propagator,
renormalized by the particle-hole excitations of medium. The vertices
of $N\pi B$ interaction in nuclear matter are denoted as $\Gamma_B$.
Integration over momenta of the nucleons of the matter is limited by
condition $p\leq p_F$. Upper index $(0)$ denotes the vacuum functions.
The second term on the RHS of Eq.~(\ref{174}) subtracts the terms,
which are already included in the physical nucleon. For
$D^{(m)}=D^{(0)}; \Gamma_B=\Gamma^{(0)}_B$ the RHS of Eq.~(\ref{174})
turns to that of Eq.~(\ref{114}). The contribution is illustrated by
Fig.~22.

The pion propagator $D^{(m)}(k)$ includes the multi-nucleon effects. Thus,
Eq.~(\ref{174}) includes the many-body interactions. However, the
rigorous calculation of its RHS is a part of a self-consistent problem.

\section{Self-consistent scenario}

We have seen that the shape of the density dependence of the quark
condensates $\kappa(\rho)$ is very important for the hadronic physics.
It is also important for description of the matter as a whole, since
characterizes a degree of restoration of the chiral symmetry with the
growing density. On the other hand, we saw that it determines the main
density dependence of the effective mass
$m^*(\rho)\approx m^*(\kappa(\rho))$.

The linear part of the density behavior of $\kappa(\rho)$ is related
to observables and can be calculated in a model-independent way.
However, the nonlinear contribution, expressed by Eq.~(\ref{174})
depends on several hadron parameters. If the intermediate baryon in
Eq.~(\ref{174}) is a nucleon, these are the nucleon effective mass $m^*$
and the nucleon pion coupling constant $g_{\pi NN}(\rho)$. The latter
can be expressed in terms of the pion decay constant $f_\pi(\rho)$ and
the nucleon isovector axial coupling $g_A(\rho)$ via the
Goldberger--Treiman relation \cite{126}
 $$
 g_{\pi NN}/2m\ =\ g_A/f_\pi\,.
 $$
 The density dependence of $g_A$ and $f_{\pi}$ can also be obtained in
 the SR approach.  One should add similar equations for the
contribution of delta-isobars to the RHS of Eq.~(\ref{174}).  Thus we
come to self-consistent set of equations which can be solved within the
QCD sum rules approach.
\begin{equation}
m^*=m^*(\kappa); \quad f_\pi=f_\pi(\kappa); \quad g_A=g_A(\kappa);
\quad \kappa=\kappa(m^*, f_{\pi}/g_A). \label{175}
\end{equation}

Of course, the SR approach should be combined with some other ones.We
have seen already that the calculation of the four-quark condensate
required a quark model for the nucleon. Also, the RHS of
Eq.~(\ref{174}) contains the in-medium pion propagator. It satisfies
the Dyson equation
$$
D^{(m)}=D^{(0)}+D^{(0)}\Pi_{\pi}D^{(m)},$$ where
the pion polarization operator $\Pi_{\pi}$ describing the baryon-hole
excitations of the matter. The operator $\Pi_{\pi}$ depends strongly on
the hadron correlations at the distances $r\geq 1/m_{\pi}$,
corresponding to small momenta. They can be included by employing of
the Finite Fermi System Theory (FFST) \cite{127}, \cite{137}. Thus, in order to
obtain the many-body effects  it is reasonable  to combine the SR
approach with the FFST.

\section{Summary}

We demonstrated that the QCD sum rules provide a consistent formalism
for solving various problems of nuclear physics.

We expressed the nucleon parameters in nuclear matter in terms of the
in-medium QCD condensates. In symmetric matter the leading
contributions to the vector and scalar self-energies are expressed in
terms of vector and scalar condensates. The vector condensate $v(\rho)$
can be calculated easily. In the gas approximation, corresponding to
inclusion only of the two-body interactions, the scalar condensate
$\kappa(\rho)$ is related to the observable pion--nucleon sigma-term.

In other words, exchange by the strongly correlated quark systems
(mesons) is expressed in terms of exchange by the system of weakly
interacting quarks with the same quantum numbers. Thus the nucleon
self-energies are obtained without employing a controversial conception
of interaction of the point-like nucleons.

Inclusion of the condensates of the lowest dimension confirmed that
the nucleon in nuclear matter can be treated as moving in superposition
of strong vector and scalar fields of the order of about 300~MeV.
The vector field is positive, while the scalar field is negative, and
a strong cancellation takes place.

This result is not altered by inclusion of the
four-quark condensates. Thus we can expect the convergence of the
series with successive inclusion of the condensates of higher
dimension. Note, however, that the calculation of the in-medium
four-quark condensates requires employing a quark model of nucleon.

We used the approach for calculation of other nucleon characteristics
in nuclear matter. We calculated the nucleon self-energies in
asymmetric nuclear matter and found their dependence on the difference
between the densities of the neutrons and protons. We calculated the
in-medium quenching of isovector axial coupling constant. We found the
neutron-proton mass difference, providing at least qualitative
explanation of the Nolen--Schiffer anomaly.

Also, we demonstrated that our approach enables to investigate
internal structure of the nucleon, providing description of some
features of the EMC effect. Such problems are unaccessible for nuclear
physics.

Inclusion of the nonlinear density dependence of the scalar condensate
$\kappa(\rho)$ corresponds to taking into account the many-body
interactions. We show that the nonlinear terms in $\kappa(\rho)$  are
provided mainly by the pion cloud. A simple inclusion of the
contributions beyond the gas approximation signals on a possible
saturation mechanism, which is due to the many-body effects. Thus it
differs from that in the Walecka model.

A more rigorous treatment of the nonlinear contributions to
$\kappa(\rho)$ requires solution of the self-consistent problem, in
which the coupling constants $g_{\pi NN}$ and $g_{\pi N\Delta}$ as well
as the effective mass of the $\Delta$ isobar should be also determined
as functions of $\kappa$, while $\kappa$ should be expressed in terms
of these parameters. This would provide the dependence of $\kappa
(\rho)$. This is an important parameter of the matter itself, since it
shows the degree of restoration of the chiral symmetry.

This analysis requires employing of the pion propagator, which is
renormalized by the particle-hole excitations of the matter. These
excitations can be included by using the Finite Fermi System Theory,
formulated and developed by A.~B.~Migdal and his colleagues several decades ago.

The work on the project is in progress.

\section{Epilogue}

Many years ago the authors of this paper participated in a very
interesting discussion on the perspectives of nuclear physics, which
took place at the traditional Petersburg (Leningrad at that time)
Winter School of Physics. In his lecture A.~B.~Migdal formulated two
questions, he wished to be answered in the future:

\begin{itemize}
\item How do the QCD condensates change in nuclear medium\,?

\item  What is the connection between the in-medium QCD condensates
and hadron parameters\,?
 \end{itemize}

 We hope that the present paper  makes the first steps to answer these questions.

\subsection*{Acknowledgements}

We thank V.~Braun, B.~L~ Ioffe, L.~Kisslinger, E.~E.~Saperstein and
C.~M.~Shakin for fruitful discussions.  We also thank G.~Stepanova for
assistance in preparation of the manuscript. The authors
acknowledge the partial support by the RFBR grant
12-02-00158 and by the grant RSGSS -  4801.2012.2.

\newpage

\begin{table}
\caption{
Nucleon parameters at the saturation value of nucleon
density. Line~1:~all radiative corrections are neglected.
Line~2:~radiative corrections are included in LLA. Line~3:~corrections
$\sim\alpha_s$ are included beyond the LLA (BLLA).}
\begin{center}
\begin{tabular}{lcccc}
\hline\hline
& $\Sigma_v$ & $m^*-m$ & $\lambda^2_m$ & $W^2_m$\\
& (MeV) & (MeV) & (GeV$^6$) & (GeV$^2$) \\
\hline

No corrections & 229 & -329 & 1.10 & 1.72 \\
LLA  &           160 & -380 & 1.10 & 1.77 \\
BLLA &           271 & -300 & 1.41 & 1.75 \\
\hline\hline
\end{tabular}
\end{center}
\end{table}

\bigskip
\begin{table}
\caption{
Dependence of the nucleon parameters on the values of
$\Lambda_{\rm QCD}$ and $\alpha_s$ at the phenomenological saturation
value of nucleon density with radiative corrections included beyond the
leading logarithmic approximation (BLLA). The results are presented for
$\Lambda_{\rm QCD}=0.23\,$GeV, $\alpha_s(\rm 1\,GeV^2)=0.47$, and
$\Lambda_{\rm QCD}=0.28\,$GeV, $\alpha_s(1\rm\,GeV^2)=0.55$. The
corresponding values of vacuum parameters are given in brackets.}
\begin{center}
\begin{tabular}{lcccc}
\hline\hline
$\alpha_s$ & $\Sigma_v$ & $m^*-m$ & $\lambda^2_m$ & $W^2_m$\\
(1 GeV$^2$) & (MeV) & (MeV) & (GeV$^6$) & (GeV$^2$) \\
\hline

0.47 & 269 & $-291(m=0.93\,$GeV) & 1.65(2.35) & 1.89(2.13)\\
0.55 & 264 & $-289(m=0.94\,$GeV) & 1.81(2.61) & 1.99(2.26)\\
\hline\hline
\end{tabular}
\end{center}
\end{table}
\clearpage

\newpage
\centerline{\bf Figure Captions}

 Fig.~1. Main contributions to the LHS of the vacuum sum
rules -- Eq.~(19).  Helix line stands for the correlator, solid lines
denote the quarks, dotted lines stand for the quarks composing the
scalar condensate.  Figs.~$a$, $b$ and $c$  correspond to the
 contributions $A_0$, $B_3$ and $A_6$.

Fig.~2. A typical radiative correction. The dashed line denotes the
 gluon.

 Fig.~3. Contributions to the singularities of the correlator in $s$
 channel $(a)$ and in $u$ channel $(b)$.

Fig.~4. Self-energy insertions to the nucleon pole contributions:
 the mean field approximation $(a)$,
 the self-energy in the direct channel $(b)$ and the exchange
 contribution $(c)$.

Fig.~5. One of the contributions, containing a higher lying
 singularity in $q^2$ at fixed $s$. Helix line stands for the
 correlator. Dashed lines stand for the mesons,  solid line is for the
 nucleon and thick the solid line denotes the nuclear matter.

Fig.~6. A singularity in $u$ channel, containing an intermediate state
 with the baryon number $B=0$. The meaning of the lines is the same as
 in Fig.~5.

Fig.~7. Contributions of the condensates of lowest dimension to the
left-hand side of the sum rules in nuclear matter. The dotted lines in
 $(a)$ stand for the quarks of the scalar  condensate. The
 dashed--dotted  lines in $(b)$ denote the quarks of the  vector quark
 condensate. The meaning of the other lines is the same as in
 previous figures.

Fig.~8. Contributions of the condensates of lowest dimension to the
 right-hand side of the sum rules in nuclear matter. The horizontal
 line denotes the nucleon, thick lines show the matter. The wavy and
 dashed lines stand for vector and scalar mesons.

Fig.~9. The dependence of the functions $f_q$, $f_P$ and $f_I$
defined by Eq.~(\ref{82}) on the Borel mass $M^2$.

Fig.~10. Density dependence of the nucleon parameters for
$\kappa_N=8$: the vector self-energy and effective mass $(a)$;
the residue $\lambda_m^2$ and the threshold $W_m^2$ $(b)$, related to their
vacuum values.  The $x$ axis corresponds to the density, related
 to its saturation value.  Solid line is for solution of Eq.(57) with
 condensates of dimensions $d=3$ and $d=4$ taken into account. Dashed
 lines show the approximate solution, corresponding to Eqs. (89)--(91).

Fig.~11. Dependence of the nucleon parameters  on the value of
 $\kappa_N$ at saturation density.  Solid line is for the effective
 mass $m^*$, dotted line shows the vector self-energy $\Sigma_V$,
dashed and dash--dotted lines are for the $\lambda_m^2$ and $W_m^2$
correspondingly. All parameters are related to their vacuum values (the
vector self-energy is related to the vacuum value of the nucleon mass)
given by Eq.(\ref{29}).

Fig.~12. Simplest contribution to the nonlinear scalar condensate.
Solid lines labeled as 1 and 2 are the nucleons of the matter.
Dashed line denotes a $\pi$ meson.

Fig.~13. Anomalous Lorentz structure of nucleon interactions with
matter.  Notation are the same as in Fig.~8.

Fig.~14. Second order interaction of the hard photon (dashed lines)
with the correlator. The dark blobs denote interaction with the matter.
Other notations are the same as in Fig.~5.

Fig.~15. The in-medium changes of the $d$ quark distribution (dashed
curve) and of the $u$ quark distribution (dot-dashed curve) of the
fraction $x$ of the momentum of the target nucleon. The solid curve
presents the function $R-1$ with the ratio $R$, defined by Eq.~(126).

Fig.~16. Interactions on the RHS of the sum rules, corresponding to
inclusion of the four-quark condensates. The wavy lines denote mesons.

Fig.~17. Density dependence of the nucleon parameters for $\kappa_N=8$
with the condensates of higher dimensions included: the vector
self-energy and effective mass $(a)$; the residue $\lambda_m^2$
and the threshold $W_m^2$ $(b)$, related to their vacuum values.  The
$x$ axis corresponds to the density, related to its saturation
value.

Fig.~18. Dependence of the nucleon parameters on the value of
$\kappa_N$ at saturation density.  Condensates of higher dimension are
included.  The meaning of the curves is the same as that for  Fig.~11.

Fig.~19.  Density dependence of the vector self energy
$\Sigma_V$ and of the scalar self energy $m^*-m$. The nuclear
matter density $\rho$ is related to its saturation value $\rho_0$.
Dotted lines: all radiation corrections are neglected. Dashed
lines: radiative corrections are included in the Leading Logarithm
Approximation (LLA). Solid lines: Corrections of the order
$\alpha_s$ are included perturbatively beyond the LLA.

Fig.~20. The density dependence of the vector self-energy $\Sigma_V$
$(a)$ and of the effective mass $m^*$ $(b)$ in isospin
asymmetric matter. The solid line is for symmetric matter. The dashed
and dotted lines are for the proton and neutron characteristics in
neutron matter ($\beta=1$).

Fig.~21. A contribution of the three-body forces to the LHS of the sum
rules. Two pairs of quarks from the four-quark condensate interact
with two different nucleons of the matter, shown by the dotted circles.

Fig.~22. Interaction of the operator $\bar qq$ (the dark blob) with
pion field. The solid line denotes the nucleon of the matter; the wavy
line stands for the pion. The bold wavy line denotes the pion
propagator renormalized by the baryon-hole excitations.

\newpage

\begin{figure} 
\centering{\epsfig{figure=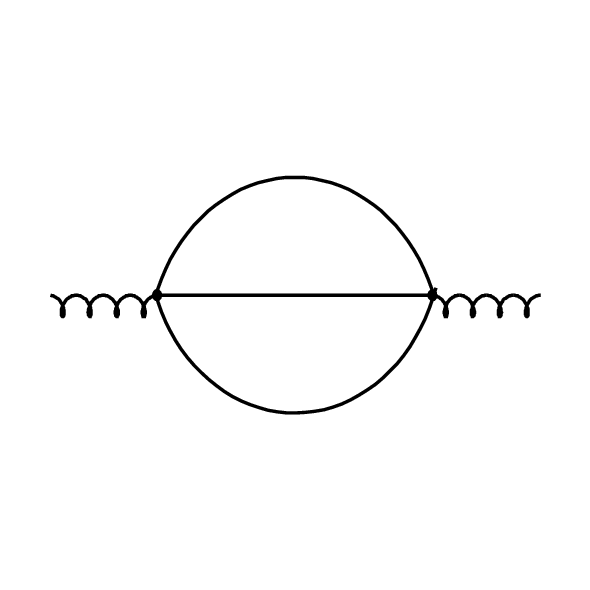,width=4.50cm}
\epsfig{figure=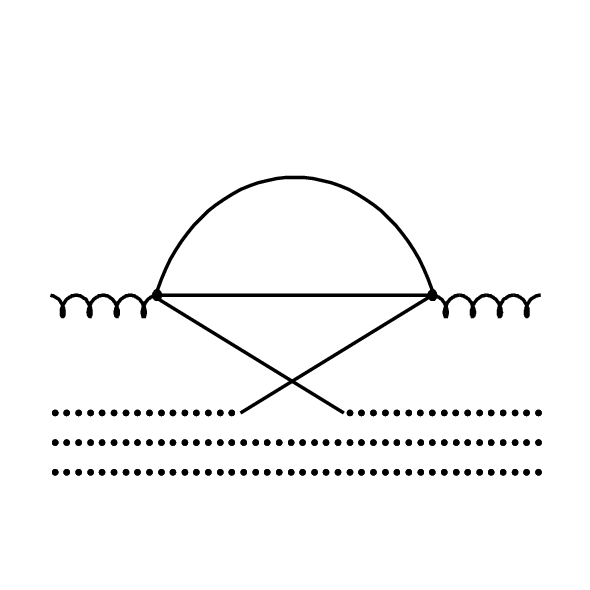,width=4.50cm}
\epsfig{figure=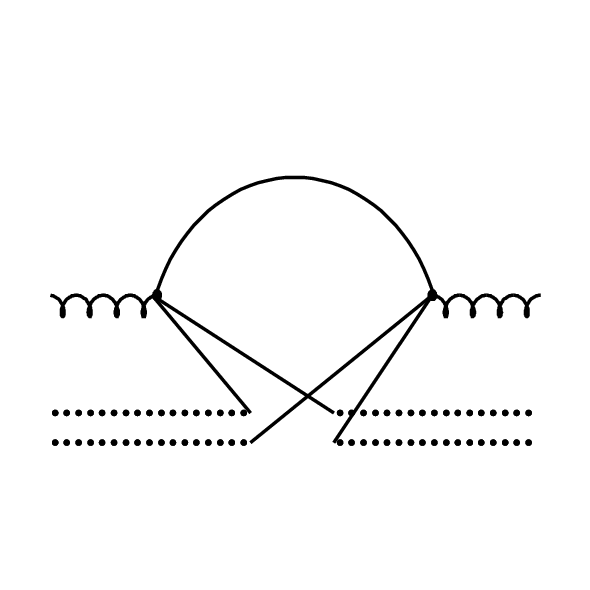,width=4.50cm}}
\vspace{-0.2cm}
{\Large
\hspace{3cm} $a$ \hspace{4cm} $b$ \hspace{4cm} $c$}
\caption{}
\end{figure}

\begin{figure} 
\centering{\epsfig{figure=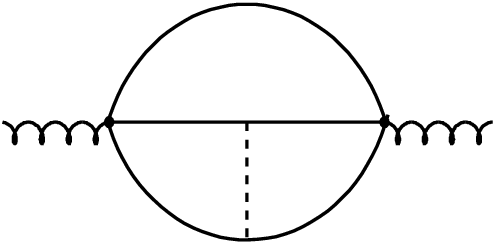,width=4.50cm}}
\caption{}
\end{figure}

\begin{figure} 
\centering{\epsfig{figure=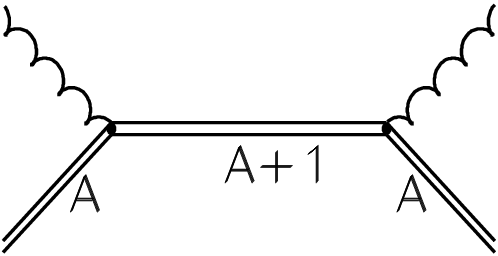,width=5.0cm}
\epsfig{figure=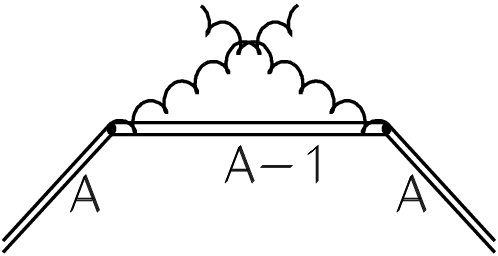,width=5.0cm}}

\vspace{0.5cm}

{\Large
\hspace{-0.5cm} $a$ \hspace{5cm} $b$}
\caption{}
\end{figure}

\begin{figure}
\centering{\epsfig{file=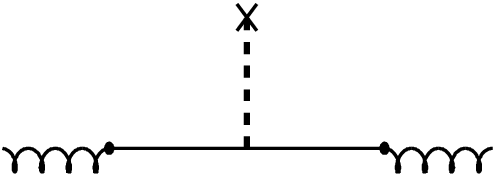,width=5cm}
\epsfig{file=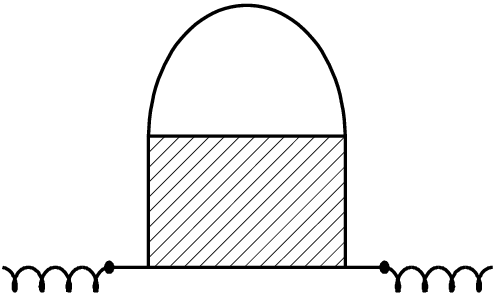,width=5cm}
\epsfig{file=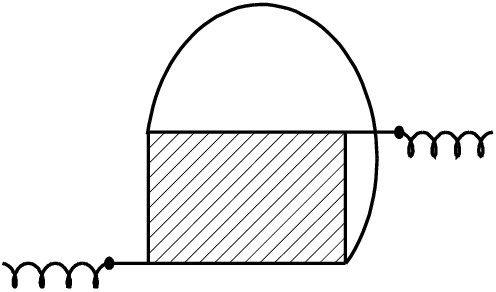,width=5cm}}

\vspace{0.5cm}

{\Large
\hspace{-0.5cm} $a$ \hspace{4cm} $b$  \hspace{4cm} $c$}
\caption{}
\end{figure}

\begin{figure} 
\centering{\epsfig{figure=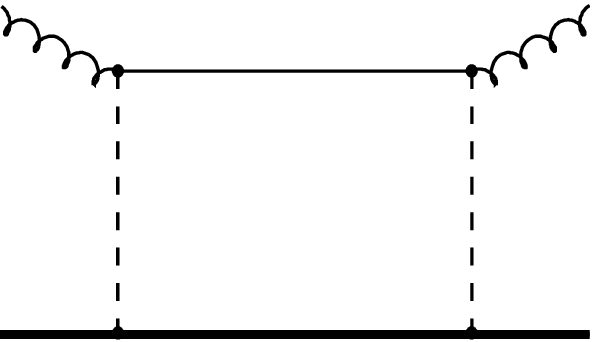,width=5.0cm}}
\caption{}
\end{figure}

\begin{figure} 
\centering{\epsfig{figure=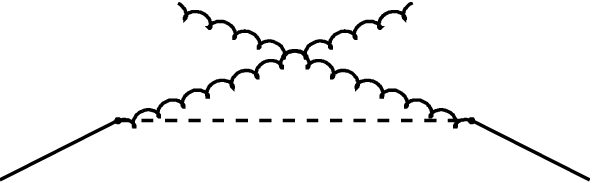,width=6.50cm}}
\caption{}
\end{figure}

\begin{figure}
\centering{\epsfig{file=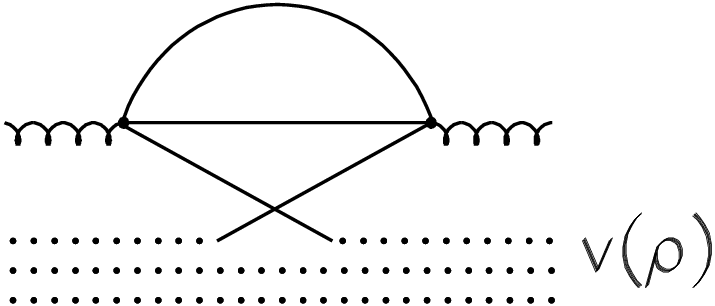,width=6cm}}
\centering{ \epsfig{file=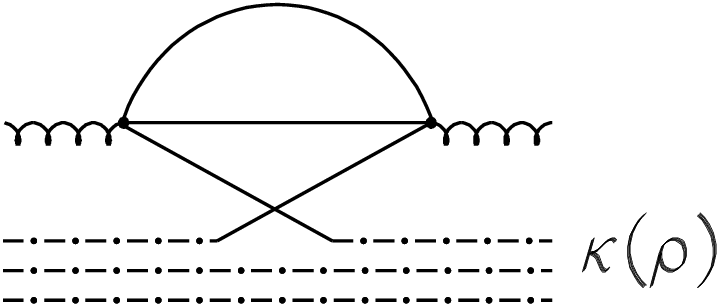,width=6cm}}

\vspace{0.5cm}

{\Large
\hspace{-1cm} $a$ \hspace{5.5cm} $b$}
 \caption{}
\end{figure}

\begin{figure} 
\centering{\epsfig{figure=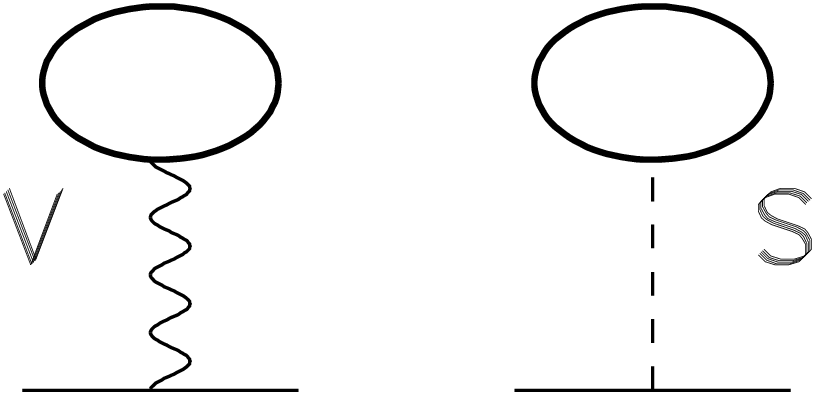,width=6.50cm}}
\caption{}
\end{figure}

\begin{figure} 
\centering{\epsfig{figure=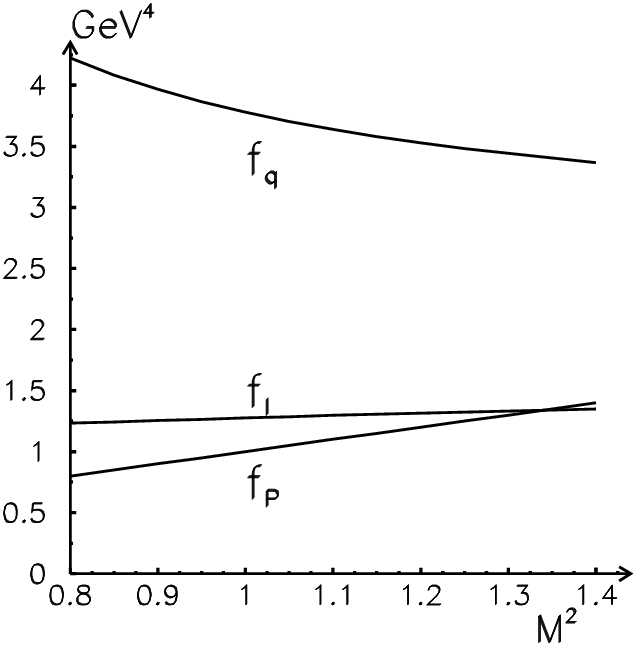,width=7.0cm}}
\caption{}
\end{figure}

\begin{figure} 
\centering{\epsfig{file=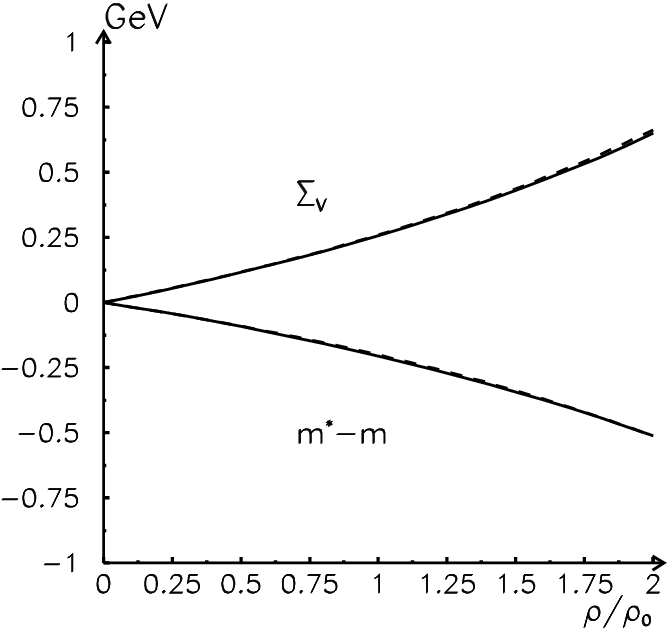,width=6.5cm} \hspace{0.5cm}
\epsfig{file=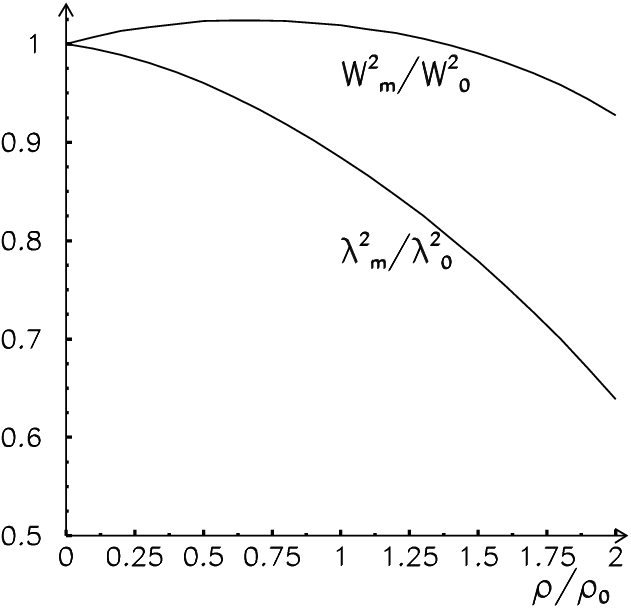,width=6.5cm}}

{\Large
\hspace{1cm} $a$ \hspace{7.0cm} $b$}

\vspace{2cm}

 \caption{}
\end{figure}

\begin{figure} 
\centering{\epsfig{figure=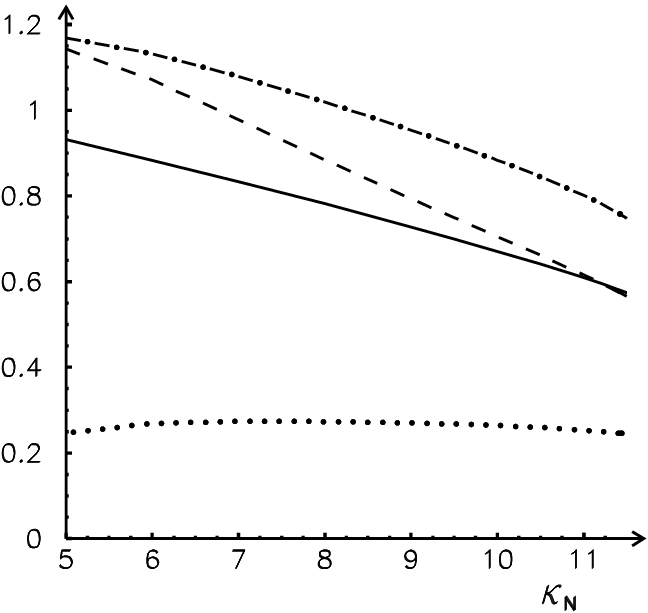,width=8.0cm}}
\caption{}
\end{figure}

\begin{figure} 
\centering{\epsfig{figure=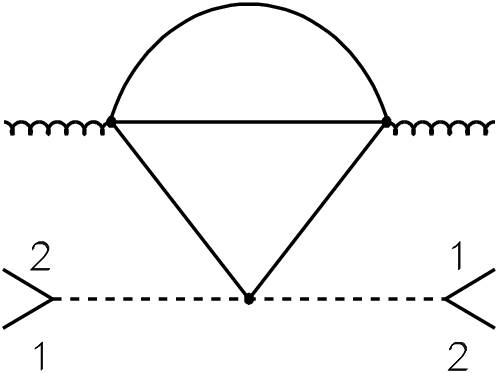,width=6.0cm}}
\caption{}
\end{figure}

\begin{figure} 
\centering{\epsfig{figure=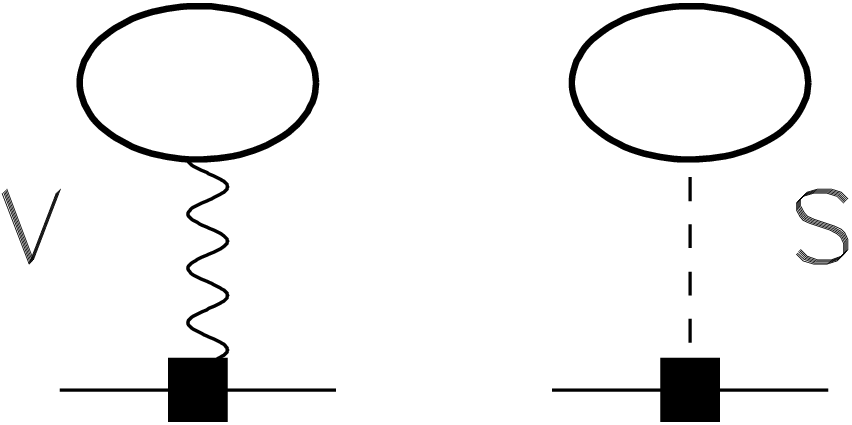,width=6.0cm}}
\caption{}
\end{figure}

\begin{figure} 
\centering{\epsfig{file=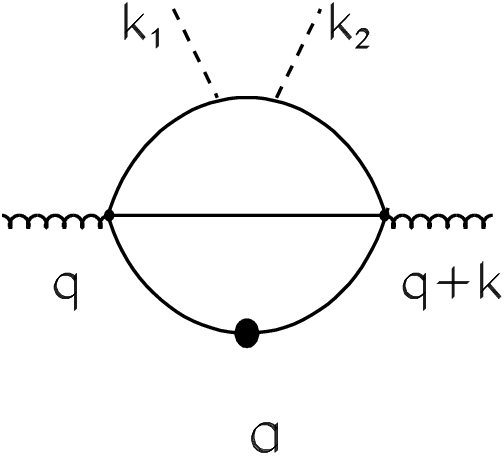,width=5cm} \hspace {0.9cm}
\epsfig{file=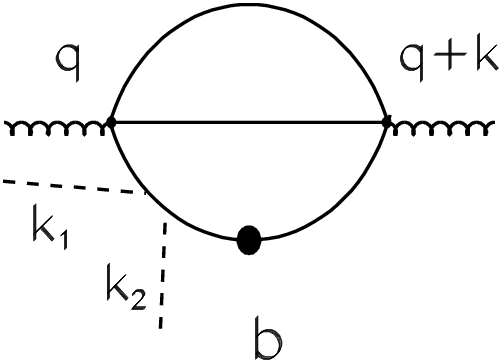,width=5cm}}
 \caption{}
\end{figure}

\begin{figure} 
\centering{\epsfig{figure=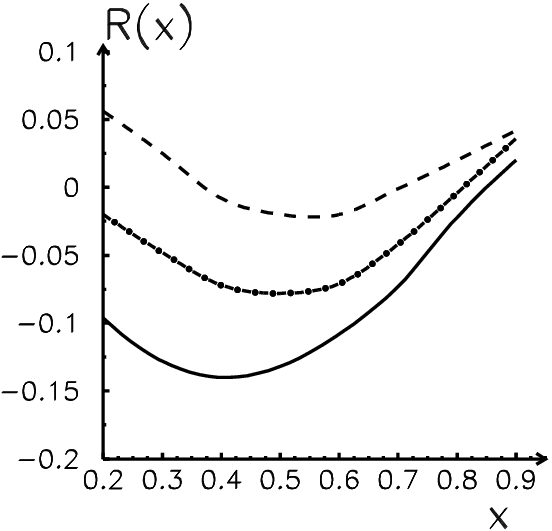,width=6.0cm}}
\caption{}
\end{figure}
\clearpage

\begin{figure} 
\centering{\epsfig{figure=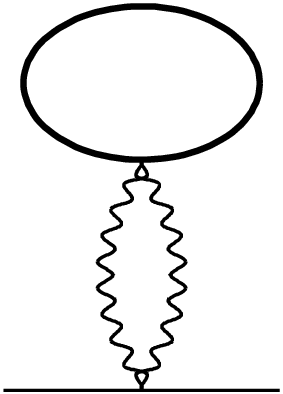,width=4.0cm}}
\caption{}
\end{figure}
\clearpage

\begin{figure} 
\centering{\epsfig{file=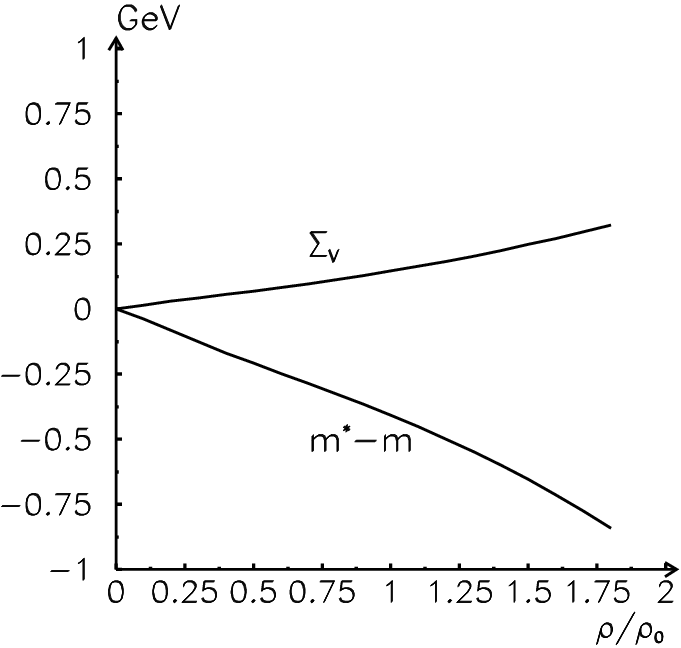,width=6.5cm} \hspace{0.5cm}
\epsfig{file=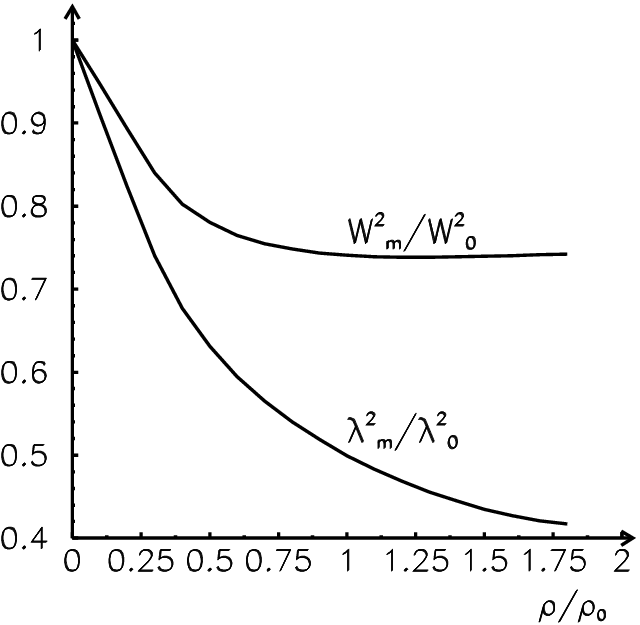,width=6.5cm}}

{\Large
\hspace{0.5cm} $a$ \hspace{7.5cm} $b$}

\vspace{2cm}

 \caption{}
\end{figure}
\clearpage

\begin{figure} 
\centering{\epsfig{figure=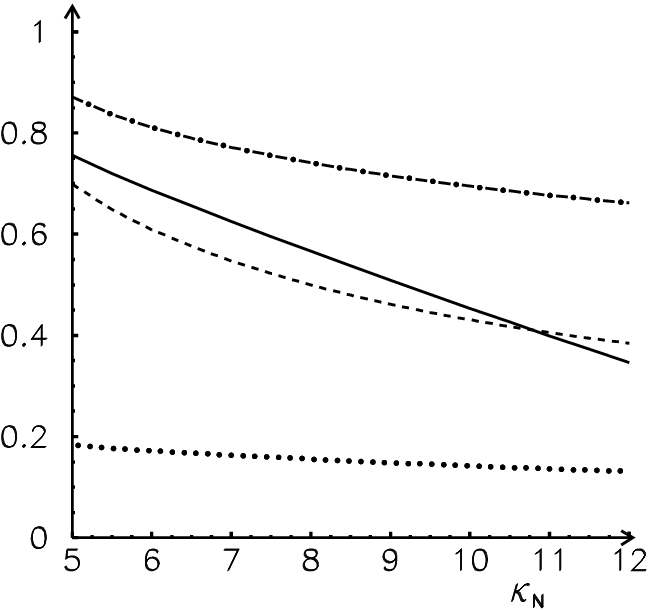,width=8.0cm}}
\caption{}
\end{figure}

\begin{figure} 

\vspace{-1.0cm}

\centering{\epsfig{file=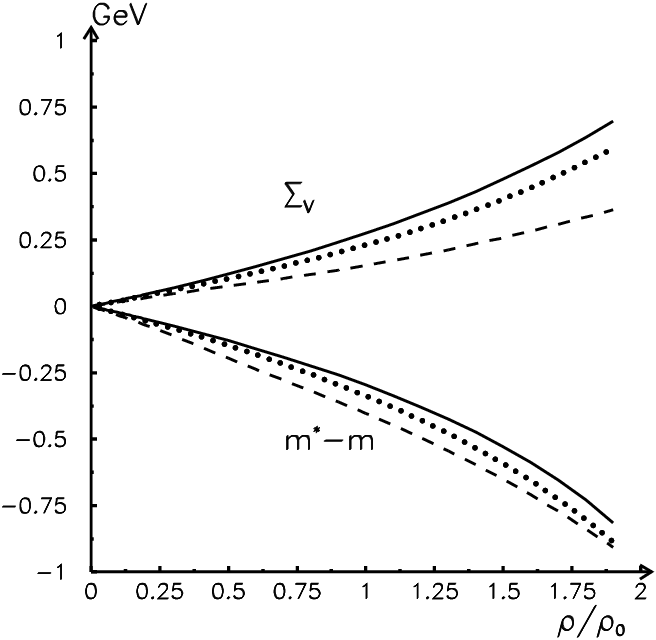,width=9cm}}
 \caption{}
\end{figure}
\clearpage

\begin{figure} 
\centering{\epsfig{file=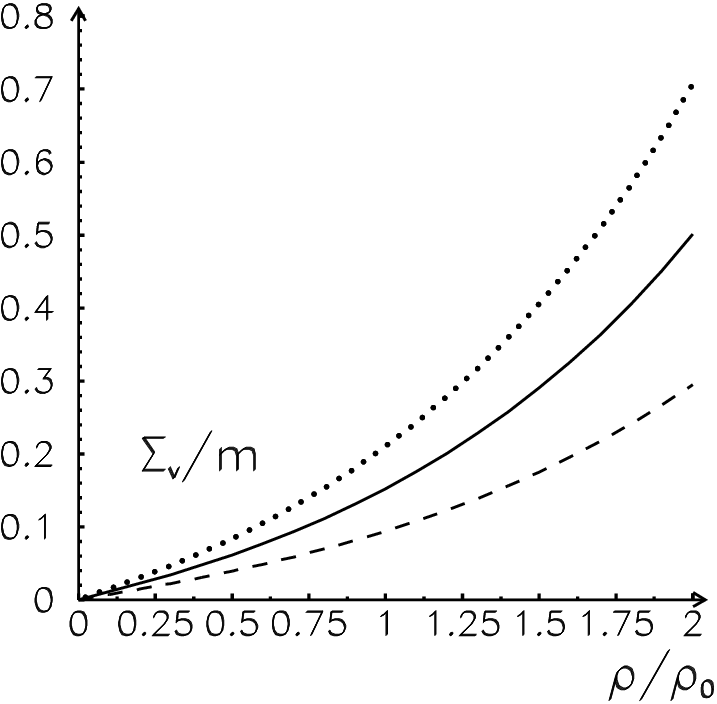,width=6.5cm} \hspace{0.5cm}
\epsfig{file=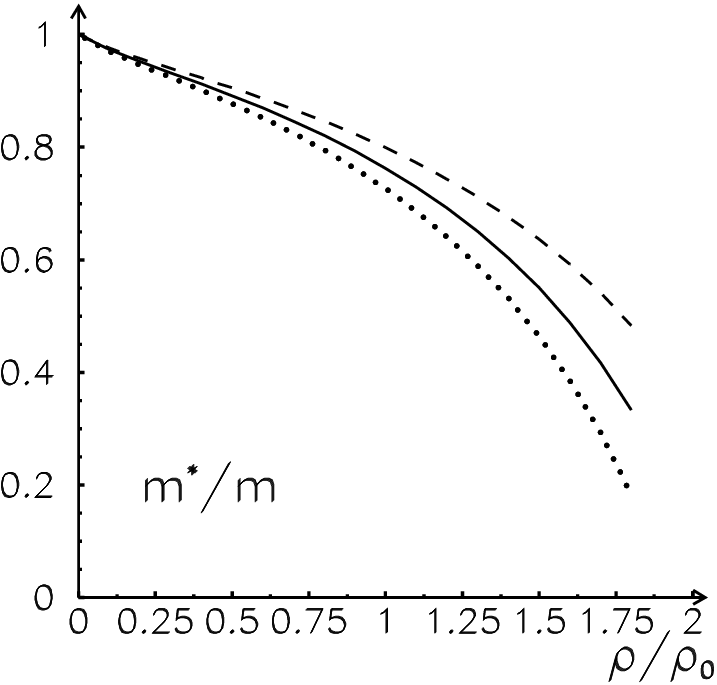,width=6.5cm}}

{\Large
\hspace{-0.5cm} $a$ \hspace{7.5cm} $b$}

 \caption{}
\end{figure}
\clearpage

\begin{figure} 
\centering{\epsfig{file=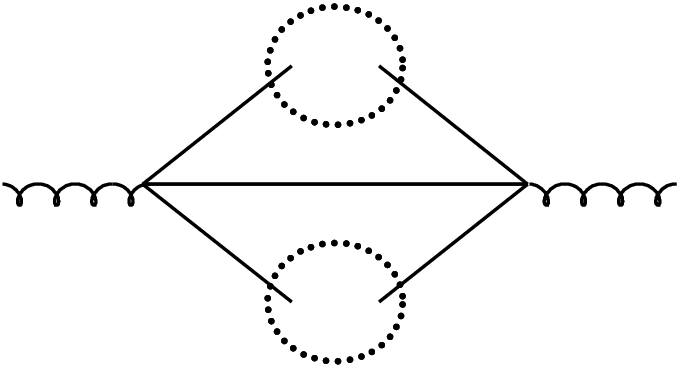,width=8.5cm}}
\caption{}
\end{figure}

\begin{figure} 
\centering{\epsfig{file=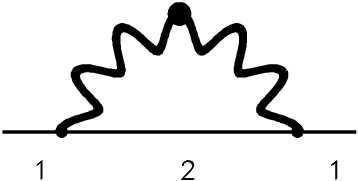,width=4.0cm}}

\vspace{1cm}

\centering{\epsfig{file=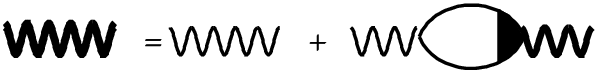,width=6.0cm}}
 \caption{}
\end{figure}
\clearpage


\newpage


\begin{thebibliography}{**}

\bibitem{1} J. D. Walecka, Ann. Phys. {\bf 83}, 491 (1974).

\bibitem{2} B. D. Serot and J. D. Walecka, Adv. Nucl. Phys. {\bf16}, 1 (1985).

\bibitem{3} L. S. Celenza and C. M. Shakin, {\it Relativistic Nuclear
Physics}, World Scientific, Phyladelphya, 1986.

\bibitem{4} R. D. Furnstahl and B. D. Serot, Phys. Rev. C~{\bf 47},
2338 (1993).

\bibitem{5} M. Stoitsov, {\em et al}., J. of Physics, Conference
Series, {\bf 180}, 012082 (2009).

\bibitem{6} J. Dobaczewski, arXiv:1009.0899 v1 [nucl-th] 5 Sept. 2010.

\bibitem{7} J. W. Negele, Comm. Nucl. Part. Phys. {\bf 14}, 1 (1985).

\bibitem{8} L. A. Sliv, M. I. Strikman and L.~L.~Frankfurt, Sov.
Phys.-Uspekhi, {\bf 28}, 281 (1985).

\bibitem{9} M. Ericson, A. Figureau, and C. Thevenet, Phys. Lett
B~{\bf45}, 19 (1973).

\bibitem{10} M. Rho, Nucl. Phys. A {\bf 231}, 493 (1974).

\bibitem{11} J. A. Miller, B. M. K. Nefkens, and I.~Slaus, Phys. Rep.
{\bf 194}, 1 (1990).

\bibitem{12} P. A. M. Guichon, Phys. Lett. B~{\bf 200}, 235 (1988).

\bibitem{13} K. Saito and A. W. Thomas, Phys. Rev. C~{\bf 51}, 2757
(1995).

\bibitem{14} T. Tsushima, K. Saito, A.~W.~Thomas and A.~Valcarce, Eur.
Phys. J.  A~{\bf31}, 626 (2007).

\bibitem{15} F. Myhrer, J. Wroldsen, Rev. Mod. Phys. {\bf 60}, 629 (1988).

\bibitem{16} S. Weinberg, Nucl. Phys. B {\bf 363}, 3 (1991).

\bibitem{17} E. Epelbaum, H. W. Hammer and Ulf-G.~Mei\ss ner, Rev. Mod.
Phys. {\bf 81}, 1773 (2009).

 \bibitem{18} R. Machleidt, arXiv: nucl-th/0609050, nucl-th/07040807;
\\ R. Machleidt and D. R. Eutem, J. Phys. G~{\bf 37}, 064041 (2010).

\bibitem{18a} B. L. Ioffe, V. S. Fadin and L.~N.~Lipatov, {\it Quantum
Chromodynamics}, Cambridge Univ. Press, 2010.

\bibitem{19} M.~A.~Shifman, A.~I.~Vainshtein, and V.~I.~Zakharov, Nucl.
Phys. B~{\bf147}, 385 (1979).

\bibitem{20} B.~L.~Ioffe, Nucl. Phys.  B~{\bf188}, 317 (1981);
E~{\bf191}, 591 (1981).

\bibitem{20a} B.~L.~Ioffe, Prog. Part. Nucl. Phys. {\bf56}, 232
(2006).

\bibitem{21} B.~L.~Ioffe, arXiv:0810.4234; Phys. At. Nucl. {\bf 72},
1214 (2009).


\bibitem{22} E. G. Drukarev and E. M. Levin, Pis'ma ZhETF. {\bf48},
307 (1988); (JETP Lett. {\bf48}, 338 (1988)); ZhETF {\bf95}, 1178 (1989)
(Sov. Phys. JETP {\bf68}, 680 (1989).

\bibitem{23}E. G. Drukarev and E. M. Levin, Nucl. Phys. A~{\bf511}, 679
(1990).

\bibitem{24}E. G. Drukarev and E. M. Levin, Prog. Part. Nucl. Phys.
 {\bf27}, 77 (1991).

\bibitem{25} T. P. Cheng, Phys. Rev. D {\bf 13}, 2161 (1976).

\bibitem{25a} T. P. Cheng and R. Dashen, Phys. Rev. Lett. {\bf 26},
594 (1971).

\bibitem{26} J. Gasser, H. Leutwyler and M.~E.~Sainio,  Phys. Lett.
B~{\bf 253}, 252, 260 (1991).

\bibitem{27} Th. Gutsche and D. Robson, Phys. Lett. B~{\bf229}, 333
(1989).

\bibitem{28} E.~G.~Drukarev, M.~G.~Ryskin, V.~A.~Sadovnikova,
V.~E.~Lyubivitskij, Th.~Gutsche, and A.~Faessler, Phys. Rev. D~{\bf
68}, 054021 (2003).

\bibitem{29} E.~G.~Drukarev, M.~G.~Ryskin, V.~A.~Sadovnikova,
Th.~Gutsche, and A.~Faessler, Phys. Rev. C~{\bf 68}, 065210 (2004).

\bibitem{30} E.~G.~Drukarev, M.~G.~Ryskin and V.~A.~Sadovnikova, Phys.
Rev. C~{\bf 70}, 065206 (2004).

\bibitem{31} V.~A.~Sadovnikova, E.~G.~Drukarev and M.~G.~Ryskin, Phys.
Rev. D~{\bf 72}, 114015 (2005).

\bibitem{32} E.~G.~Drukarev, M.~G.~Ryskin and V.~A.~Sadovnikova, Phys.
Rev. C~{\bf 80}, 045208 (2009).

\bibitem{33} E.~G.~Drukarev and M.~G.~Ryskin, Nucl. Phys. A~{\bf 578},
333 (1994).

\bibitem{34} E.~G.~Drukarev, M.~G.~Ryskin and V.~A.~Sadovnikova, Eur.
Phys. J. {\bf 4}, 171 (1999).

\bibitem{35} E.~G.~Drukarev and E. ~M. ~Levin, Nucl. Phys. A~{\bf 532},
695 (1991).

\bibitem{36} E.~G.~Drukarev and M.~G.~Ryskin, Nucl. Phys. A~{\bf572},
 560 (1994); A~{\bf 577}, 375c (1994).

\bibitem{37} E.~G.~Drukarev and  M.~G.~Ryskin, Z. Phys. A~{\bf356},
 457 (1997).

\bibitem{38} L. J.  Reinders, H. R. Rubinstein, and S.~Yazaki, Phys.
Rep. {\bf 127}, 1 (1985).

\bibitem{39} K. G. Wilson, Phys. Rev. {\bf179}, 499 (1969).

\bibitem{40} B. L. Ioffe, Z. Phys. C~{\bf18}, 67 (1983).

\bibitem{41} B. L. Ioffe and A. V. Smilga, Nucl. Phys. B~{\bf232}, 109
(1984).

\bibitem{42} M. Gell-Mann, R. J. Oakes, and B.~Renner, Phys. Rev.
{\bf175}, 2195 (1968).

\bibitem{43} A. I. Vainshtein, V. I. Zakharov and M. A.~Shifman, JETP
Letters, {\bf27}, 55 (1978).

\bibitem{d1} A. E. Dorohov, N. I. Kochelev, Z. Phys. C~{\bf 46}, 281
(1990).

\bibitem{d2} H. Forkel, M. K. Banerjee, Phys. Rev. Lett. {\bf71}, 484
(1993).



\bibitem{44} V.~A.~Sadovnikova, E.~G.~Drukarev and M.~G.~Ryskin, Yad.
Phys. {\bf71}, 1459 (2008) [Phys. At. Nucl. {\bf71}, 1431 (2008)].

\bibitem{45} E.~G.~Drukarev, M.~G.~Ryskin and V.~A.~Sadovnikova, Phys.
Rev. D~{\bf 80}, 014008 (2009).

\bibitem{46} Y.~Chung, H.~G.~Dosch, M.~Kremer and D.~Schall, Phys.
Lett. B~{\bf 102}, 175 (1981); Nucl. Phys. B~{\bf 197}, 55 (1982).

\bibitem{47} H.~G.~Dosch, M.~Jamin and S.~Narison, Phys. Lett.
B~{\bf220}, 251 (1989).

\bibitem{48} M. E. Peskin, Phys. Lett. B~{\bf88}, 128 (1979).

\bibitem{49} A. A. Ovchinnikov, A.~A.~Pivovarov, and L.~R.~Surguladze,
Int. J. Mod. Phys. A~{\bf6}, 2025 (1991).

\bibitem{50} M. Jamin, Z. Phys. C~{\bf37}, 635 (1988).

\bibitem{51} K. Nakamura, {\em et al.} (Particle Data Group), J. Phys.
G~{\bf 37}, 075502 (2010).

\bibitem{52} D. B.  Leinweber, Ann. Phys.
(NY) {\bf 254}, 328 (1997).

\bibitem{53} E. M.  Henley and J.~Pasupathy, Nucl. Phys. A~{\bf556},
467 (1993).

\bibitem{54} H. Lehmann, Nuovo Cimento, {\bf 11}, 342 (1954).

\bibitem{54a} A. A. Abrikosov, L. P. Gorkov, and I.~E.~Dzyaloshinskii,
``{\it Methods of Quantum Field Theory in Statistical Physics}",
Prentice-Hall, Inc., Englewood Cliffs, N.J. 1963.

\bibitem{55} R. J. Furnstahl, D.~K.~Grigel and T.~D.~Cohen, Phys. Rev.
C~{\bf46}, 1507 (1992).

\bibitem{56} X. Jin, M. Nielsen, T. D. Cohen, R.~J.~Furnstahl and
D.~K.~Grigel, Phys. Rev. C~{\bf 49}, 464 (1994).

\bibitem{56a} X. Jin,  Phys. Rev. C~{\bf 51}, 2260 (1995).

\bibitem{57} X. Jin and R. J. Furnstahl, Phys. Rev. C~{\bf 49}, 464
(1994).

\bibitem{58} T. D. Cohen, R. J. Furnstahl, D.~K.~Grigel and X.~Jin,
Prog. Part. Nucl. Phys. {\bf 35}, 221 (1995).

\bibitem{60} C. J. Horowitz and B. D. Serot, Nucl. Phys. A~{\bf464},
613 (1987).

\bibitem{61} C. Itzykson and J.-B. Zuber, ``{\it Quantum Field Theory}",
(McGraw-Hill, NY, 1980).

\bibitem{62} A. Gl\"{u}ck, E. Reya and A. Vogt, Eur. Phys. J. C~{\bf5},
461 (1998).

\bibitem{63} Yo-xin Liu, Dong-fen Gao, and Hua~Guo, Phys. Rev. C~{\bf
68}, 035204 (2003).

\bibitem{64} E.~G.~Drukarev, M.~G.~Ryskin and V.~A.~Sadovnikova, Prog.
Part. Nucl. Phys. {\bf47}, 73 (2001).

\bibitem{65}J. Gasser and M. E. Sainio, hep-ph/0002283;\\ M. E. Sainio,
hep-ph/0110413; $\pi$N Newsletter {\bf 16}, 138 (2002).

\bibitem{66} R. Koch, Z. Phys. C {\bf 15}, 161 (1982).

\bibitem{67} M.M. Pavan, I. I. Strakovsky, R. L. Workman, and R. A.
Arndt, $\pi$N Newsletter {\bf 16}, 110 (2002).

\bibitem{68} P. Schweitzer, Eur. Phys. J. A~{\bf 22}, 89 (2004).

\bibitem{69} G. E. Hite, W. B. Kaufmann, and R. J. Jacob, Phys. Rev.
C~{\bf71}, 065201 (2005).

\bibitem{70} E. Reya, Rev. Mod. Phys. {\bf 46}, 545 (1974).

\bibitem{71} M. Procura, Th. R. Hemmert and W. Weise, Phys. Rev.
D~{\bf69}, 034505 (2004).

\bibitem{71a} C. C. Barros Jr, and M.~B.~Robilotta, Eur. Phys. J.
C~{\bf45}, 445 (2006).

\bibitem{72} M.~A.~Shifman, A.~I.~Vainshtein, and V.~I.~Zakharov,
 Phys. Lett. B~{\bf78}, 443 (1978).

\bibitem{K1} K. Gottfried, Phys. Rev. Lett. {\bf 18}, 1174 (1967).

\bibitem{K2} A. L. Kataev, arXiv: hep-ph/0311091.

\bibitem{76} S. Weinberg, in ``{\it A Festschrift for I.~I.~Rabi}",
L.~Motz, ed. (NY, 1977).

\bibitem{77} M. Anselmino and S. Forte, Z. Phys. C~{\bf61}, 453 (1994).

\bibitem{78} S. Forte, Phys. Rev. D {\bf 47}, 1842 (1993).

\bibitem{79} A. Bulgac, G. A. Miller, and M.~I.~Strikman, Phys. Rev.
C~{\bf56}, 3307 (1997).

\bibitem{79a} B. L. Birbrair and E. L. Kryshen, Yad. Phys. {\bf72},
1092 (2009); [Phys. At. Nucl. {\bf72}, 1154 (2009)].

\bibitem{80a} M. Lutz, D. Friman, Ch. Appel, Phys. Lett. B~474, 7 (2000).

\bibitem{80}E.~G.~Drukarev, M.~G.~Ryskin and V.~A.~Sadovnikova, Z.
Phys. A~{\bf353}, 455 (1996).

\bibitem{80b} F. Tondeur, Nucl. Phys. A~{\bf303}, 185 (1978).

\bibitem{81} V. M. Belyaev and Ya.~I.~Kogan, Phys. Lett. B~{\bf136}, 273
(1984).

\bibitem{82} C. B. Chiu, J.~Pasupathy and S.~J.~Wilson, Phys. Rev.
D~{\bf 32}, 1786 (1985).

\bibitem{83} E. M. Henley, W-Y. P. Hwang and L.~S.~Kisslinger,  Phys.
Rev. D~{\bf 46}, 431 (1992).

\bibitem{84} V. A. Novikov, M.~A.~Shifman, A.~I.~Vainshtein,
M.~B.~Voloshin and V.~I.~Zakharov, Nucl.
Phys. B~{\bf237}, 525 (1984).

\bibitem{85} B. Parthasarathy and J. Pasupathy,  Phys. Rev. D~{\bf 37},
2140 (1988).

\bibitem{86} D. H. Wilkinson, Phys. Rev. C {\bf 7}, 930 (1973).

\bibitem{87} G. D. Alkhazov, S. A. Artamonov, V.~I.~Isakov,
K.~A.~Mezilev and Yu.~N.~Novikov, Phys. Lett. B~{\bf 198}, 37 (1987).

\bibitem{88} S. Shlomo, Rep. Prog. Phys. {\bf41}, 957 (1978).

\bibitem{89} J. A. Nolen, J. P. Schiffer,  Annu. Rev. Nucl. Part. Sci.
{\bf19}, 471 (1969).

\bibitem{90} E. M. Henley and G. Krein, Phys. Rev. Lett. {\bf62}, 2586
(1989).

\bibitem{91} U. G. Mei\ss ner and H. Weigel,  Phys. Lett. {\bf267},
167 (1991).

\bibitem{92} K. Saito and A. W. Thomas,  Phys. Lett. B~{\bf 335}, 17
(1994).

\bibitem{93} U. G. Mei\ss ner, A. M. Rakhimov, A.~Wirzba and
U.~T.~Yakshiev, Eur. Phys. J.  A~{\bf31}, 357 (2007).

\bibitem{94} U. G. Mei\ss ner, A. M. Rakhimov, A.~Wirzba and
U.~T.~Yakshiev, Eur. Phys. J.  A~{\bf36}, 37 (2008).

\bibitem{95} D.~Espiru, P.~Pascual and R.~Tarrach, Nucl. Phys.
B~{\bf214}, 285 (1983).

\bibitem{96} C. Adami, E. G. Drukarev, and B. l. Ioffe,  Phys. Rev.
D~{\bf 48}, 2304 (1993).

\bibitem{96a} K. C. Yang, W-Y.~P.~Hwang, E.~M.~Henley, and
L.~S.~Kisslinger,  Phys. Rev. D~{\bf 47}, 3001 (1993).

\bibitem{97} T. Hatsuda, H. H{\o}gaasen and M.~Prakash, Phys. Rev.
Lett. {\bf 66}, 2851 (1991).

\bibitem{98} C. Adami and G. Brown, Z. Phys. A~{\bf340}, 93 (1991).

\bibitem{99} T. Schafer, V. Koch, and G. Brown, Nucl. Phys. A~{\bf
562}, 644 (1993).

\bibitem{100} A. V. Kolesnichenko, Sov. J. Nucl. Phys. {\bf 39}, 968 (1982).

\bibitem{101} V. M. Belyaev and B. Yu. Blok, Z. Phys. C.~{\bf30}, 279
(1986).

\bibitem{102} V. M. Braun and A. V. Kolesnichenko, Nucl. Phys.
B~{\bf283}, 723 (1987).

\bibitem{103} V. M. Belyaev and B. L. Ioffe, Nucl. Phys. B~{\bf310},
548 (1988).

\bibitem{104} V. Braun, P. Gornicki, and L. Mankiewicz, Phys. Rev.
D~{\bf 51}, 6036 (1995).

\bibitem{105} J. J. Aubert, Phys. Lett. B~{\bf123}, 275 (1983).

\bibitem{106} J. Seely {\em et al}., Phys. Rev. Lett. {\bf103}, 202301
(2009).

\bibitem{107} D. W. Higinbotham, J. Gomez, E.~Piasetsky, ArXiv:
1003.4497 v2 [hep-ph].

\bibitem{108} R. L. Jaffe, Phys. Rev. Lett. {\bf 50}, 228 (1983).

\bibitem{109} L. Frankfurt and M. Strikman, Phys. Rep. {\bf 160}, 235 (1988).

 \bibitem{110} V. L. Eletsky and B. L. Ioffe, Phys. Rev. Lett. {\bf
78}, 1010 (1997).

\bibitem{111} L. S. Celenza, C. M. Shakin, W. D. Sun, and J. Szweda,
Phys. Rev. C~{\bf51}, 937 (1995).

\bibitem{112} V. E. Lyubovitskij, Th. Gutsche, and A. Faessler, Phys.
Rev. C~{\bf64}, 065203 (2001).

\bibitem{113} Th. Gutsche, V. E. Lyubovitskij, and A. Faessler, Prog.
Part. Nucl. Phys. {\bf50}, 235 (2003).

\bibitem{114} V. E. Lyubovitskij, Th. Gutsche, A. Faessler, and E. G.
Drukarev, Phys. Rev. D~{\bf63}, 054026 (2001).

\bibitem{115} E.~G.~Drukarev, M.~G.~Ryskin, V.~A.~Sadovnikova, and A.
Faessler,  Phys. Rev. D~{\bf65}, 074015 (2002); 66, 039903 (E).

\bibitem{116} D. P. Roy, J. Phys. G~{\bf 30}, R113 (2004).

\bibitem{123} S. Groote, J. G. K\"{o}rner,  and A. A. Pivovarov,  Phys.
Rev. D~{\bf78}, 034039 (2008).

\bibitem{117} V. Greco, M. Colonna, M.~Di~Toro, G.~Fabbri, and
E.~Matera,  Phys. Rev. C~{\bf64}, 045203 (2001).

\bibitem{118} F. de Jong and H. Lenske,  Phys. Rev. C~{\bf 57}, 3099
(1998).

\bibitem{119} H. Huber, F. Weber, and M. K. Weigel, Phys. Rev. C~{\bf
51}, 1790 (1995).

\bibitem{120} I. Bombaci and U. Lombardo, Phys. Rev. C~{\bf 44}, 1892
(1991).

\bibitem{121} W. Zuo, I. Bombaci and U. Lombardo, Phys. Rev. C~{\bf 60},
024605 (1999).

\bibitem{122} Chen Lie Wen, arhiv: 0910.008 v1 [nucl-th].

\bibitem{124} Huang Dong, T. T. S. Kuo, and R. Machleidt, Phys. Rev.
C~{\bf80}, 065803 (2009).

\bibitem{125} K. Hebeler and A. Schwenk, Phys. Rev. C~{\bf 82}, 014314
(2010).

\bibitem{126} M. L. Goldberger and S. B. Treiman, Phys. Rev.
{\bf110}, 1478 (1958).

\bibitem{127} A. B. Migdal,
{\it Theory of Finite Fermi Systems and Application to
Atomic Nuclei}, Wiley, NY, 1969.

\bibitem{137}
A. B. Migdal, D. N. Voskresensky, E.~E.~Saperstein, M.~A.~Troitsky,
Phys. Rep. {\bf192}, 179 (1990);\\
A. B. Migdal, D. N. Voskresensky, E.~E.~Saperstein, M.~A.~Troitsky,
{\it Pion degrees of freedom in nuclear medium}, Nauka, Moscow, 1991.

\end{thebibliography}
\end{document}